\documentclass[aps,superscriptaddress,showpacs,nofootinbib,11pt]{revtex4-1}
\usepackage{mathrsfs}
\usepackage{bigints}
\usepackage{latexsym,bm}
\usepackage{graphicx}
\usepackage{indentfirst}
\usepackage{slashed}
\usepackage{amssymb}
\usepackage{amsmath}
\usepackage{bbm}
\usepackage{color}
\usepackage[dvipsnames]{xcolor}
\usepackage{epsfig}
\usepackage[titletoc]{appendix}
\usepackage{rotating}
\usepackage{epstopdf}
\usepackage{extarrows}
\usepackage{tabularx}
\usepackage{soul} 
\usepackage{xcolor}
\usepackage{lipsum}
\usepackage[colorlinks=true,allcolors=BlueViolet,hyperfootnotes=false]{hyperref}

\usepackage[utf8]{inputenc}

\newcommand{\be}{\begin{equation}} \newcommand{\ee}{\end{equation}}
\newcommand{\ba}{\begin{array}{c}} \newcommand{\ea}{\end{array}}
\newcommand{\bea}{\begin{eqnarray}} \newcommand{\eea}{\end{eqnarray}}

\newcommand{\mbs}{M_{B_s}}
\newcommand{\mds}{M_{D_s}}
\newcommand{\mdss}{M_{D_s^*}}

\newcommand\tth{t_\text{th}}

\newcommand\zhpqcd{\tilde z}

\newcommand\tstrut{\rule{0pt}{2.9ex}}       

\begin{document}

\title{\Large Study of new physics effects in $\bar B_s\to D^{(*)}_s\tau^-\bar\nu_\tau$ semileptonic decays 
using lattice QCD form factors and heavy quark effective theory}

\author{Neus Penalva}
\affiliation{Instituto de F\'{\i}sica Corpuscular (centro mixto CSIC-UV), 
Institutos de Investigaci\'on de Paterna,
C/Catedr\'atico Jos\'e Beltr\'an 2, E-46980 Paterna, Valencia, Spain}

\author{Jonathan M. Flynn}\affiliation{Physics \& Astronomy,
  University of Southampton, Southampton SO17 1BJ, UK}

\author{Eliecer Hern\'andez}
\affiliation{Departamento de F\'\i sica Fundamental 
  e IUFFyM,\\ Universidad de Salamanca, Plaza de la Merced s/n, E-37008 Salamanca, Spain}
\author{ Juan Nieves}
\affiliation{Instituto de F\'{\i}sica Corpuscular (centro mixto CSIC-UV), 
Institutos de Investigaci\'on de Paterna,
C/Catedr\'atico Jos\'e Beltr\'an 2, E-46980 Paterna, Valencia, Spain}


\date{\today}
%
\begin{abstract}
We benefit from the lattice QCD determination by the HPQCD  of the Standard Model (SM) form 
factors for the  $\bar B_s\to D_s$ [Phys. Rev. D 101, 074513 (2020)] and the SM and tensor ones 
for the $\bar B_s\to D_s^{*}$ (arXiv:2304.03137 [hep-lat]) semileptonic decays, and the heavy quark effective theory (HQET) relations
 for the analogous $B\to D^{(*)}$ decays obtained by F.U. Bernlochner et al. in Phys. Rev. D 95, 115008 (2017),
  to extract the leading and sub-leading Isgur-Wise functions for the
$\bar B_s\to D_s^{(*)}$ decays. Further use of the HQET relations allows us to evaluate the corresponding scalar, pseudoscalar and tensor form factors needed for a phenomenological study of new physics (NP) effects on the $\bar B_s\to D_s^{(*)}$ semileptonic decay. 
At present, the experimental values for the ratios ${\cal R}_{D^{(*)}}=\Gamma[\bar B\to D^{(*)}\tau^-\bar\nu_\tau]/\Gamma[\bar B\to
D^{(*)}e^-(\mu^-)\bar\nu_{e(\mu)}]$ are the best signal in favor of lepton flavor universality violation (LFUV) seen in charged current (CC) $b\to c$ decays.  In this work
we conduct a study of NP effects on the $\bar B_s\to D_s^{(*)}\tau^-\bar\nu_\tau$ semileptonic decays by comparing  tau spin, angular and spin-angular asymmetry distributions obtained within the SM and three different NP scenarios. As expected from SU(3) light-flavor symmetry, we get  results close to the ones found in a similar analysis of the $\bar B\to D^{(*)}$ case. The measurement of the $\bar B_s\to D_s^{(*)}\ell\bar\nu_\ell$ semileptonic decays, which is within reach of present experiments,   could  then be of relevance in helping to establish or rule out LFUV in CC $b\to c$ transitions. 
\end{abstract} 

%


\maketitle
\section{Introduction}
%

Present experimental data on $\bar B\to D^{(*)}$ semileptonic decays points to the possibility of lepton flavor universality violation (LFUV) 
 that will affect charged-current (CC) $b\to c\tau^-\bar\nu_\tau$ semileptonic transitions. The ratios ${\cal R}_{D}=\Gamma(\bar B\to D\tau^-\bar\nu_\tau)/\Gamma(\bar B\to D\mu^-\bar\nu_\mu)$ and  ${\cal R}_{D^*}=\Gamma(\bar B\to D^*\tau^-\bar\nu_\tau)/\Gamma(\bar B\to D^*\mu^-\bar\nu_\mu)$ have been measured by the 
BaBar~\cite{BaBar:2012obs, BaBar:2013mob}, Belle~\cite{Belle:2015qfa, 
Belle:2016ure,Belle:2016dyj,Belle:2019rba} and LHCb~\cite{LHCb:2015gmp, 
LHCb:2017smo,LHCb:2017rln,LHCb:2023zxo} experiments and their combined analysis by the HFLAV
collaboration indicates a $3\sigma$ tension with SM 
predictions~\cite{HFLAV:2019otj,LHCbseminar}. 

 LFUV requires the existence of new physics (NP) beyond the Standard Model (SM) and, if confirmed, would have a tremendous impact in particle physics. This  makes the study of as many analogous CC decays as possible timely and necessary in order to confirm or rule out LFUV. 
 The 
${\cal R}_{J/\psi}=
  \Gamma(\bar B_c\to J/\psi\tau^-\bar\nu_\tau)/\Gamma(\bar B_c\to
   J/\psi\mu^-\bar\nu_\mu)$ ratio has  been measured by the LHCb 
   collaboration~\cite{LHCb:2017vlu} finding a $1.8\,\sigma$  discrepancy with SM 
   results~\cite{Anisimov:1998uk,Ivanov:2006ni,
Hernandez:2006gt,Huang:2007kb,Wang:2008xt,Wen-Fei:2013uea, Watanabe:2017mip, Issadykov:2018myx,Tran:2018kuv,
Hu:2019qcn,Leljak:2019eyw,Azizi:2019aaf,Wang:2018duy}. 
 Another reaction where a similar behavior was to be expected is
 the baryon $\Lambda_b \to \Lambda_c\ell\bar\nu_\ell$ decay. However, in this case, the recent measurement of the 
${\cal
R}_{\Lambda_c}=\Gamma(\Lambda_b\to\Lambda_c\tau^-\bar\nu_\tau)/
\Gamma(\Lambda_b\to\Lambda_c\mu^-\bar\nu_\mu)$ ratio by the LHCb 
collaboration~\cite{LHCb:2022piu} is
in agreement,  within errors,
with the SM prediction~\cite{Detmold:2015aaa}. In this experiment, the $\tau^-$
lepton was reconstructed using  the  $\tau^-\to\pi^-\pi^+\pi^-(\pi^0)\nu_\tau$ hadronic decay. It is then of great interest to see whether the current ${\cal R}_{\Lambda_c}$ experimental value is confirmed or not using the muonic reconstruction channel. Such an analysis is under way~\cite{Marco}.

LHCb has very recently~\cite{LHCb:2023zxo} presented the
first simultaneous measurement in hadron collisions of ${\cal R}_{D^*}$ and ${\cal R}_{D^0}$, identifying the tau lepton from its the decay mode $\tau^-\to \mu^-\nu_\tau\bar\nu_\mu$. The measured values
are ${\cal R}_{D^*}=0.281\pm 0.018\pm 0.024$ and ${\cal R}_{D^0}=0.441\pm 0.060\pm 0.066$, where the correlation between
these measurements is $-0.43$. The result for the former ratio supersedes the higher value previously reported in~\cite{LHCb:2015gmp} and it is now in better agreement with the SM. LHCb earlier measured ${\cal R}_{D^*}= 0.291 \pm 0.019 \pm  0.026 \pm  0.013$ \cite{LHCb:2017smo,LHCb:2017rln} using hadronic tau decays, but a new result was reported in \cite{LHCb:2023cjr}, ${\cal R}_{D^*}=0.257 \pm 0.012 \pm  0.014 \pm  0.012$, obtained after combining the previous results from  Refs.~\cite{LHCb:2017smo,LHCb:2017rln} and a new one from a partial Run 2 data sample~\cite{LHCb:2023cjr}. The final LHCb result  for ${\cal R}_{D^*}$ from hadronic tau decays is in closer agreement with the SM expectation.  Nevertheless combined global results for ${\cal R}_{D^*}$ and ${\cal R}_{D}$ from different experiments and detection techniques remain around $3\sigma$ away from the SM expectation (HFLAV Winter 2023 update~\cite{HFLAV:2022pwe} presented in \cite{LHCbseminar}).

One would also expect to see LFUV effects in  $\bar B_s\to D_s^{(*)}$ semileptonic decays which are SU(3) analogues of the $\bar B\to D^{(*)}$ ones. A measurement of ${\cal R}_{D_s}$ by LHCb~\cite{LHCbseminar} is also underway, making the study of these reactions timely. The theoretical analysis of NP effects in those decays requires however knowledge of beyond-the-SM (BSM) form factors that have not   yet been determined. The HPQCD lattice QCD (LQCD) collaboration has evaluated the SM form factors for the $\bar B_s\to D_s$ and $\bar B_s\to D_s^{*}$ semileptonic transitions in Refs.~\cite{McLean:2019qcx} and \cite{Harrison:2021tol}, respectively. More recently, HPQCD  has given  updated  $\bar B_s\to D_s^*$  SM form-factors~\cite{Harrison:2023dzh}. This latter work also provides the form factors that expand the matrix elements  of the NP $\bar c \sigma^{\mu\nu} b$  operator for initial $\bar B_s$ and final $D_s^*$ states.

 On the other hand, the approximate heavy quark spin symmetry (HQSS) of QCD  allows one to construct an effective field theory (HQET) to compute these form-factors. Indeed, the HQET expressions for them can be obtained up to next-to-leading (NLO)  ${\cal O}(\alpha_s,
\Lambda_{\rm QCD}/m_{c,b})$ and  next-to-next-to-leading (NNLO) ${\cal O}(\alpha_s\Lambda_{\rm QCD}/m_{c,b},
\Lambda^2_{\rm QCD}/m^2_{c,b})$ orders from Refs.~\cite{Bernlochner:2017jka} and \cite{Bernlochner:2022ywh}, respectively\footnote{The partial NNLO ${\cal O}(\Lambda^2_{\rm QCD}/m^2_c)$ corrections were previously studied in  refs.~\cite{Bordone:2019vic,Bordone:2019guc}.  }. 
One can use this information to fit the leading and sub-leading HQSS Isgur-Wise (IW) functions, which describe the $\bar B_s\to D_s^{(*)}$  form factors, to the lattice data of Refs.~\cite{McLean:2019qcx} and \cite{Harrison:2023dzh}. First, from the comparison of results to those available for $\bar B \to D^{(*)}$ decays, one could in principle estimate the size of the SU(3) light-flavor breaking corrections.  Second, and more interesting,  once the IW functions are known, the scalar, pseudoscalar and tensor\footnote{As already mentioned, LQCD tensor form factors were computed in \cite{Harrison:2023dzh} for the $\bar B_s \to D_s^{*}$ transition. However, there is no LQCD input on the tensor matrix element in the case of the $\bar B_s \to D_s$ decay mode.  } form factors that are needed, in addition to the SM ones, for an analysis of possible NP effects on the $\bar B_s\to D_s^{(*)}\tau\bar\nu_\tau$ decays can be obtained from their HQET expressions.   Thus,  we will show results for  tau spin, angular and spin-angular asymmetry distributions for these decays obtained within the SM and three different NP scenarios, and analyze the role that different tau asymmetries in the $\bar B_s\to D_s^{(*)}\tau^-\bar\nu_\tau$ decay could play,  not only in establishing the existence of NP, but also in distinguishing between different NP extensions of the SM. We will also study partially integrated  angular and energy distributions of the charged particle produced in the subsequent $\tau^-
 \to\pi^-\nu_\tau,\,\rho^-\nu_\tau,e^-(\mu^-)\bar\nu_{e(\mu)}\nu_\tau$  decays. The latter differential decay widths have a better statistics than the asymmetries themselves and they could also help in establishing the presence of NP beyond the SM.

 The $\bar B_s\to D^{(*)}_s$ form-factors have also been  studied 
using HQET and sum rules in \cite{Bordone:2019guc}. In that work, additional constraints are found which allow the authors to go beyond the assumption of $SU(3)$ flavor symmetry, and SM lepton-flavour universality ratios are reported. NP effects on both $\bar B_s\to D_s$  and $\bar B_s\to D^{*}_s$ decays have been discussed in \cite{Dutta:2018jxz,Das:2021lws} using the SM LQCD form-factors computed in Refs.~\cite{Monahan:2017uby} and \cite{Harrison:2021tol}, respectively. The works of Refs.~\cite{Dutta:2018jxz,Das:2021lws} do not consider NP tensor operators and make use of the equations of motion to estimate the NP scalar and pseudo-scalar form-factors. The distribution of the tau-decay products is not studied in either of the two works and hence they do not have access to the full set of  tau angular, 
tau spin and tau angular-spin asymmetries that can be extracted  by measuring 
the $\tau$ in a general polarization state.  Moreover those papers do not address the distribution of the  tau-decay products. However, different angular distributions from the decay products of the outgoing $D_s^*$ are studied in \cite{Das:2021lws}.

In Ref.~\cite{Bernlochner:2022ywh}, the NNLO corrections were computed  introducing (postulating) a supplemental power counting within HQET. The authors of that work claimed that the the postulated truncation leads to
 small, highly constrained set of second-order power corrections, compared to the standard approach. Nevertheless, 
there appears a plethora of free parameters, a number considerably larger than in the NLO case of  Ref.~\cite{Bernlochner:2017jka}. Though, it seems the scheme followed in  Ref.~\cite{Bernlochner:2022ywh} provides excellent fits to the available $\bar B\to D^{(*)}$ LQCD predictions and experimental data, we have found that the NNLO parameters cannot be determined reliably from the available $\bar B_s\to D_s$ and $\bar B_s\to D_s^{*}$ LQCD form-factor data, given the statistical and systematic precision with which they are currently obtained. For that reason, we will limit this work to NLO HQET corrections, except in the case of the form factors that are protected from
${\cal O}(\Lambda_{\rm QCD}/m_c)$ corrections at zero recoil~\cite{Luke:1990eg}, for which we will  include NNLO
${\cal O}[\Lambda_{\rm QCD}^2/m_c^2]$ terms. These form factors do not vanish in the heavy quark limit and turn out to be the best determined in the LQCD simulations, in particular near zero recoil, making it necessary to consider some sub-leading corrections in addition to those induced by short-distance physics. 

This work is organized as follows. In Sec.~\ref{sec:SMHQSS} we describe the fitting procedure to obtain the IW functions, with some auxiliary details collected in the Appendix.
  A thorough analysis of NP effects, based on observables that can be measured by the analysis of the visible kinematics of the subsequent hadronic $\tau^-\to
  \pi^-\nu_\tau, \tau^-\to \rho^-\nu_\tau$ and leptonic $\tau^-\to \ell^-\bar\nu_\ell\nu_\tau$ decays, is conducted in 
  Sec.~\ref{sec:NP}. Finally in Sec.~\ref{sec:summary} we summarize the main findings.

\section{HQET fit of the $\bar B_s\to D^{(*)}_s$ semileptonic-decay LQCD form factors and SM distributions}
\label{sec:SMHQSS}
In this section we will describe how we fit the LQCD form-factor data   from
 Refs.~\cite{Harrison:2023dzh,McLean:2019qcx} to their expressions deduced from 
   NLO HQET and derived in Ref.~\cite{Bernlochner:2017jka}. A comparison of both sets of form factors will be shown below in Fig.~\ref{fig:dsdsstarcomp}.  We will also show the SM predictions from both sets  for differential decay widths and tau spin, angular and spin-angular asymmetry distributions. Further use of HQSS will allow us  to predict BSM form factors not evaluated in the lattice, and that  are needed  to test  possible NP effects in $ \bar B_s\to D^{(*)}_s\tau^-\bar\nu_\tau$ semileptonic decays, something we will do in the next section.
\subsection{LQCD form factors}

We will use the LQCD results from HPQCD for the SM form factors of the $\bar B_s\to
D_s$ decay~\cite{McLean:2019qcx}  and the SM and tensor form factors of the  $\bar
B_s\to D^*_s$ decay~\cite{Harrison:2023dzh}.
\subsubsection{$\bar B_s\to
D_s$}
For the $\bar B_s\to D_s$ semileptonic decay, the form-factor
decomposition in Ref.~\cite{McLean:2019qcx} is
\be
\langle D_s;\vec p\,'\,|\bar c(0)\gamma^\mu b(0)|\bar B_s;\vec p\,\rangle=
f_+(q^2)\Big[p^\mu+p^{\prime\mu}-\frac{\mbs^2-\mds^2}{q^2}\,q^\mu\Big]
+f_0(q^2)\frac{\mbs^2-\mds^2}{q^2}\,q^\mu,
\ee
with the constraint
\be
f_0(0)=f_+(0).
\label{eq:dsconst}
\ee
The form factors are parametrized  as~\cite{McLean:2019qcx}
\bea
f_0(q^2)&=&\frac1{1-q^2/M^2_{B_{c0}}}\sum_{n=0}^2
 \tilde a^0_n \zhpqcd^n,\nonumber\\
f_+(q^2)&=&\frac1{1-q^2/M^2_{B^*_{c}}}\sum_{n=0}^2
 \tilde a^+_n \Big[\zhpqcd^n
-\frac{(-1)^{n-3}n}{3}\zhpqcd^3\Big],
\eea
where $q^\mu$ is the four-momentum transfer to the leptons and 
\be
\zhpqcd(q^2)= z(q^2;\tth,0), \qquad \tth=(\mbs+\mds)^2.
\ee
where
\be
z(q^2;\tth,t_0)=\frac{\sqrt{\tth-q^2}-\sqrt{\tth-t_0}}{\sqrt{\tth-q^2}+\sqrt{\tth-t_0}}, \label{eq:defz}
\ee
The constraint in Eq.~(\ref{eq:dsconst}) imposes $\tilde a_0^0=\tilde
a_0^+$.

\begin{figure}
\begin{center}
\makebox[0pt]{\includegraphics[scale=0.8]{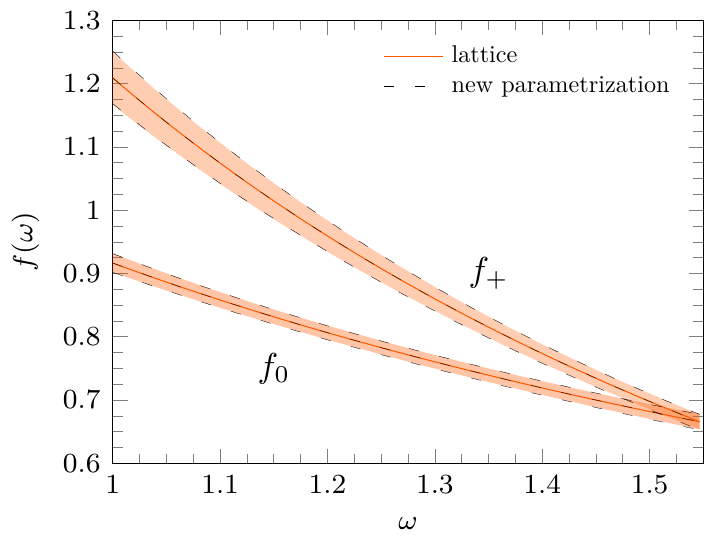}}\\
\caption{ Comparison of the original SM-LQCD form factors for  $\bar B_s\to D_s$~\cite{McLean:2019qcx} and their description in this work using the parametrization of Eq.~(\ref{eq:f0p_newpar}). 
Both central values and 68\% confidence level (CL) bands show excellent agreement.}\label{fig:comp}
\end{center}
\end{figure}

To improve the quality of our  HQSS  form-factor fit,  we change the parametrization
above and  symmetrize the range of $z$ corresponding to $0 \leq q^2 \leq
t_-$ where $t_-=(\mbs-\mds)^2$. Thus, we use
\bea
f_0(q^2)=\frac1{1-q^2/M^2_{B_{c0}}}\sum_{n=0}^2 a^0_n z^n\ \ ,\ \
f_+(q^2)=\frac1{1-q^2/M^2_{B^*_{c}}}\sum_{n=0}^2 a^+_n z^n,
\label{eq:f0p_newpar}
\eea
with
\be
z(q^2)= z(q^2;\tth,t_0), \qquad \tth=(M_B+M_D)^2\ ,\
t_0=\tth-\sqrt{\tth(\tth-t_-)}.\label{eq:def2z}
\ee
The central values and errors of the new expansion coefficients,
together with the corresponding correlation matrix, are collected in 
Table~\ref{tab:f0fp} of the Appendix. Note that we use  Eq.~\eqref{eq:dsconst} to fix $a^+_2$ for $\bar
B_s\to D_s$. The quality of this
new expansion can be seen in Fig.~\ref{fig:comp} where we compare the new parameterization with the original one in Ref.~\cite{McLean:2019qcx}. The agreement is
excellent.
\subsubsection{$B_s\to D_s^*$}
In this case, in Ref.~\cite{Harrison:2023dzh} they use the HQET basis to expand the matrix elements of the vector, axial and tensor quark-current operators
\bea
\langle D^*_s;\vec p\,',r|\bar c(0)\gamma^\mu b(0)|\bar B_s;\vec p\,\rangle&=&i\sqrt{\mbs\mdss}\,h_V(q^2)\epsilon^{\mu\nu\rho\sigma}\epsilon^*_{\nu}(\vec p\,',r)v'_\rho v_\sigma,\nonumber\\
\langle D^*_s;\vec p\,',r|\bar c(0)\gamma^\mu \gamma_5 b(0)|\bar B_s;\vec p\,\rangle&=&\sqrt{\mbs\mdss}\,\big\{h_{A_1}(q^2)(\omega+1)\epsilon^{*\mu}(\vec p\,',r)-
h_{A_2}(q^2)[\epsilon^{*\mu}(\vec p\,',r)\cdot v] v^\mu\nonumber\\
&&\hspace{2.15cm} {}-
h_{A_3}(q^2)[\epsilon^{*\mu}(\vec p\,',r)\cdot v] v^{\prime\mu}\big\}\nonumber\\
\langle D^*_s;\vec p\,',r|\bar c(0)\sigma^{\mu\nu} b(0)|\bar B_s;\vec p\,\rangle&=&
-\sqrt{\mbs\mdss}\{\,h_{T_1}(q^2)\epsilon^{*}_\alpha(\vec p\,',r)
(v+v^{\prime})_\beta+h_{T_2}(q^2)\epsilon^{*}_\alpha(\vec p\,',r)
(v-v^{\prime})_\beta\nonumber\\
&&\hspace{2.15cm} {}+h_{T_3}(q^2)[\epsilon^{*}(\vec p\,',r)\cdot v]
v_\alpha v^{\prime}_\beta\}\,\epsilon^{\mu\nu\alpha\beta},
\label{eq:hqetha123vt123}
\eea
where $\epsilon^{*}(\vec p\,',r)$ is the polarization vector of the final $D_s^*$ meson, $v(v')$ is the four-velocity of the $\bar B_c (D^*_s)$ meson and $\omega=v\cdot v'$. The  convention $\epsilon_{0123}=+1$ is used.

\begin{figure}
\begin{center}
\makebox[0pt]{\includegraphics[scale=0.6]{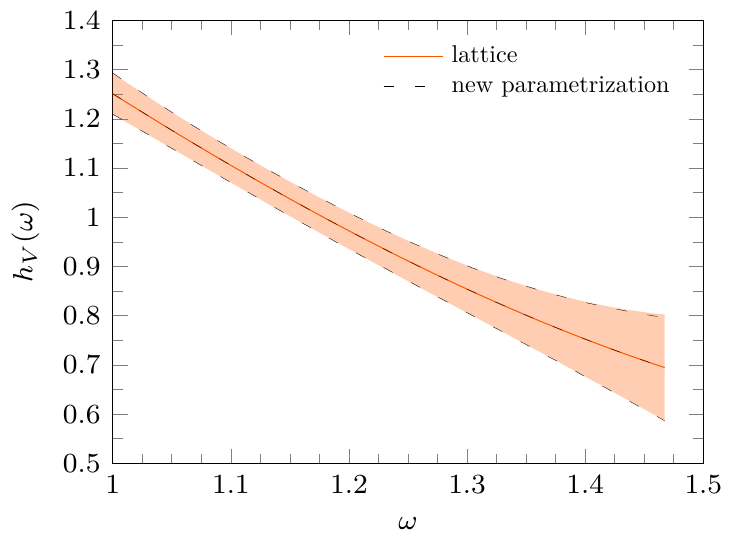}}\\
\includegraphics[scale=0.6]{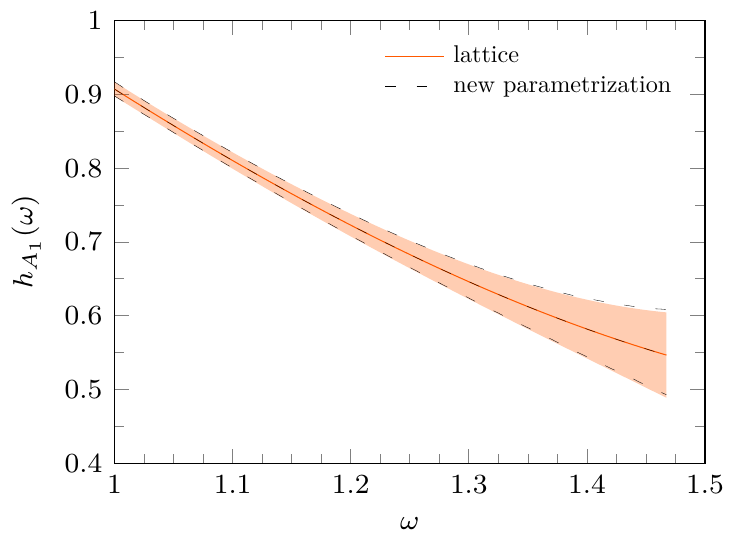}\includegraphics[scale=0.6]{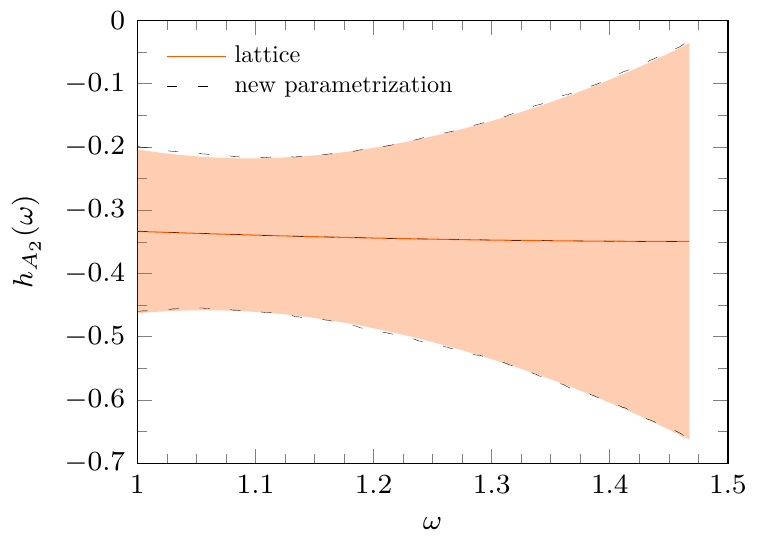}
\includegraphics[scale=0.6]{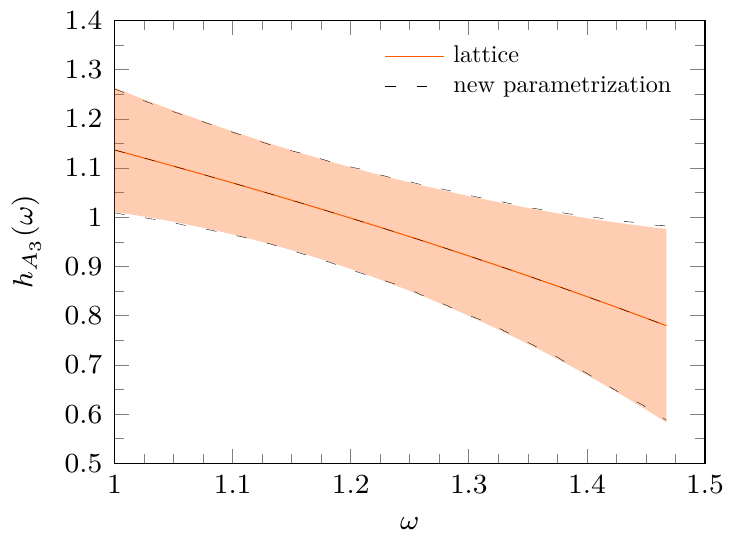}\\
\includegraphics[scale=0.6]{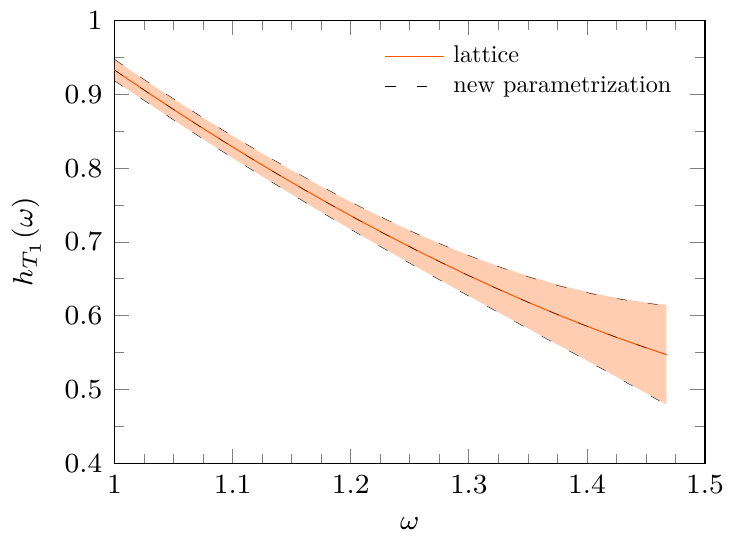}\includegraphics[scale=0.6]{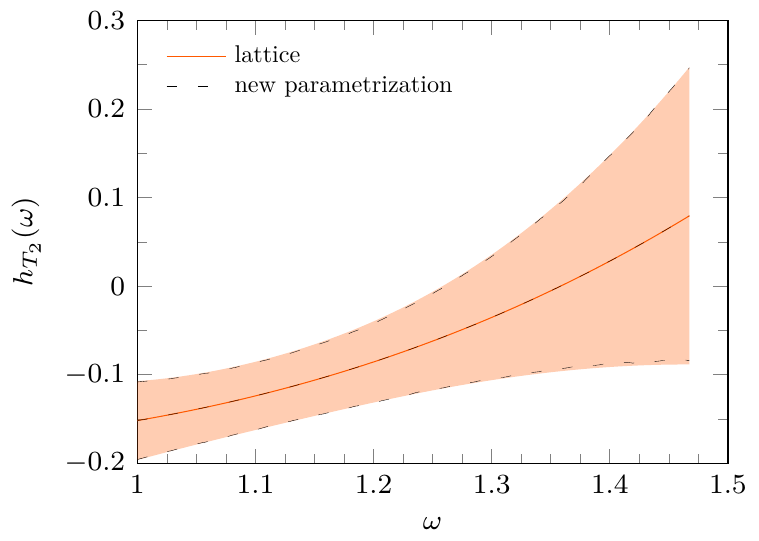}
\includegraphics[scale=0.6]{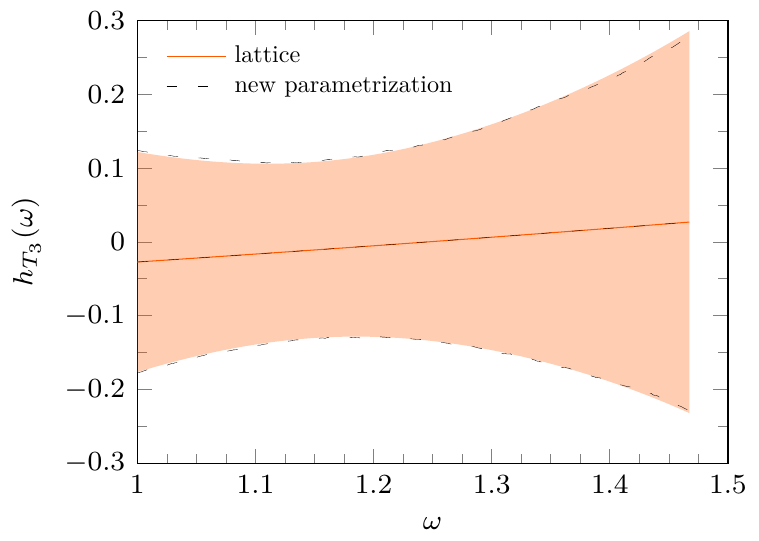}
\caption{ Comparison of the original LQCD form factors for  $\bar B_s\to D^*_s$~\cite{Harrison:2023dzh} 
and their description in this work using the parametrization of Eq.~(\ref{eq:newparDsstar}). 
Both central values and 68\% confidence level (CL) bands show excellent agreement.}\label{fig:compDsstar}
\end{center}
\end{figure}

In Ref.~\cite{Harrison:2023dzh}, the above form factors include a third-degree polynomial in $\omega-1$, logarithms  determined from staggered chiral perturbation theory and some extra analytical dependence on $M_{\pi(K)}^2$. The continuum-limit values  can be extracted from the supplemental material available in the source file provided in Ref.~\cite{Harrison:2023dzh}. Since the logarithms have a very mild dependence on $\omega-1$, and in order to facilitate the further HQSS form-factor fit that we are going to conduct, we have made a description of the lattice form factors  just as a third-degree polynomial in $\omega-1$. Thus, we use
\bea
h_F(\omega) =\sum_{n=0}^3a_n^F(\omega-1)^n.
\label{eq:newparDsstar}
\eea
In the appendix, we give the central values and errors of the new expansion coefficients   in Tables~\ref{tab:ha123vdstar_newpar} and \ref{tab:ht123dstar_newpar}  while the correlation matrix is compiled in Tables~\ref{tab:ha123} to \ref{tab:ht123ha123v}. Again, the quality of these
new expansions can be seen in Fig.~\ref{fig:compDsstar}, where we compare the results of the simpler
 parametrization in Eq.~\ref{eq:newparDsstar} with the original lattice values in Ref.~\cite{McLean:2019qcx}. The agreement is once again
excellent.
\subsection{HQSS form factors}

In HQET, for $\bar B_s\to D_s$, one normally uses the following form-factor decomposition of the 
transition-current matrix elements~\cite{Bernlochner:2017jka}
\bea
\langle D_s;\vec p\,'\,|\bar c(0)\gamma^\mu b(0)|\bar B_s;\vec p\,\rangle&=&
\sqrt{\mbs\mds}\,[\,h_+(\omega)(v^\mu+v^{\prime\mu})+h_-(\omega)(v^\mu-v^{\prime\mu})],
\eea
with $h_{\pm}$ related to $f_+$ and $f_0$ above through
\bea
f_+&=&\frac1{2\sqrt{\mbs\mds}} [(\mbs+\mds)\,h_+-(\mbs-\mds)\,h_-]\nonumber\\
f_0&=&\sqrt{\mbs\mds}\Big[\frac{\omega+1}{\mbs+\mds}h_+-\frac{\omega-1}{\mbs-\mds}h_-\Big].
\eea
For $\bar B_s\to D^*_s$, the corresponding expressions for the vector-, axial- and tensor-current matrix elements have already been given in  Eq.~(\ref{eq:hqetha123vt123}).

In Ref.~\cite{Bernlochner:2017jka}, all the  above form factors have been computed in the effective field theory, up to  ${\cal O}(\alpha_s,
\Lambda_{\rm QCD}/m_{c,b})$ corrections, for the analogous $\bar B\to D^{(*)}$ semileptonic decays. We take advantage of this study and  use the findings of Ref.~\cite{Bernlochner:2017jka} to describe the $\bar B_s\to D^{(*)}_s$ form-factors. In the infinite heavy quark 
mass limit the form factors are given by the leading IW function  $\xi(\omega)$ or they are zero. It is
 thus convenient to factor out the IW function and define  $\hat h_i(\omega)=h_i(\omega)/\xi(\omega)$, 
 which, up to ${\cal O}(\alpha_s,\Lambda_{\rm QCD}/m_{c,b})$ corrections, read~\cite{Bernlochner:2017jka}
\bea
\hat h_{A_1}&=&1+\hat\alpha_s C_{A_1}+\epsilon_c\Big(\hat L_2-\hat L_5\frac{\omega-1}{\omega+1}\Big)+\epsilon_b\Big(\hat L_1-\hat L_4\frac{\omega-1}{\omega+1}\Big),\nonumber\\
\hat h_{A_2}&=&\hat\alpha_s C_{A_2}+\epsilon_c(\hat L_3+\hat L_6),\nonumber\\
\hat h_{A_3}&=&1+\hat\alpha_s(C_{A_1}+C_{A_3})+\epsilon_c\, (\hat L_2-\hat L_3+\hat L_6-\hat L_5)+\epsilon_b\,(\hat L_1-\hat L_4),\nonumber\\
\hat h_V&=&1+\hat\alpha_s C_{V_1}+\epsilon_c\,(\hat L_2-\hat L_5)+\epsilon_b\,(\hat L_1-\hat L_4),
\nonumber\\
\hat h_{T_1}&=&1+\hat\alpha_s\Big[C_{T_1}+\frac{\omega-1}2(C_{T_2}-C_{T_3})\Big]+\epsilon_c\hat L_2+\epsilon_b\hat L_1\nonumber\\
\hat h_{T_2}&=&\hat\alpha_s\frac{\omega+1}2(C_{T_2}+C_{T_3})+\epsilon_c\hat L_5-\epsilon_b\hat L_4\nonumber\\
\hat h_{T_3}&=&\hat\alpha_sC_{T_2}+\epsilon_c\,(\hat L_6-\hat L_3)
\\\nonumber\\
\hat h_+&=&1+\hat\alpha_s\Big[
C_{V_1}+\frac{\omega+1}2(C_{V_2}+C_{V_3})\Big]+(\epsilon_c+\epsilon_b)\hat L_1,\nonumber\\
\hat h_-&=&\hat\alpha_s\frac{\omega+1}2(C_{V_2}-C_{V_3})+
(\epsilon_c-\epsilon_b)\hat L_4.
\eea
The terms proportional to $\hat\alpha_s=\alpha_s/\pi$ are perturbative corrections computed by matching QCD to the HQET and, although dependent on $\omega$, they are independent of the light degrees of freedom. The different $C_{A,V,T}$ functions can be found in Appendix A of Ref.~\cite{Bernlochner:2017jka}. 
In addition, $\epsilon_{c,b}$ are given by
$\epsilon_{c,b}=\bar\Lambda/(2m_{c,b})$, with $\bar\Lambda$ 
a low energy constant (LEC) of order ${\cal O}(\Lambda_{\rm QCD})$ for which we take the value  quoted in 
Ref.~\cite{Bernlochner:2017jka}.  The six $\omega$-dependent $\hat L_j$ functions can be 
written in terms of just three sub-leading IW functions $\hat\chi_{2,3}$ and $\eta$ 
(see Eq.~(8) in Ref.~\cite{Bernlochner:2017jka}) for which  the  following near zero-recoil ($\omega=1$) 
expansions are used\footnote{In the case of $\hat \chi_3$ one has that $\hat\chi_3(1)=0$ from Luke's theorem~\cite{Luke:1990eg}. }
\bea
\hat\chi_2(\omega)=\hat\chi_2(1)+\hat\chi'_2(1)(\omega-1),\ \
\hat\chi_3(\omega)=\hat\chi'_3(1)(\omega-1),\ \ 
\eta(\omega)=\eta(1)+\eta'(1)(\omega-1).
\eea
 Strictly speaking, $\bar\Lambda$ depends on the light-quark degrees of freedom. Thus, one expects some SU(3) breaking
  that will modify its value compared to that used in 
 Ref.~\cite{Bernlochner:2017jka} for $\bar B\to D^{(*)}$ decays. 
 By keeping it the same, we reabsorb this change into the sub-leading IW functions which, together with the leading one, also suffer from SU(3) breaking effects.

 For the leading IW function $\xi$ we shall take the parametrization in 
 Ref.~\cite{Murgui:2019czp}, where one has that
\be 
\xi(\omega)=1-8\rho^2\hat z+(64c-16\rho^2)\hat z^2+(256 c-24\rho^2+512 d)\hat z^3
\ee
and
\be 
\hat z(\omega)=\frac{\sqrt{\omega+1}-\sqrt2}{\sqrt{\omega+1}+\sqrt2}.
\ee
In addition, following Ref.~\cite{Murgui:2019czp}, we include the 
${\cal O}[(\Lambda_{\rm QCD}/m_c)^2]$ corrections introduced in 
Ref.~\cite{Jung:2018lfu}, which affect the form factors that are protected from
${\cal O}(\Lambda_{\rm QCD}/m_c)$ corrections at zero recoil. In our case, not only  $\hat h_+$ 
and $\hat h_{A_1}$ but also $h_{T_1}$. We shall use
\bea
\hat h_+\to\hat h_++\epsilon_c^2\, l_1(1),\ \ \hat h_{A_1}\to\hat h_{A_1}+\epsilon_c^2\, l_2(1)
,\ \ \hat h_{T_1}\to\hat h_{T_1}+\epsilon_c^2\, l_3(1)
\eea

\subsection{Fit of the SM-LQCD form factors to their HQSS/HQET expressions. }
\label{sec:SMHQSSfit}
\begin{figure}
\begin{center}
\includegraphics[scale=0.6]{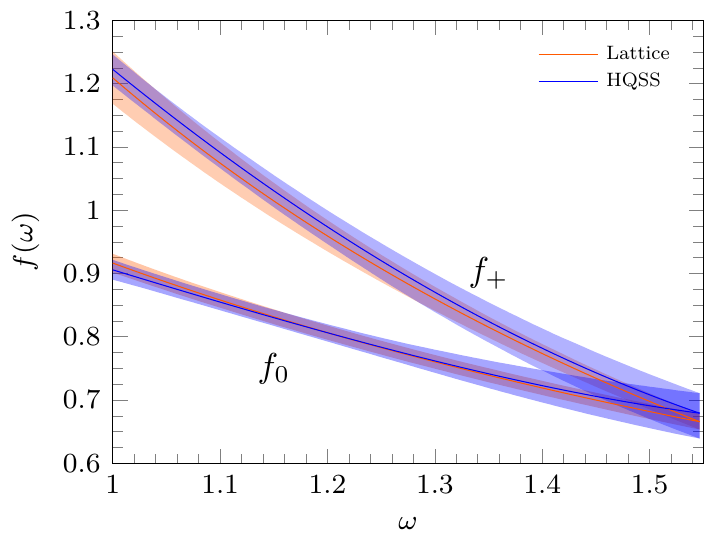}\includegraphics[scale=0.6]{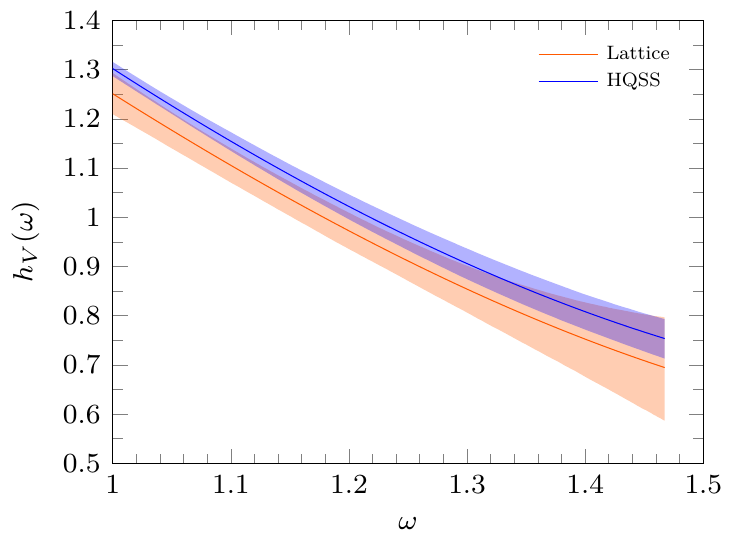}\\
\includegraphics[scale=0.6]{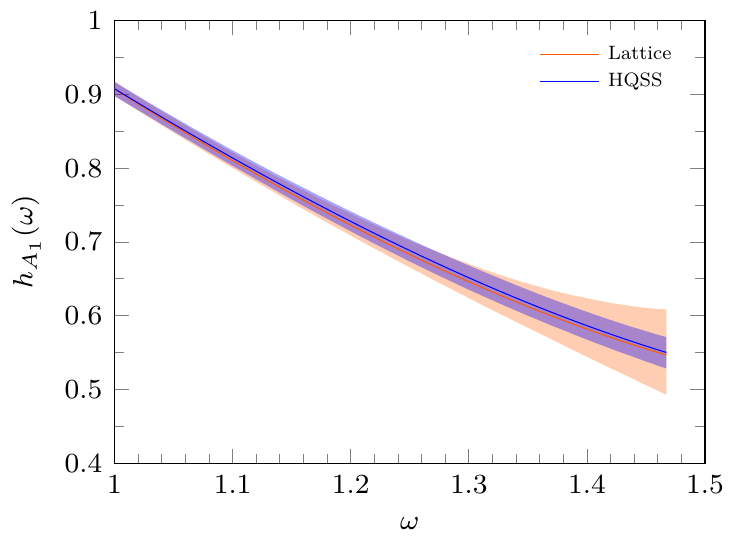} \includegraphics[scale=0.6]{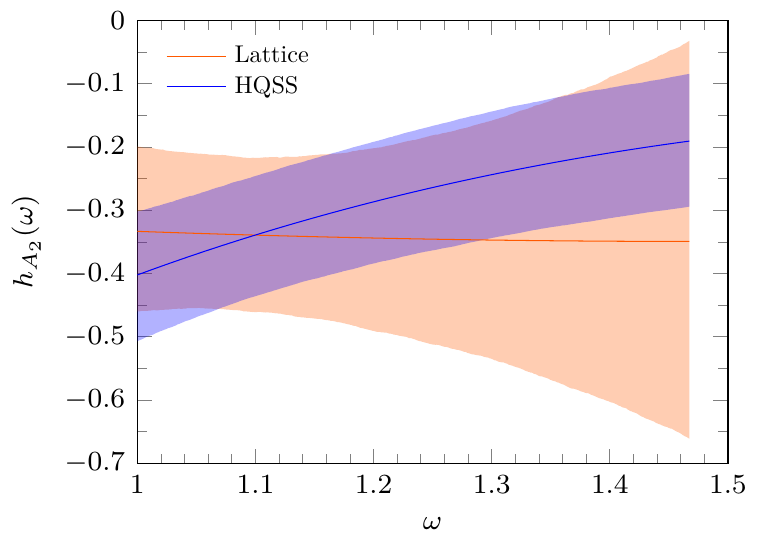} \includegraphics[scale=0.6]{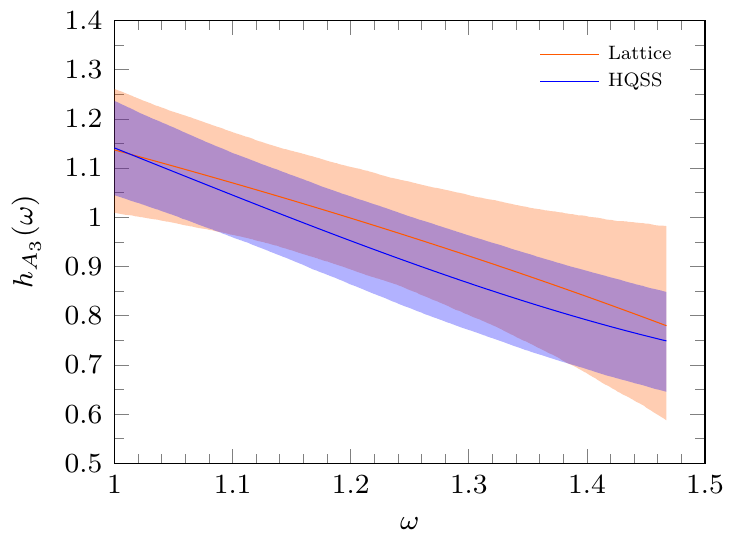}\\
\includegraphics[scale=0.6]{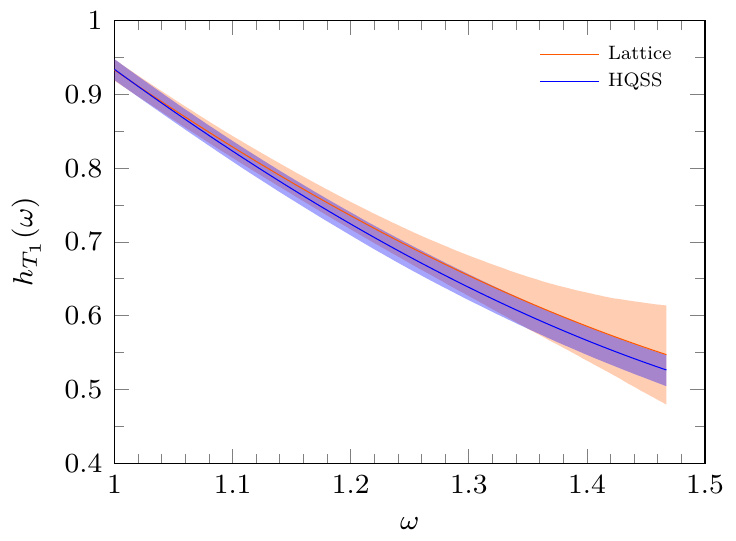} \includegraphics[scale=0.6]{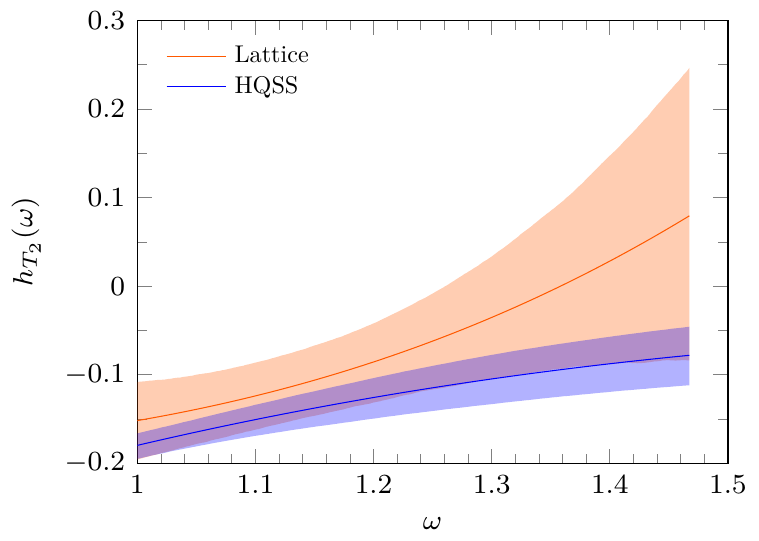} \includegraphics[scale=0.6]{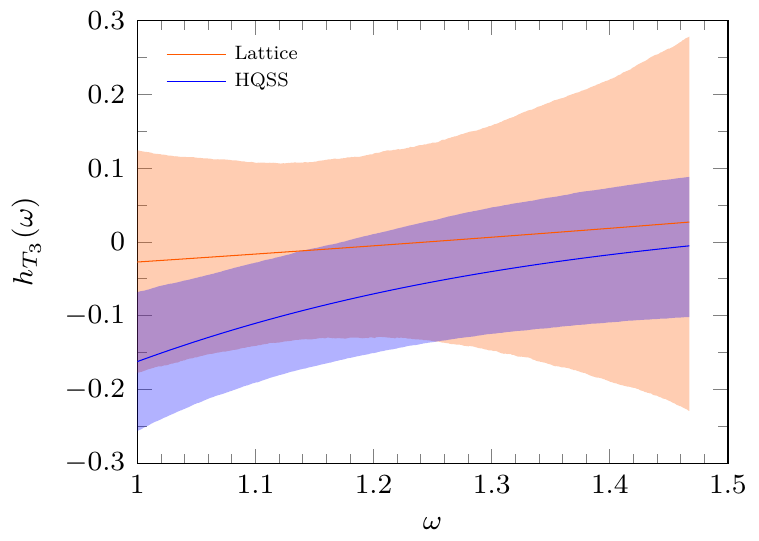}
\caption{ Comparison of the original LQCD  form factors 
for  $\bar B_s\to D_s$~\cite{,McLean:2019qcx} and $\bar B_s\to D^*_s$~\cite{Harrison:2021tol} semileptonic decays and the HQET predictions after the fitting procedure described in the main text.}
\label{fig:dsdsstarcomp}
\end{center}
\end{figure}
\begin{table}[t]
\begin{center}
\begin{tabular}{c|c|c|c|c}\hline\hline\tstrut
& $\bar B_s \to D_s^{(*)}$ & $\bar B_s \to D_s^{(*)}$ (unc) & $\epsilon[\bar B_s \to D_s^{(*)}]$& $\bar B \to D^{(*)}$ \cite{Murgui:2019czp}\\\hline\tstrut
 $\rho^2$& \hspace{.25cm}$1.29\pm 0.06$ & \hspace{.25cm}$1.26\pm0.07 $ & 0.07& \hspace{.25cm}$1.32 \pm 0.06$\\
 $c$&\hspace{.25cm}$0.63\pm 0.24$ &\hspace{.25cm}$0.53\pm 0.20$ & 0.26&\hspace{.25cm}$1.20 \pm 0.12$\\
 $d$&\hspace{.25cm}$0.15\pm0.42$ &\hspace{.25cm}$0.20\pm 0.37$ & 0.42&$-0.84 \pm 0.17$  \\
 $\hat\chi_2(1)$&$-0.15\pm0.13$&$-0.16\pm0.11$ & 0.13&  $-0.058 \pm 0.020$\\
 $\hat\chi'_2(1)$&$-0.15\pm0.29$&$-0.50\pm 0.39$& 0.45&\hspace{.25cm}$0.001\pm 0.020$\\
 $\hat\chi'_3(1)$&$-0.03\pm0.05$&$-0.04\pm 0.05$&0.05 &\hspace{.25cm}$0.036\pm0.020$ \\
 $\eta(1)$&\hspace{.25cm}$0.07\pm0.12$ &\hspace{.25cm}$0.13\pm0.17$ & 0.13&\hspace{.25cm}$0.355 \pm 0.040$ \\
 $\eta'(1)$&$-0.81\pm 0.28$&\hspace{.25cm}$0.28\pm0.80$&1.13 &$-0.03\pm 0.11$\\
 $l_1(1)$&\hspace{.25cm}$0.04\pm0.53$ &\hspace{.25cm}$0.16\pm0.53$ & 0.54&\hspace{.25cm}$0.14\pm 0.23$\\
 $l_2(1)$&$-1.77\pm0.30$&$-1.75\pm0.30$ & 0.30&$-2.00\pm0.30$\\
 $l_3(1)$&$-2.86\pm0.44$&$-2.91\pm0.44$&0.44&
 \\\hline\hline
\end{tabular}
    \caption{Second column: Mean values and uncertainties  of the $\rho^2,c, d,\hat\chi_2(1),\hat\chi'_2(1),\hat\chi'_3(1),\eta(1)$ and $\eta'(1)$  LECs obtained by fitting the $\bar B_s \to D_s^{(*)}$ LQCD form factors from 
    Refs.~\cite{McLean:2019qcx,Harrison:2023dzh} to their ${\cal O}(\alpha_s,
\Lambda_{\rm QCD}/m_{c,b})$ HQET expressions given in \cite{Bernlochner:2017jka}. The first three parameters determine the leading IW function, while the last five enter in the $1/m_{c,b}$ sub-leading corrections. In addition, $l_1(1),l_2(1)$ and $l_3(1)$ account for ${\cal O}[(\Lambda_{\rm QCD}/m_c)^2]$ contributions~\cite{Jung:2018lfu}, which affect the $\hat h_+, \hat h_{A_1}$ and 
$\hat h_{T_1}$ form factors, respectively,  which are protected from
${\cal O}(\Lambda_{\rm QCD}/m_c)$ corrections at zero recoil. Third column: Results from the totally uncorrelated fit, where we consider only the diagonal elements of the matrix $C$ in the definition of the merit function of Eq.~\eqref{eq:chi2}. Fourth column: Final total errors considered on the fitted LECs and used in the evaluation of the uncertainty bands for derived observables. They are computed by combining in quadrature  the errors from the central fit (second column) with the magnitudes of the differences between the mean values of the central and uncorrelated fits. Fifth column: Results for the analogous SU(3) fit carried out in Ref.~\cite{Murgui:2019czp} to $\bar B \to D^{(*)}$ form-factor LQCD and experimental inputs. Note a typo (global sign) in the numerical value of $l_2(1)$ given in the original Table 1 of Ref.~\cite{Murgui:2019czp}. }
   \label{tab:hqssfit}
   \end{center}
\end{table}
\begin{table}[t]
\begin{ruledtabular}
\begin{center}
\begin{tabular}{lccccccccccc}
& $\rho^2$& $c$&$d$&$\hat\chi_2(1)$&$\hat\chi'_2(1)$&$\hat\chi'_3(1)$&$\eta(1)$&$\eta'(1)$&$l_1(1)$&$l_2(1)$&$l_3(1)$
\\ \hline \tstrut
$\rho^2$  & 1.000&	  0.595& 	$-0.288$&     $-0.241$ &   \hspace{0.25cm}0.231&   \hspace{0.25cm}0.286 &     \hspace{0.25cm}0.074 & $-0.063$& $-0.015$& $-0.015$&$-0.017$\\
  $c$  & &1.000&    $-0.839$& $-0.221$& \hspace{0.25cm}0.440 &  $-0.057$&    $ -0.022$&      \hspace{0.25cm}0.051  &   \hspace{0.25cm}0.102 &   \hspace{0.25cm}0.021&      \hspace{0.25cm}0.015\\
$d$  & &&\hspace{0.25cm}1.000&   \hspace{0.25cm}0.082&    \hspace{0.25cm}0.042&      \hspace{0.25cm}0.330&   \hspace{0.25cm}0.037& 	 $-0.050$& 	$-0.287$ &    $-0.018$&     $-0.015$\\
 $\hat\chi_2(1)$  & &&&\hspace{0.25cm}1.000&	 $-0.200$& 	\hspace{0.25cm}0.641&     $-0.222$ &    $-0.179$&     \hspace{0.25cm}0.034 &   $ -0.009$&     $-0.009$\\
  $\hat\chi'_2(1)$ & &&&&\hspace{0.25cm}1.000 &   \hspace{0.25cm}0.284 &    $-0.054$&     $-0.053$& 	 $-0.263$& 	\hspace{0.25cm}0.029& 	\hspace{0.25cm}0.021\\
  $\hat\chi'_3(1)$ & &&&&&\hspace{0.25cm}1.000&	\hspace{0.25cm}0.007&    \hspace{0.25cm}0.001 &    $-0.263$ &    $-0.033$&     $-0.030$\\
  $\eta(1)$ & &&&&&&\hspace{0.25cm}1.000 &   \hspace{0.25cm}0.211&  \hspace{0.25cm}0.125& 	 $-0.112$& 	$-0.085$\\
  $\eta'(1)$ & &&&&&&&\hspace{0.25cm}1.000 &   $-0.103$&     $-0.026$ &    $-0.019$\\
  $l_1(1)$ & &&&&&&&& \hspace{0.25cm}1.000 &   $-0.014$& 	 $-0.010$\\
  $l_2(1)$  &&&&&&&&&& \hspace{0.25cm}1.000 &  \hspace{0.25cm}0.293\\
  $l_3(1)$ & &&&&&&&&&&\hspace{0.25cm}1.000\\
\end{tabular}
    \caption{Correlation matrix of the  $\rho^2$, $c$, $d$, $\hat\chi_2(1)$, $\hat\chi'_2(1)$, $\hat\chi'_3(1)$,$\eta(1)$, $\eta'(1)$, $l_1(1)$, $l_2(1)$ and $l_3(1)$ bestfit parameters after fitting the LQCD form factors from Refs.~\cite{McLean:2019qcx,Harrison:2021tol} to their ${\cal O}(\alpha_s,
\Lambda_{\rm QCD}/m_{c,b})$ HQET expressions.}
   \label{tab:hqssfitcovmat}
   \end{center}
   \end{ruledtabular}
\end{table}

Treating the eleven HQET LECs $\rho^2,c,d,\hat\chi_2(1),\hat\chi'_2(1),\hat\chi'_3(1),\eta(1), \eta'(1),l_1(1),l2(1)$ and $l_3(1)$ introduced above as free parameters, we can fit the LQCD form factors to their HQSS expressions. We fit the thirty-three independent coefficients  $a_i^F$, with $F=+,0,{A_{1,2,3}},{V}, {T_{1,2,3}}$, that expand the LQCD form factors. The fit minimizes a $\chi^2$ function, that in a simplified notation we can write as
\be
\chi^2=\sum_{j}\sum_k(a_j-f_j)C^{-1}_{jk}(a_k-f_k).\label{eq:chi2}
\ee
Here, the sum is over all the expansion coefficients, for which the $a's$ represent their central values,
and the $f's$ stand for the expressions of the corresponding expansion coefficients in terms of the
$\rho^2$, $c$, $d$, $\hat\chi_2(1)$, $\hat\chi'_2(1)$, $\hat\chi'_3(1)$,  $\eta(1)$, $\eta'(1)$, $l_1(1), l_2(1)$ and $l_3(1)$ 
best fit LECs.  The $f_j$ terms are obtained in the following way: For the $D_s$ case we first multiply the HQSS form factors by the corresponding pole factors in  Eq.~(\ref{eq:f0p_newpar}), and then we expand the result in powers of the $ z$ variable defined in Eq.~(\ref{eq:def2z}). For the  $D^*_s$ case, we directly expand the  HQSS form factors in powers of $\omega-1$. 
The covariance matrix $C$ is block diagonal, built from the separate $D_s$ and
$D^*_s$ covariance matrices compiled in Tables~\ref{tab:f0fp} for $\bar B_s\to D_s$ and
Tables~\ref{tab:ha123} to \ref{tab:ht123ha123v} for $\bar B_s\to D_s^*$ transitions respectively. 

Since the LQCD results come from simulations on the same ensembles, with the same lattice actions and the same treatment of the chiral and continuum limits, we expect correlations between as well as within them. Lacking information on the former, we also tried fits with the $D^*_s$ results taken as either fully  correlated or fully anti-correlated with the $D_s$ ones. That is, we augmented the correlation matrix corresponding to $C$ with off-diagonal blocks for these two extreme cases with all entries taken to be either 1 or $-1$. However, the new $C$ matrices constructed in this way had negative eigenvalues. We also explored partially correlated scenarios (all matrix elements of the $D_s-D^*_s$ off-diagonal blocks set to $r$, with $|r|\le 1$),  but we found positive definite covariance matrices only for very small correlations $|r|$, of the  percent order. Finally, we carried out a totally uncorrelated fit, where we considered only the diagonal elements of the matrix $C$ in the definition of the merit function of Eq.~\eqref{eq:chi2}. That is to say, in this fit we also switched off  the separate $D^*_s$ and $D_s$ correlations.
The results for the central fitted parameters and errors are given in the second column of Table~\ref{tab:hqssfit}, while the corresponding correlation matrix appears in Table~\ref{tab:hqssfitcovmat}. The fit has $\chi^2/\text{dof}=0.44$. 
We see the $c$ and $d$ coefficients are not determined with precision, which is probably a reflection of the large uncertainties in the lattice $\bar B_s\to D_s^*$ form factors  at high $\omega$ values. Large uncertainties are also seen in the parameters of the sub-leading IW functions  and  $l_1(1)$.
The next  column shows the results from the totally uncorrelated fit (diagonal $C$ matrix) which has $\chi^2/\text{dof}=0.32$. The two fits give compatible results. We take the magnitude of the differences between the mean values of the central fit and those obtained in the uncorrelated (diagonal $C$ matrix) fit as a further systematic error that we will combine in quadrature with the errors from the central fit to get our final error estimate for each of the parameters. Their values are presented in the next-to-last column of Table~\ref{tab:hqssfit}. We retain the correlation matrix from the central fit (Table~\ref{tab:hqssfitcovmat}). Using these ingredients we construct Gaussian distributions which are then used to compute 68\% confidence level bands for derived observables.

In the final column of Table~\ref{tab:hqssfit}, we include  Table 1 of \cite{Murgui:2019czp}, which contains the results for the analogous fit carried out in that work to $\bar B \to D^{(*)}$ LQCD and experimental form-factor inputs. For the two parameters that are better determined, $\rho^2$ and $l_2(1)$, we see  small variations, compatible with the expected SU(3) light-flavor breaking corrections ($\sim 25-30\% $). For the others, all of them evaluated here  with sizeable uncertainties, the differences between the central values in both fits are large. However, due to those sizable uncertainties,  an interpretation as  genuine unexpectedly-large SU(3)-breaking effects is very much limited. 

A comparison of the original LQCD form factors from Refs.~\cite{McLean:2019qcx,Harrison:2023dzh} and the HQET predictions after the fitting procedure just described is shown in  Fig.~\ref{fig:dsdsstarcomp}. The LQCD error bands are notably much wider in most cases and we see a good agreement, within uncertainties, for all form-factors. The exception is  $h_V$ for $\omega$ values below $1.15$ where the error bands hardly overlap.

In the next subsection, we show the different $q^2-$distributions that fully determine the semileptonic $\bar B_s \to D^{(*)}_s\tau^-\bar\nu_\tau$ transitions for polarized final tau-leptons~\cite{Penalva:2021gef,Penalva:2021wye}. 

\subsection{Visible kinematics of the sequential 
$H_b\to H_c\tau^-(\pi^-\nu_\tau,\rho^-\nu_\tau,\ell^-\bar\nu_\ell\nu_\tau)\bar\nu_\tau$ decays}
\label{sec:assy}
If the spins of the $H_{b,c}$ hadrons are not measured, the ideal experiment to obtain the maximum information would be one in which both the momentum and spin (or helicity) state of the $\tau$ lepton could be established. This is however not possible since the $\tau$ is very short-lived. Thus, information about the $H_b\to H_c\tau^-\bar\nu_\tau$ parent decay has to be accessed via the visible kinematics of the $\tau$ decay products. 

We have considered the three $\tau$ decay channels 
$\tau^-\to\pi^-\nu_\tau$, $\rho^-\nu_\tau$ and $\ell^-\bar\nu_\ell \nu_\tau$, with 
$\ell=\mu,e$, that account for up to 70\% of the total $\tau$ decay width. Of the $\tau$-decay products, only the charged 
particle $d=\pi^-,\,\rho^-$ or $\ell^-$ will be observed and, in the zero $\tau$-width limit, one can write the differential
 decay width~\cite{Alonso:2016gym,Alonso:2017ktd, Asadi:2020fdo,Penalva:2021wye}
\bea
\frac{d^3\Gamma_d}{d\omega  d\xi_d d\cos\theta_d} & = & {\cal B}_{d}
\frac{d\Gamma_{\rm SL}}{d\omega} \Big\{  F^d_0(\omega,\xi_d)+ F^d_1(\omega,\xi_d) 
\cos\theta_d + F^d_2(\omega,\xi_d)P_2(\cos\theta_d)\Big\}.\label{eq:visible-distr}
\eea
As already mentioned,  $\omega$ is 
the product of the four-velocities of the $H_b$ and $H_c$ hadrons, which is related to
the four-momentum transfer squared $q^2$ through the relation $q^2=M^2+M^{\prime 2}-2MM'\omega$, with
$M(M')$ the  mass of the $H_b(H_c)$ hadron. In addition, $\xi_d$ is the ratio of the
$d$ charged particle  and $\tau$ energies measured in the $\tau^-\bar\nu_\tau$ 
center of mass frame (CM), while $\theta_d$ is the angle made by the three-momenta of the
$d$ charged particle and the $H_c$ final hadron, also measured in the CM frame (for the kinematics, see for instance Fig.~1 of Ref.~\cite{Penalva:2022vxy}). ${\cal B}_d$ is the branching ratio for the
corresponding $\tau$ decay mode  and $P_2$ 
stands for  the Legendre polynomial 
of order two. In addition, $d\Gamma_{\rm SL}/d\omega$ accounts for the
unpolarized $H_b\to H_c\tau^-\bar\nu_\tau$ decay width that can be written
as~\cite{Penalva:2022vxy}
\be
\frac{d\Gamma_{\rm SL}}{d\omega}=\frac{G_F^2|V_{cb}|^2M^{\prime3}M^2}{24\pi^3} \label{eq:n0w}
\sqrt{\omega^2-1}\Big(1-\frac{m_\tau^2}{q^2}\Big)^2n_0(\omega),
\ee
with $G_F$ the Fermi decay constant and $V_{cb}$ the corresponding 
Cabibbo-Kobayashi-Maskawa matrix element. The  $n_0(\omega)$ function contains
 all the dynamical information, including possible NP effects.
Finally, the $F^d_{0,1,2}(\omega,\xi_d)$  functions read~\cite{Penalva:2021wye}
 \begin{eqnarray}
 F^d_0(\omega,\xi_d) &=& C_n^d(\omega,\xi_d)+C_{P_L}^d(\omega,\xi_d)\,\langle P^{\rm CM}_L\rangle(\omega), \nonumber \\
 F^d_1(\omega,\xi_d) &=& C_{A_{FB}}^d(\omega,\xi_d)\,A_{FB}(\omega)+C_{Z_L}^d(\omega,\xi_d)\,Z_L(\omega)
 + C_{P_T}^d(\omega,\xi_d)\,\langle P^{\rm CM}_T\rangle(\omega), \nonumber \\ 
 F^d_2(\omega,\xi_d) &=& C_{A_Q}^d(\omega,\xi_d)\,A_{Q}(\omega)+
 C_{Z_Q}^d(\omega,\xi_d)\,Z_Q(\omega)+ C_{Z_\perp}^d(\omega,\xi_d)\,Z_\perp(\omega).
 \label{eq:coeff}
\end{eqnarray}
with  $C^d_a(\omega,\xi_d)$  kinematical coefficients that are decay-mode dependent and whose expressions can be found in
  Appendix G of Ref.~\cite{Penalva:2021wye}.  The rest of the observables in Eq.~(\ref{eq:coeff}) represent
  spin ($\langle P^{\rm CM}_{L,T}\rangle(\omega)$), angular  
 ($A_{FB,Q}(\omega)$) and spin-angular ($Z_{L,Q,\perp}(\omega)$)  asymmetries of the 
 $H_b\to H_c\tau\bar\nu_\tau$ parent decay~\cite{Penalva:2021wye}.
In the absence of CP-odd contributions, 
 these asymmetries, together with $d\Gamma_{\rm SL}/d\omega$, encode  
 the maximal information obtainable if one could directly analyze the polarized 
 $H_b\to H_c \tau \bar\nu_\tau$  transitions (see Ref.~\cite{Penalva:2021gef} and especially 
 Eq.~(3.46) of Ref.~\cite{Penalva:2021wye} and the related discussion). 
\begin{figure}
\begin{center}
\includegraphics[scale=0.75]{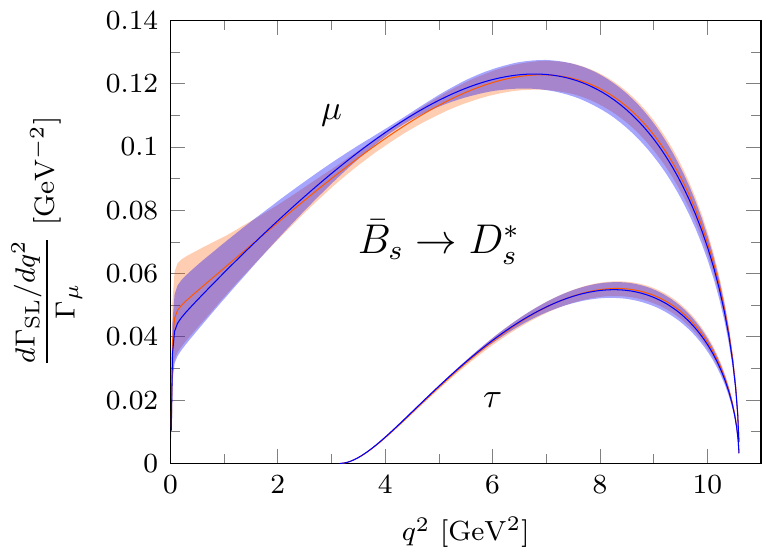}\ \includegraphics[scale=0.75]{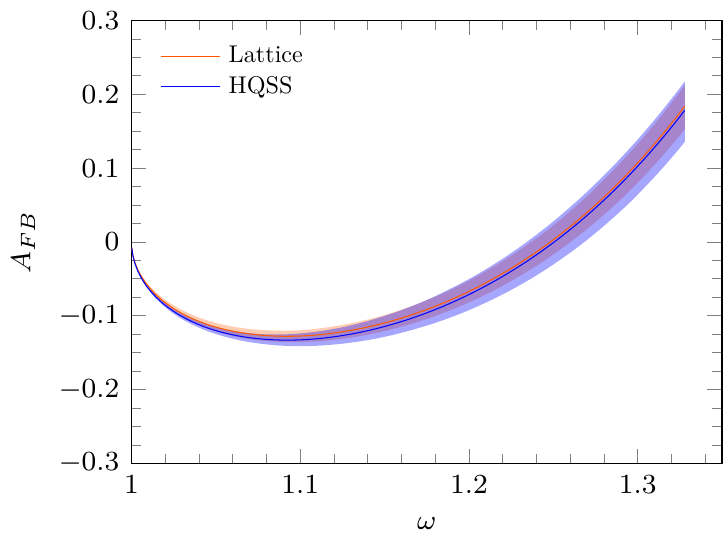}\\
\includegraphics[scale=0.75]{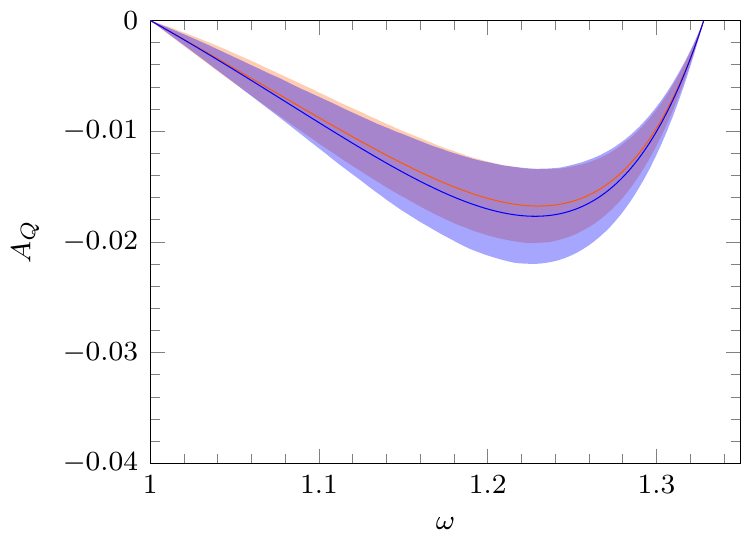}\includegraphics[scale=0.75]{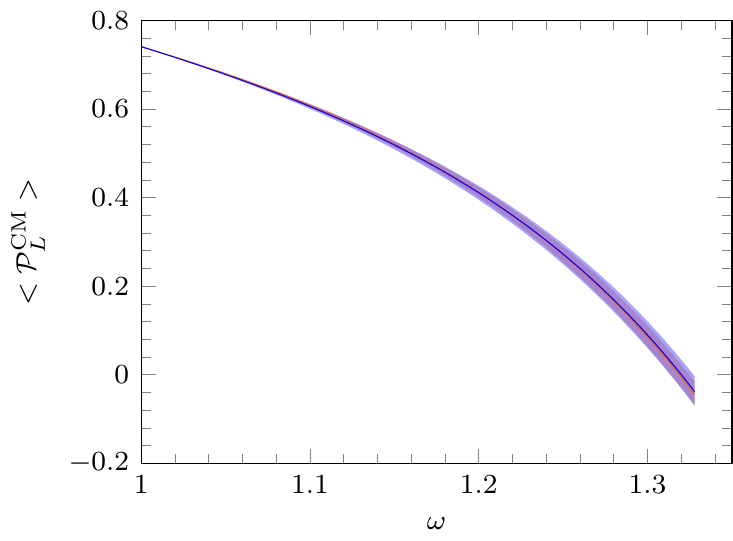}\\
\includegraphics[scale=0.75]{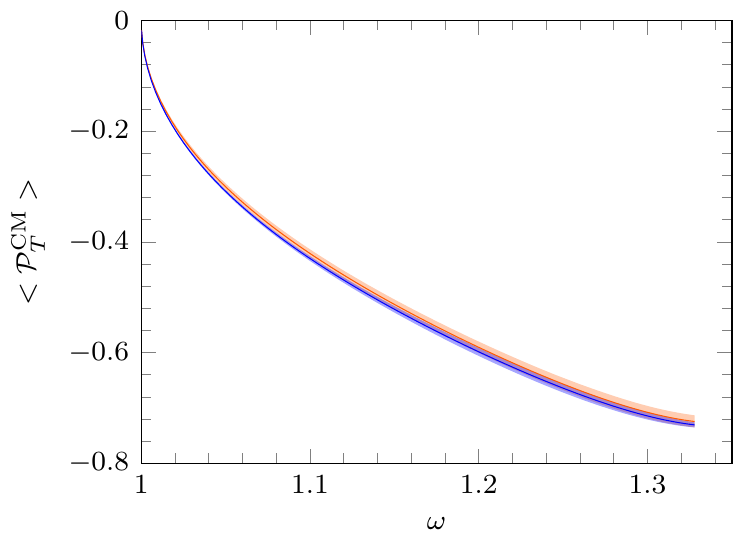}\includegraphics[scale=0.75]{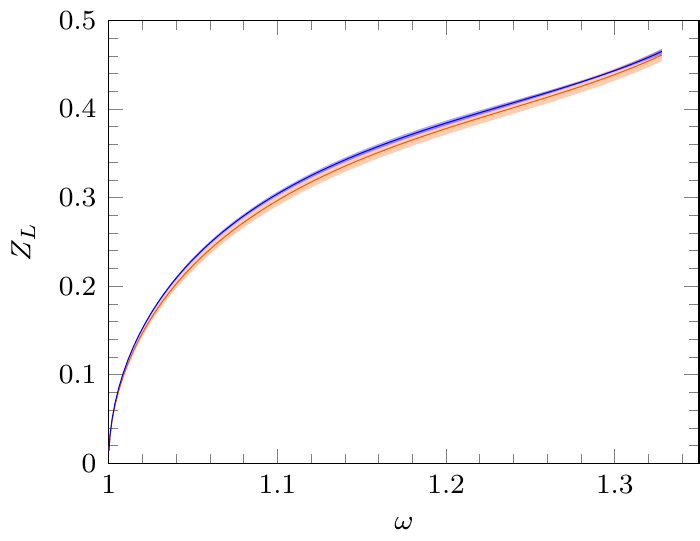}\\
\includegraphics[scale=0.75]{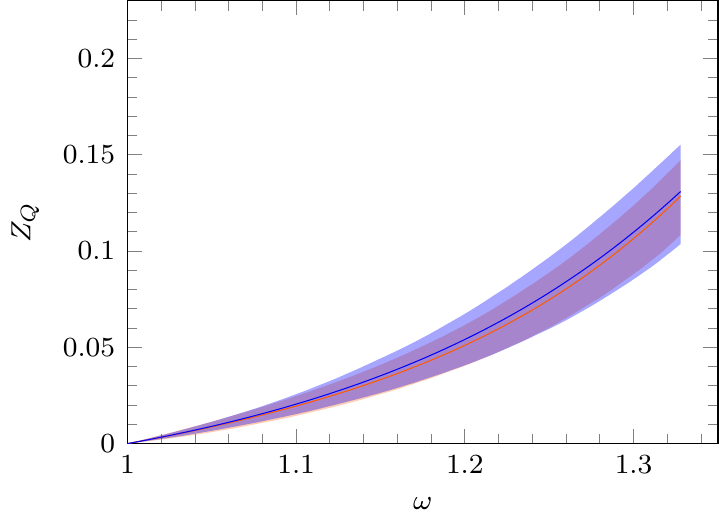}\includegraphics[scale=0.75]{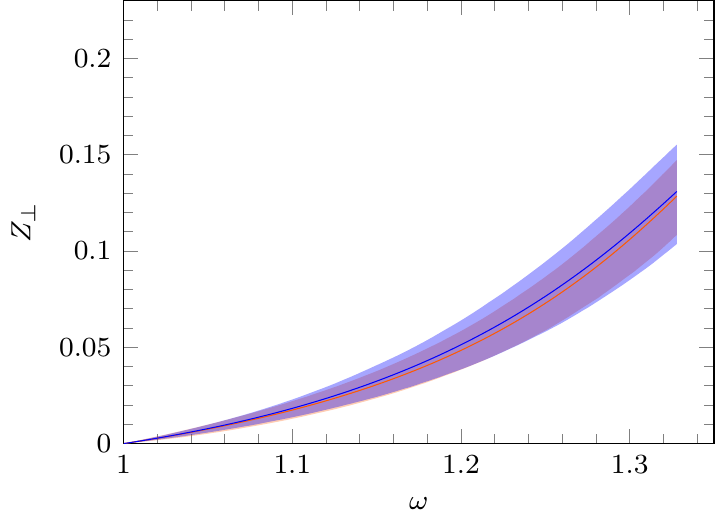}
\caption{ $d\Gamma_{\rm SL}/dq^2$ differential decay width, divided by $\Gamma_\mu=\Gamma(\bar B_s\to D_s^*\mu^-\bar\nu_\mu)$, and  the different tau-asymmetries
introduced in Eq.~(\ref{eq:coeff}) for the semileptonic $\bar B_s\to D_s^* \tau \bar\nu_\tau$ decay. We compare  the results evaluated with the SM-LQCD form factors from 
Refs.~\cite{Harrison:2021tol,McLean:2019qcx} and  with the  SM-HQET form factors obtained after the
fitting procedure described in Subsec.~\ref{sec:SMHQSSfit}.}
\label{fig:dsstarsmasi}
\end{center}
\end{figure}%
\begin{figure}
\begin{center}
\includegraphics[scale=0.75]{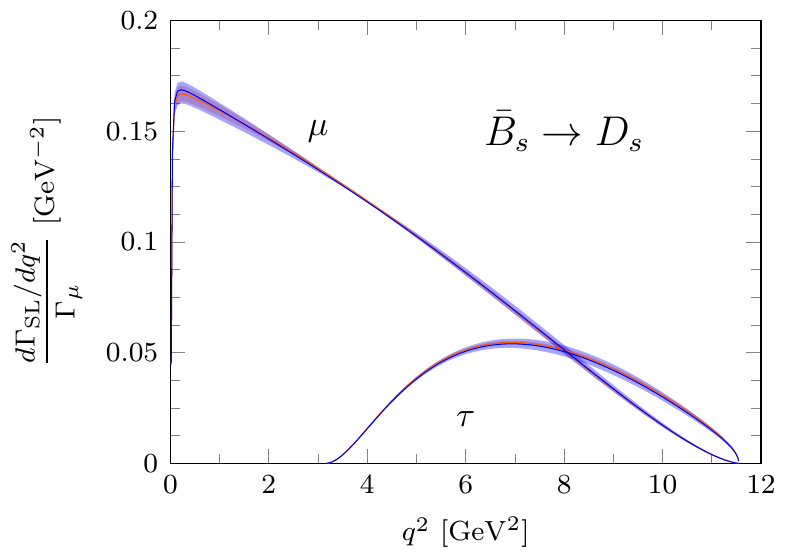}\ \includegraphics[scale=0.75]{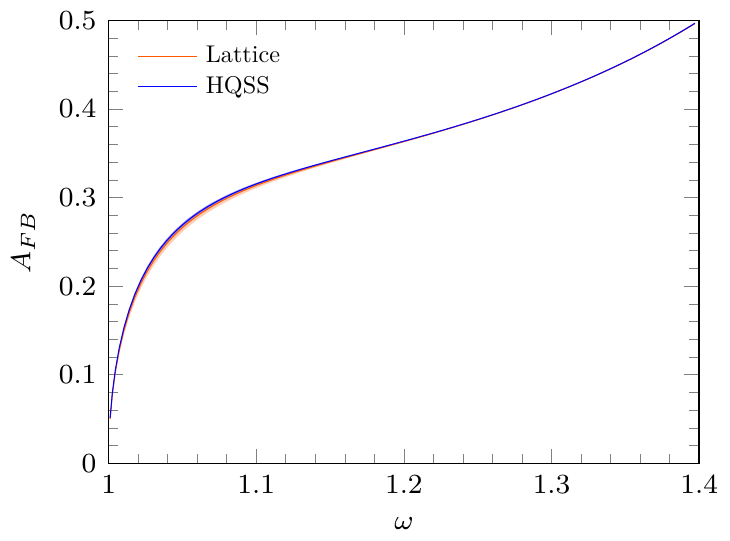}\\
\includegraphics[scale=0.75]{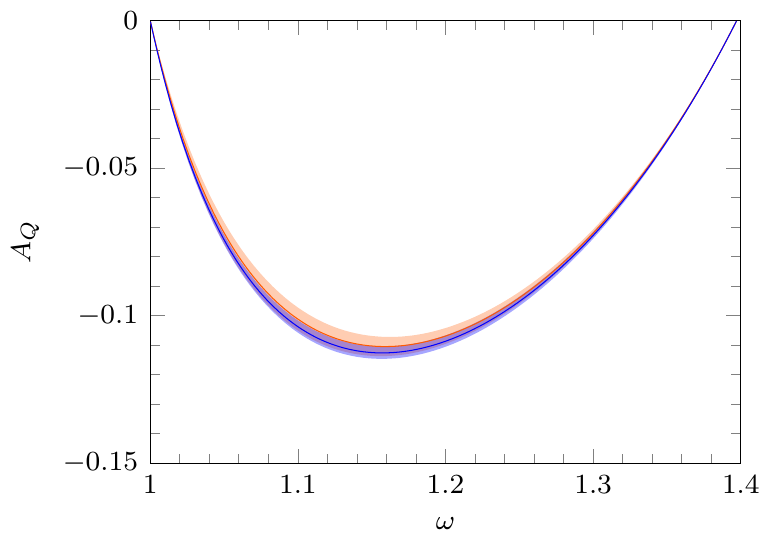}\includegraphics[scale=0.75]{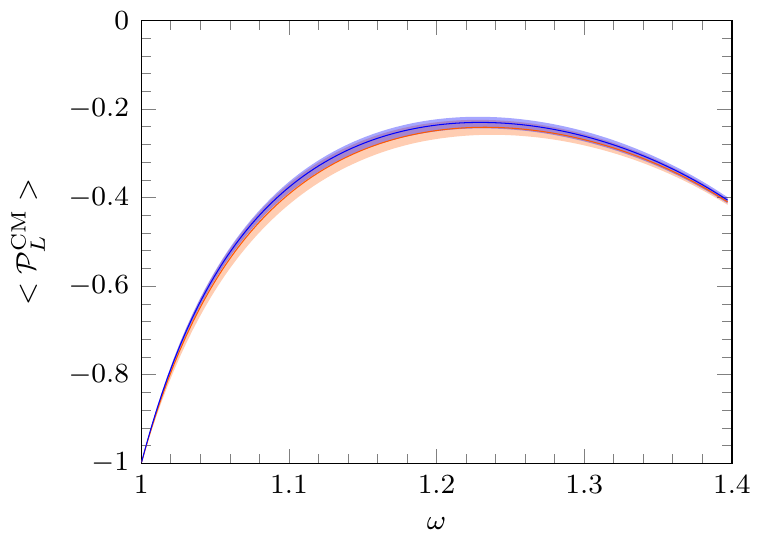}\\
\includegraphics[scale=0.75]{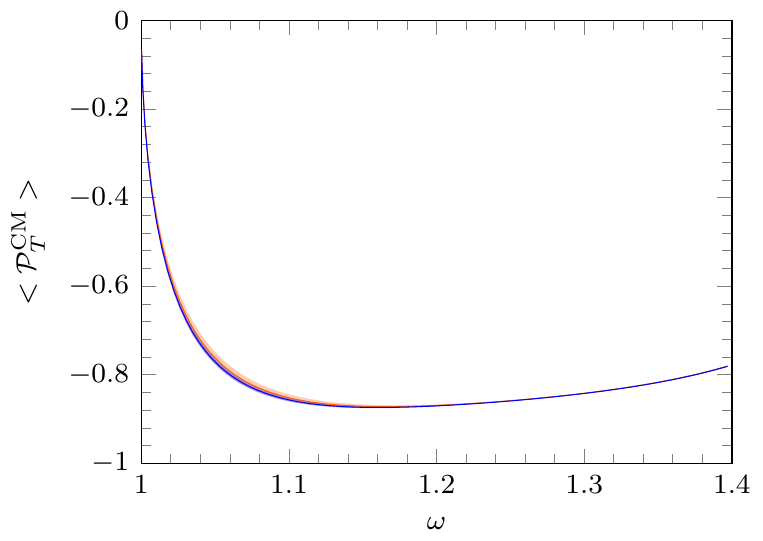}\includegraphics[scale=0.75]{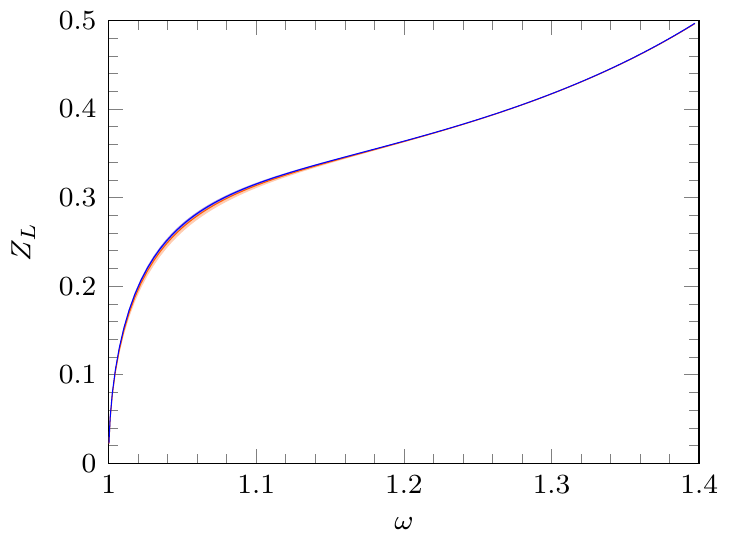}\\
\includegraphics[scale=0.75]{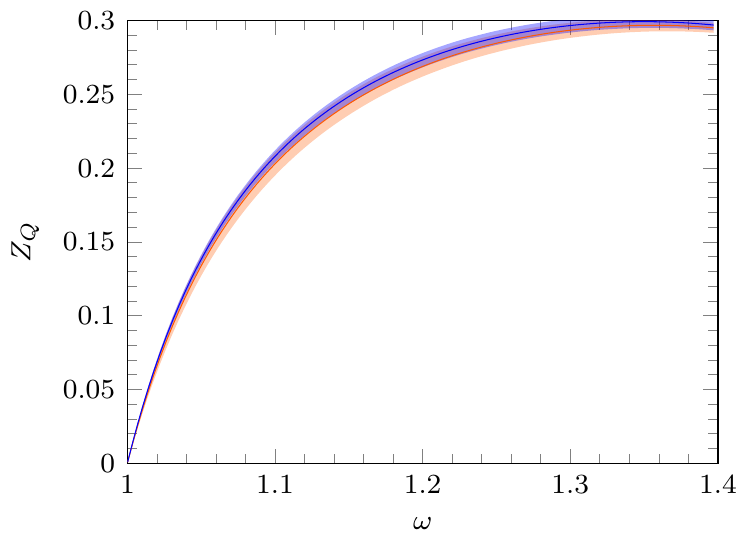}\includegraphics[scale=0.75]{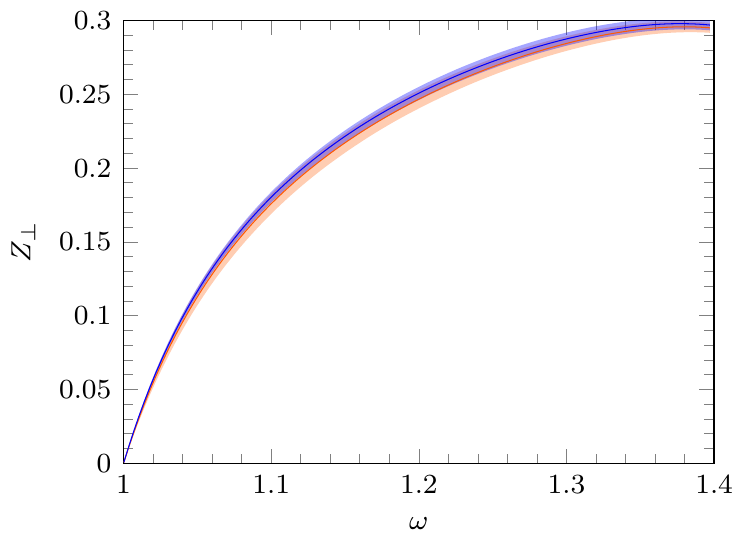}
\caption{ Same as Fig.~\ref{fig:dsstarsmasi} for the $\bar B_s\to D_s$ semileptonic decay.}
\label{fig:dssmasi}
\end{center}
\end{figure}%
 All the above observables ($n_0,\langle P^{\rm CM}_{L,T}\rangle, A_{FB,Q}$ and $ Z_{L,Q,\perp}$) are determined by the matrix elements of the $b\to c$ current between the initial $(H_b)$ and final $(H_c)$ hadrons. After summing over hadron polarizations the hadron tensors can be expressed  in terms of  Lorentz scalar  structure functions, which depend on $q^2$ or equivalently on $\omega$, the hadron masses and some   Wilson coefficients if physics beyond the SM is considered.  Lorentz, parity and time-reversal  transformations of the hadron currents and states limit their number, as  discussed in detail in Ref.~\cite{Penalva:2020xup}. The discussion of Subsec.\ 2.2  of Ref.~\cite{Penalva:2021wye} shows how to get the unpolarized $d\Gamma_{\rm SL}/d\omega$ distribution and the tau spin, angular and spin-angular asymmetries in terms of general structure functions which can be obtained from the matrix elements of the relevant hadron operators. The matrix elements are in turn parametrized in terms of form-factors. The findings of Refs.~\cite{Penalva:2020xup,Penalva:2021wye} are quite general and can be applied not only to the SM but also to any extension  of the SM based on the low-energy effective Hamiltonian comprising the full set of dimension-6 semileptonic $b\to c \tau \bar\nu_\tau$ operators with left- and right-handed neutrino fields.

For pseudoscalar meson decay into pseudoscalar or vector mesons, the relations between structure functions and form factors can be found in Appendix~B of Ref.~\cite{Penalva:2020ftd}. 

In Figs.~\ref{fig:dsstarsmasi} and \ref{fig:dssmasi} we show, for the
$\bar B_s\to D_s^*$ and $\bar B_s\to D_s$ semileptonic decays respectively, the results for the SM $d\Gamma/dq^2$ differential decay width and the different asymmetries, introduced above, that can be obtained from the measurement of the visible kinematics of the charged $\tau$-decay product.  
Only the differential $d\Gamma/dq^2$ distribution was shown in the original LQCD work of Ref.~\cite{McLean:2019qcx} (\cite{Harrison:2023dzh}) for $\bar B_s \to D_s$ ($\bar B_s \to D^*_s$). The tau forward-backward angular $A_{FB}$ and spin  $\langle P^{\rm CM}_L\rangle$ asymmetries were also presented for $\bar B_s \to D^*_s$ in Ref.~\cite{Harrison:2021tol} but with lattice form factors that have now been  superseded by the new ones evaluated in Ref.~\cite{Harrison:2023dzh}. 
The rest of the observables are shown here for the very first time for the SM in Figs.~\ref{fig:dsstarsmasi} and \ref{fig:dssmasi} and for some extensions of the SM in the next section. As for Figs.~\ref{fig:dsstarsmasi} and \ref{fig:dssmasi}, they have been evaluated both with the SM-LQCD form factors from 
Refs.~\cite{Harrison:2021tol,McLean:2019qcx} and  with the SM-HQET form factors obtained in Subsec.~\ref{sec:SMHQSSfit}. The two results  agree within uncertainties in all cases. 

All this gives us confidence in the quality of the fitted HQET IW functions so that we can go a step further and use the relations in Ref.~\cite{Bernlochner:2017jka} to obtain in addition the HQSS scalar and pseudoscalar  form factors of the two $\bar B_s\to D_s^{(*)}$ semileptonic transitions and the tensor one for the $\bar B_s\to D_s$ decay . Such a scheme relies on Eqs.~(14) and (15) of Ref.~\cite{Bernlochner:2017jka}  and it also properly includes short-distance and $1/m_{c,b}$ [${\cal O}(\alpha_s,\Lambda_{\rm QCD}/m_{c,b})$] corrections for the NP form-factors.
Using the full set of HQSS form factors we can address, in the next section, the possibility of NP effects in these two decays.

\section{New physics effects in $\bar B_s\to D_s^{(*)}\tau^-\nu_\tau$ semileptonic decays}
\label{sec:NP}

Following Ref.~\cite{Fajfer:2012vx}, to account for NP effects in a model independent way, we shall take a phenomenological effective field theory approach in which we consider all dimension-six $b\to c\tau\bar\nu_\tau$ semileptonic operators (see Sec.~\ref{sec:eh} below). These effective low energy operators are assumed to be generated by BSM physics that enters at a much higher energy scale. Their strengths  are governed by Wilson coefficients (WCs)
 that can be fitted to experimental data.  This data typically includes the 
 ${\cal R}_{D^{(*)}}=\Gamma(B\to D^{(*)}\tau^-\bar\nu_\tau)/
 \Gamma(B\to D^{(*)}\mu^-\bar\nu_\mu)$ ratios, the tau longitudinal polarization  
asymmetry  and the  longitudinal  $D^*$ polarization (also  measured
by Belle~\cite{Belle:2016dyj, Belle:2019ewo}), the $\tau$ forward-backward
asymmetry and the upper bound for the $\bar B_c\to \tau\bar\nu_\tau$  decay rate~\cite{Alonso:2016oyd}. There have been a large number of calculations along these lines, for the  
 $\bar B \to D^{(*)}$~\cite{Nierste:2008qe, Tanaka:2012nw, Fajfer:2012vx, 
Duraisamy:2013pia,Duraisamy:2014sna,Becirevic:2016hea, Ligeti:2016npd, Ivanov:2017mrj,
Bernlochner:2017jka, Blanke:2018yud, Bhattacharya:2018kig, Colangelo:2018cnj,Murgui:2019czp, 
Shi:2019gxi, Alok:2019uqc, Mandal:2020htr, Kumbhakar:2020jdz,Iguro:2020cpg, 
Bhattacharya:2020lfm,  Penalva:2021gef,Penalva:2020ftd},   
$\bar B_c\to J/\psi,\eta_c$~\cite{Dutta:2017xmj,Tran:2018kuv,Leljak:2019eyw,Harrison:2020nrv, Penalva:2020ftd}, $\Lambda_b \to \Lambda_c$ ~\cite{Dutta:2015ueb,Shivashankara:2015cta, Li:2016pdv,Datta:2017aue,Ray:2018hrx,
Blanke:2018yud,Gutsche:2018nks,Bernlochner:2018bfn,DiSalvo:2018ngq,Blanke:2019qrx,
Boer:2019zmp,Murgui:2019czp,Mu:2019bin,Hu:2020axt, Penalva:2019rgt, 
Penalva:2020xup, Penalva:2021gef,Bernlochner:2022hyz} and\footnote{The
 isoscalar $\Lambda_c(2595)$ and $\Lambda_c(2625)$, with $J^P=1/2^-$ and $3/2^-$ respectively, are promising candidates for the lightest heavy-quark-spin doublet
 of negative-parity charmed-baryon  resonances~\cite{Leibovich:1997az,Papucci:2021pmj,Du:2022rbf}, although some reservations are given in~\cite{Nieves:2019nol}.  Experimental distributions for the semileptonic decay of the ground-state bottom baryon $\Lambda_b$ into both excited states would definitely help shed light on this issue~\cite{Du:2022rbf}.} $\Lambda_b \to \Lambda_c(2595), \Lambda_c(2625)$~\cite{Leibovich:1997az,Boer:2018vpx,Gutsche:2018nks,Nieves:2019kdh,Meinel:2021rbm,Papucci:2021pmj,Du:2022ipt}  semileptonic decays.

Here, profiting from the lattice determination of the  $\bar B_s\to D_s$ SM form factors~\cite{McLean:2019qcx} and the $\bar B_s\to D_s^*$ SM and tensor form factors~\cite{Harrison:2023dzh}, together with the 
 HQET study of $\bar B\to D^{(*)}$ form factors in Ref.~\cite{Bernlochner:2017jka}, we have  
obtained all the  $\bar B_s\to D^{(*)}_s$ form factors needed for a similar study of the $\bar B_s\to D^{(*)}_s\tau\bar\nu_\tau$ semileptonic decays. If NP is responsible for LFUV, one would expect to see its effects in these reactions at a level similar to that found in the analogous $\bar B\to D^{(*)}$ decays. In addition to the  ${\cal R}_{D^{(*)}_s} =\Gamma(\bar B\to D^{(*)}_s\tau\bar\nu_\tau)/\Gamma(\bar B_s\to D^{(*)}_s\ell\bar\nu_{\ell})$ ratios, we will investigate the role that the different asymmetries presented in Subsec.~\ref{sec:assy} could play in establishing the presence of LFUV and, if experimentally confirmed, to distinguish between different extensions of the SM.

\subsection{$H_b\to H_c\ell^-\bar\nu_\ell$ Effective Hamiltonian}
\label{sec:eh}
The effective low-energy Hamiltonian that we use follows Ref.~\cite{Mandal:2020htr} and  it 
includes  all possible dimension-six semileptonic $b\to c$ operators with both left-handed (L) 
 and right-handed (R)
neutrino fields,
\bea
H_{\rm eff}&=&\frac{4G_F V_{cb}}{\sqrt2}\left[(1+C^V_{LL}){\cal O}^V_{LL}+
C^V_{RL}{\cal O}^V_{RL}+C^S_{LL}{\cal O}^S_{LL}+C^S_{RL}{\cal O}^S_{RL}
+C^T_{LL}{\cal O}^T_{LL}\right.\nonumber \\
&&+\left. C^V_{LR}{\cal O}^V_{LR}+
C^V_{RR}{\cal O}^V_{RR}+C^S_{LR}{\cal O}^S_{LR}+C^S_{RR}{\cal O}^S_{RR}
+C^T_{RR}{\cal O}^T_{RR} \right]+h.c. .
\label{eq:hnp}
\eea
Here, the $C^X_{AB}$ ($X= S, V,T$ and 
$A,B=L,R$) are, complex in general, Wilson coefficients that parameterize 
the deviations from the SM.  They can 
 be lepton and flavor dependent although they  are generally assumed to be nonzero 
only for the third 
quark and lepton generation. The dimension six operators read
\begin{align}
\label{eq:hnp2L}
{\cal O}^V_{(L,R)L} &= (\bar c \gamma^\mu b_{L,R}) 
(\bar \ell \gamma_\mu \nu_{\ell L}), & {\cal O}^S_{(L,R)L} &= 
(\bar c\,  b_{L,R}) (\bar \ell \, \nu_{\ell L}), & {\cal O}^T_{LL} &= 
(\bar c\, \sigma^{\mu\nu} b_{L}) (\bar \ell \sigma_{\mu\nu} \nu_{\ell L}),\\
\label{eq:hnp2R}
{\cal O}^V_{(L,R)R} &= (\bar c \gamma^\mu b_{L,R}) 
(\bar \ell \gamma_\mu \nu_{\ell R}), &  {\cal O}^S_{(L,R)R} &= 
(\bar c\,  b_{L,R}) (\bar \ell \, \nu_{\ell R}), & {\cal O}^T_{RR} &= 
(\bar c\, \sigma^{\mu\nu} b_{R}) (\bar \ell \sigma_{\mu\nu} \nu_{\ell R}),
\end{align}
with $\psi_{R,L}= (1 \pm \gamma_5)\psi/2$.  The effective Hamiltonian can be rewritten as~\cite{Penalva:2021wye}
\bea
H_{\rm eff}&=&\frac{4G_F V_{cb}}{\sqrt2}\sum_{\chi=L,R}\Big[
\bar c(C_\chi^V\gamma^\mu+h_\chi C_\chi^A\gamma^\mu\gamma_5)b\ \bar l\gamma_\mu
\nu_{l\chi}+\bar c\,(C_\chi^S+h_\chi C_\chi^P\gamma_5)b\ \bar l\gamma_\mu
\nu_{l\chi}\nonumber\\
&&\hspace{3cm}+C_\chi^T\,\bar c\,\sigma^{\mu\nu}(1+h_\chi\gamma_5)b\ \bar l\sigma_{\mu\nu}\nu_{l\chi}\big]
\label{eq:eh2}
\eea
with $h_L=-1,h_R=+1$ and
\be
\begin{aligned}
C^V_L &= (1+C^V_{LL}+C^V_{RL}), & C^A_L &= (1+C^V_{LL}- C^V_{RL}), \\
C^S_L &= (C^S_{LL}+ C^S_{RL}), &
C^P_L &= (C^S_{LL}- C^S_{RL}), & C^T_L &= C^T_{LL}, \\
C^V_R &= (C^V_{LR}+ C^V_{RR}), & C^A_R &=-(C^V_{LR}- C^V_{RR}),\\
C^S_R &= (C^S_{LR}+ C^S_{RR}), &
C^P_R &=-(C^S_{LR}- C^S_{RR}), & C^T_R &= C^T_{RR}, 
\end{aligned}
\label{eq:WCR}
\ee

We shall compare results obtained in the SM and in three different NP extensions. The latter correspond to the L Fit 7 of Ref~\cite{Murgui:2019czp}, where only left-handed neutrino operators are considered, the R S7a scenario of Ref.~\cite{Mandal:2020htr}
 with only right-handed neutrino operators, and the left-handed neutrino  L $R_2$ leptoquark model of Ref.~\cite{Shi:2019gxi}, for which the two nonzero WCs ($C^S_{LL}$ and $C^T_{LL}$) are complex\footnote{The numerical values that we use for  these two WCs can be found at the beginning of Subsec. 4.2.1 of Ref.~\cite{Penalva:2021gef}.}. In this latter case
the effective Hamiltonian  violates CP. 

None of the observables $d\Gamma_{\rm SL}/d\omega,\langle P^{\rm CM}_{L,T}\rangle, A_{FB,Q}$ and $ Z_{L,Q,\perp}$
entering Eqs.~\eqref{eq:visible-distr} and \eqref{eq:coeff} are   sensitive to CP-symmetry breaking terms~\cite{Penalva:2021gef,Penalva:2021wye}. Hence, we will also show results for the L $R_2$ leptoquark model of  Ref.~\cite{Shi:2019gxi} for other distributions, related to the tau polarization
component ($P_{TT}$) along an axis perpendicular to the hadron-tau plane~\cite{Penalva:2021gef},
which could be accessed if one could further measure the azimuthal  angle ($\phi_d$)  of the  charged  $d$ particle (see Fig.~1 of Ref.~\cite{Penalva:2022vxy}). Note that in the differential distribution given in Eq.~\eqref{eq:visible-distr} this angle has been integrated out since measuring $\phi_d$ would require a full reconstruction of the tau three-momentum.
    The latter can be circumvented  through the analysis
  of  distributions that also involve the decay products of the $H_c$ hadron. Thus,
  some CP-odd observables have been presented for $\bar B \to D^*$  and $\Lambda_b\to
   \Lambda_c$ decays  in Refs.~\cite{Duraisamy:2013pia, 
Duraisamy:2014sna, Ligeti:2016npd, Bhattacharya:2020lfm}  and Refs.~\cite{Boer:2019zmp,Hu:2020axt} respectively.

As already mentioned, we refer the reader to   Ref.~\cite{Penalva:2021wye}, and references therein,  for a full account of our formalism.

 \subsection{Partially integrated sequential 
$H_b\to H_c\tau^-(\pi^-\nu_\tau,\rho^-\nu_\tau,\ell^-\bar\nu_\ell\nu_\tau)\bar\nu_\tau$ decay distributions}

The feasibility of NP studies can be severely limited, however, by the 
 statistical precision in the measurement of the triple differential decay width of Eq.~\eqref{eq:visible-distr}. One can increase statistics, at the expense of losing information in some of the observables,  by integrating over one or more of the 
$\omega,\xi_d$ and $\theta_d$ variables,  although in this case not 
 all observables entering in Eq.~\eqref{eq:coeff} can be extracted. In this way one can obtain the distributions~\cite{Penalva:2022vxy}
\bea
\frac{d^2\Gamma_d}{d\omega  d\xi_d} & = & 2{\cal B}_{d}
\frac{d\Gamma_{\rm SL}}{d\omega}\Big\{  C_n^d(\omega,\xi_d)+C_{P_L}^d(\omega,\xi_d)\,\langle 
P^{\rm CM}_L\rangle(\omega)\Big\},\label{eq:wE}
\eea
from which only $d\Gamma_{\rm SL}/d\omega$ and  the CM $\tau$ longitudinal 
polarization can be extracted, or
\bea
\frac{d^2\Gamma_d}{d\omega  d\cos\theta_d}  =  {\cal B}_{d}
\frac{d\Gamma_{\rm SL}}{d\omega} \Big[ 
\frac12+ \widetilde F^d_1(\omega) 
\cos\theta_d +\widetilde  F^d_2(\omega)P_2(\cos\theta_d)\Big],
\label{eq:visible-distr_theta_d}
\eea
with
\bea
\widetilde F^d_1(\omega)&=&C^d_{A_{FB}}(\omega)\,A_{FB}(\omega)+C^d_{Z_L}(\omega)
\,Z_L(\omega)+C^d_{P_T}(\omega)\,\langle P_T^{\rm CM}\rangle(\omega),
\label{eq:F1} \\
\widetilde F^d_2(\omega)&=&C^d_{A_Q}(\omega)\,A_Q(\omega)+C^d_{Z_Q}(\omega)
\,Z_Q(\omega)+C^d_{Z_\perp}(\omega)\,Z_\perp(\omega),\label{eq:F2}
\eea
which retains information on $d\Gamma_{\rm SL}/dq^2$ and six out of the seven original asymmetries. The latter cannot, however, be extracted from the knowledge of $\widetilde F^d_1$ and $\widetilde F^d_2$ alone.

One can further integrate over $\omega$ to obtain~\cite{Penalva:2022vxy}
\bea
\frac{d\Gamma_d}{d\cos\theta_d}={\cal B}_d\Gamma_{\rm SL}\Big[
\frac12+\widehat F_1^d\cos\theta_d+\widehat F_2^d\, 
P_2(\cos\theta_d)\Big], \quad  \widehat F_{1,2}^d=\frac1\Gamma_{\rm SL}\int_1^{\omega_{\rm max}}
\frac{d\Gamma_{\rm SL}}{d\omega}\widetilde F_{1,2}^d(\omega)\,d\omega.\label{eq:Gcos}
\eea
and
\bea
\frac{d\Gamma_d}{dE_d} & = & 2{\cal B}_{d} \int_{\omega_{\rm inf}(E_d)}^{\omega_{\rm sup}(E_d)} d\omega
\frac{1}{\gamma m_\tau}\frac{d\Gamma_{\rm SL}}{d\omega}\Big\{ 
C_n^d(\omega,\xi_d)+C_{P_L}^d(\omega,\xi_d)\,\langle P^{\rm CM}_L\rangle(\omega)\Big\},
\label{eq:EdG}
\eea
where $\gamma=(q^2+m_\tau^2)/(2m_\tau\sqrt{q^2})$ and,  in the latter case, the appropriate $\omega$ limits can be found in 
Ref.~\cite{Penalva:2022vxy}.

 Although the information on the individual asymmetries is now completely lost,
 the above two distributions could still be useful observables in the search for NP beyond the SM.

\subsection{NP results and discussion}
\subsubsection{LFUV ratios, unpolarized differential decay widths and  tau angular, spin and spin-angular asymmetries} 
\begin{figure}
\begin{center}
\includegraphics[scale=0.75]{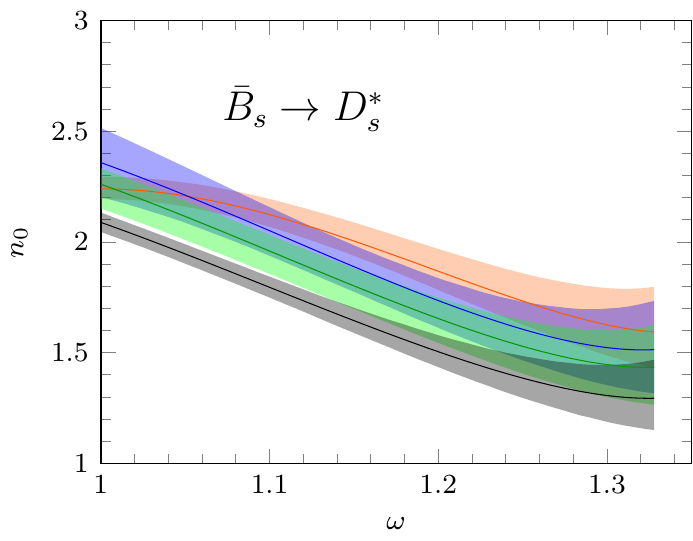}\ \ 
\includegraphics[scale=0.75]{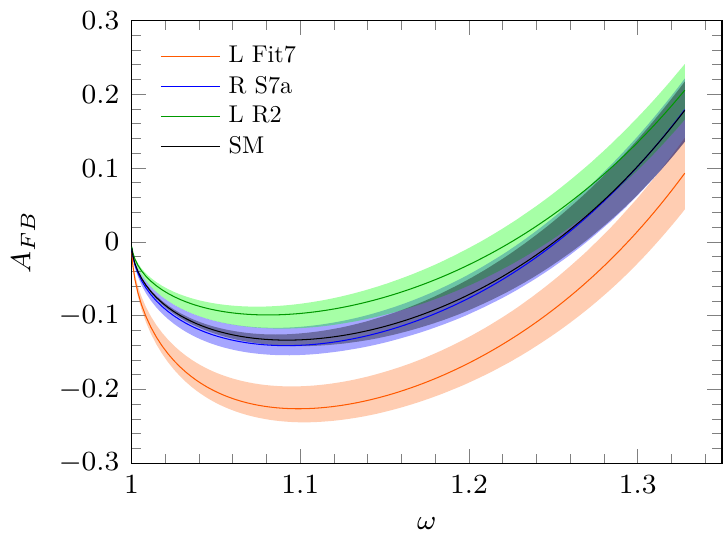}\\ 
\includegraphics[scale=0.75]{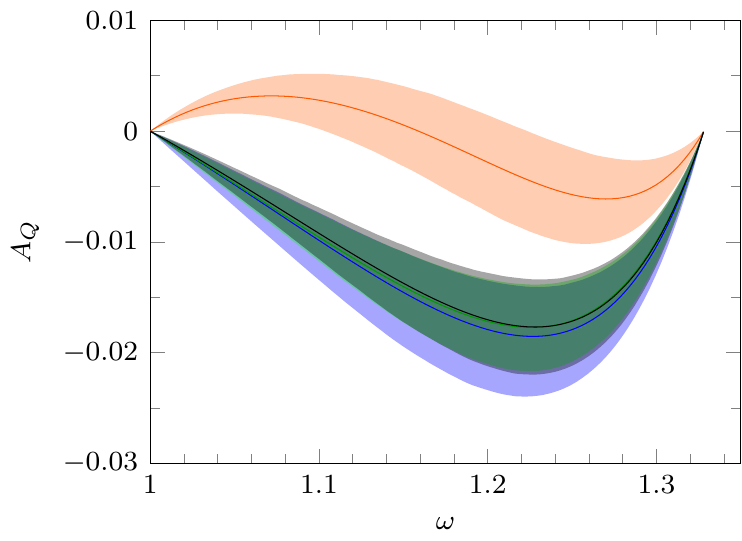}\ \ \includegraphics[scale=0.75]{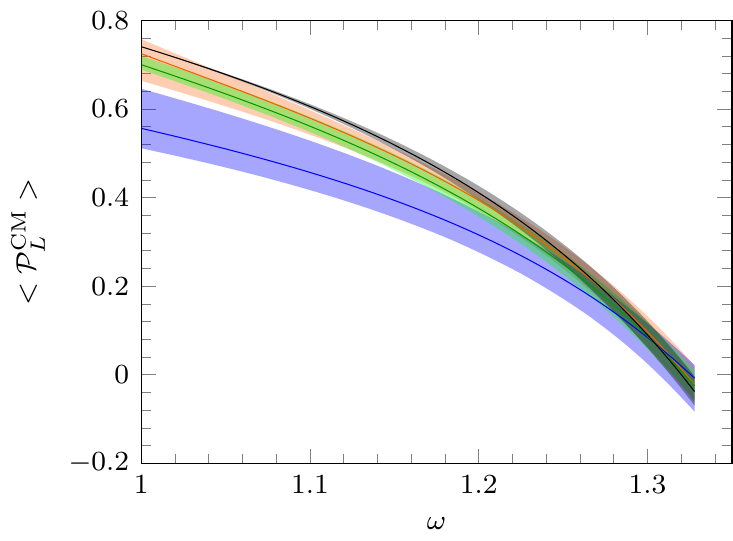}
\includegraphics[scale=0.75]{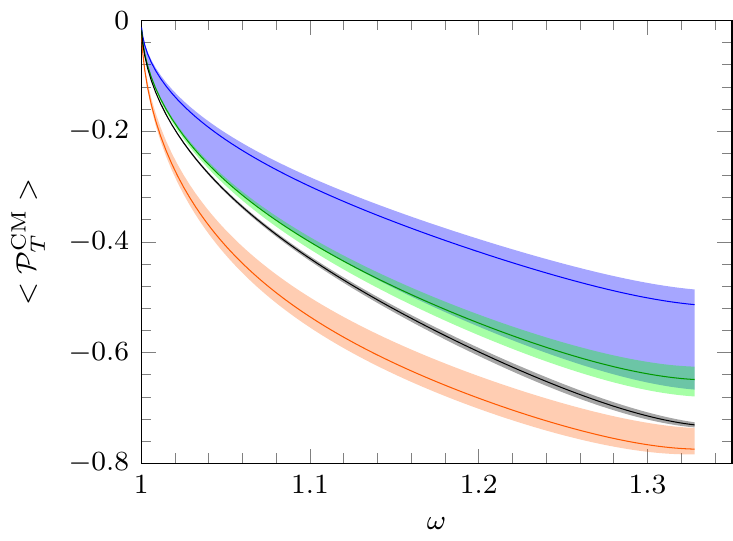}\ \ \includegraphics[scale=0.75]{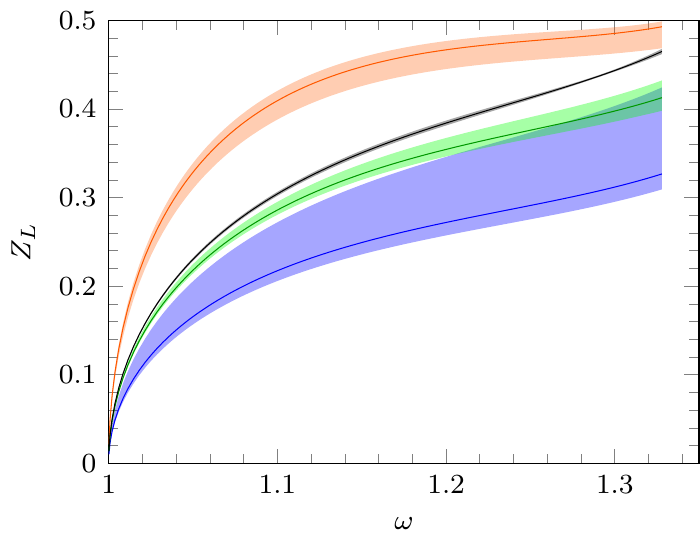}\\
\includegraphics[scale=0.75]{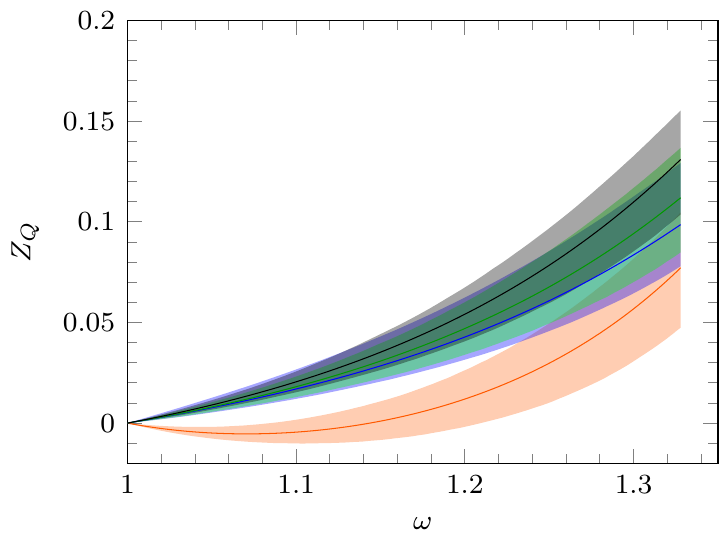}\  \ \includegraphics[scale=0.75]{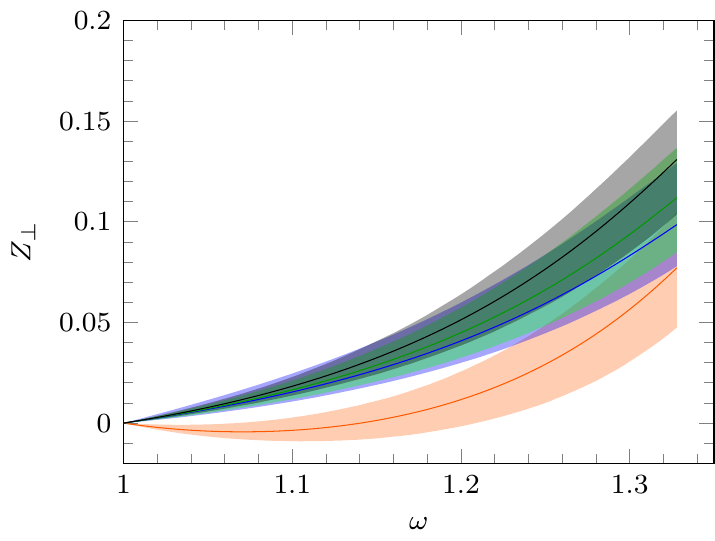}
\caption{ Distribution $n_0$ from Eq.~(\ref{eq:n0w}) and the tau asymmetries
introduced in Eq.~(\ref{eq:coeff}) for the $\bar B_s\to D^*_s\tau\bar\nu_\tau$ decay. We compare the results for these observables obtained in the SM and  the NP models L Fit 7, R S7a and
L $R_2$ of Refs~\cite{Murgui:2019czp}, \cite{Mandal:2020htr} and ~\cite{Shi:2019gxi}, respectively. We use the HQET form-factors derived from the LQCD form factors obtained in Refs.~\cite{Harrison:2023dzh,McLean:2019qcx}. }
\label{fig:dsstarasi}
\end{center}
\end{figure}

\begin{figure}
\begin{center}
\includegraphics[scale=0.75]{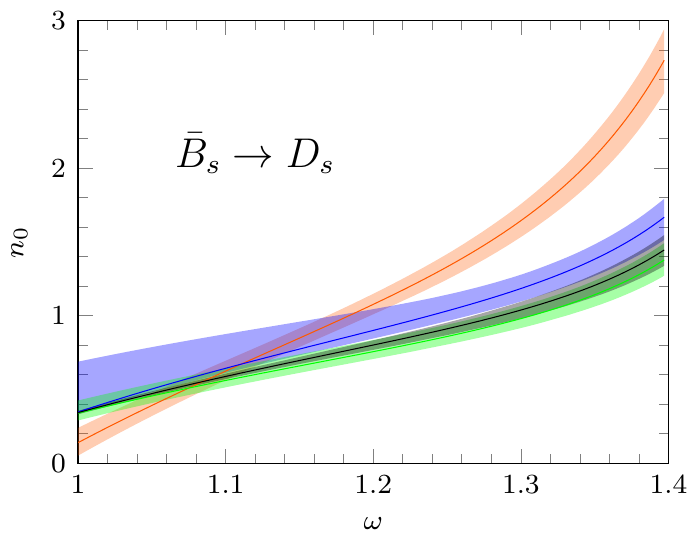}\ \ 
\includegraphics[scale=0.75]{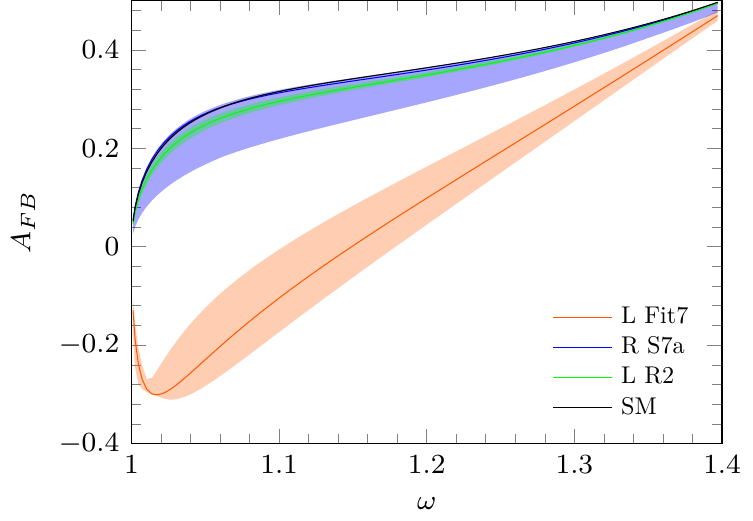}\\ 
\includegraphics[scale=0.75]{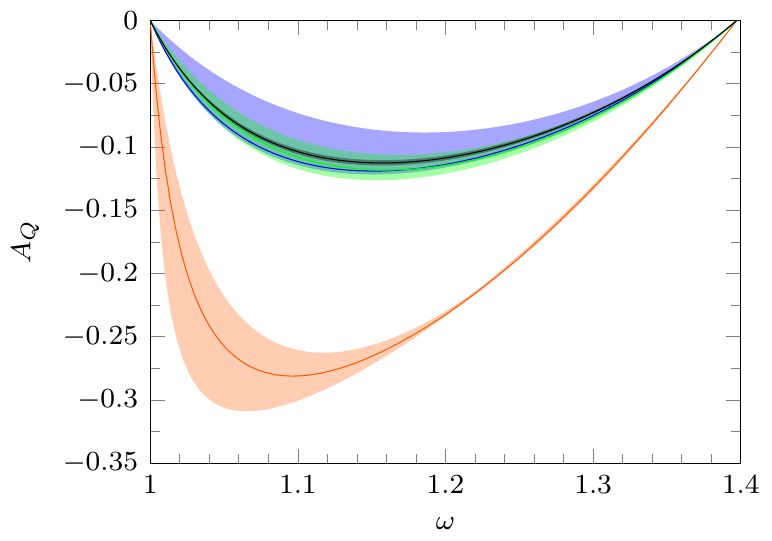}\ \ \includegraphics[scale=0.75]{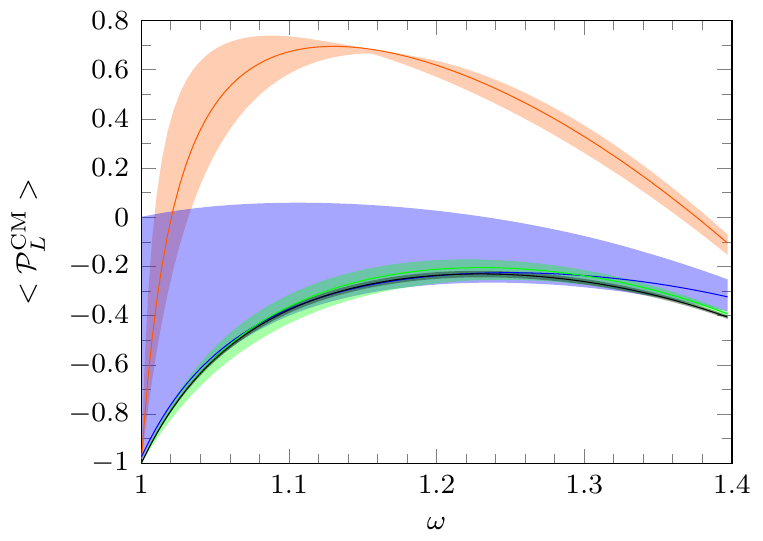}
\includegraphics[scale=0.75]{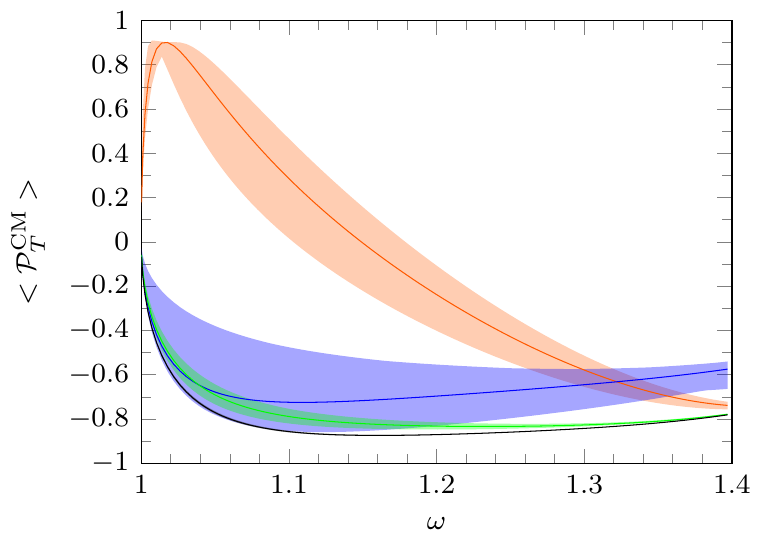}\ \ \includegraphics[scale=0.75]{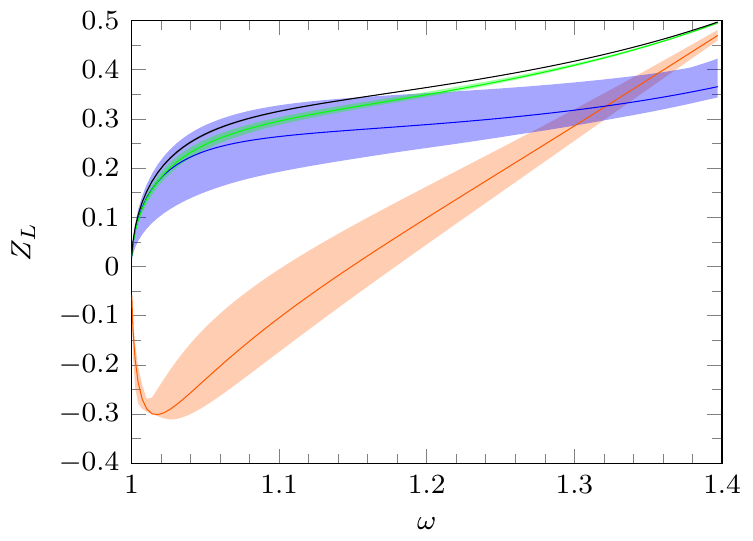}\\
\includegraphics[scale=0.75]{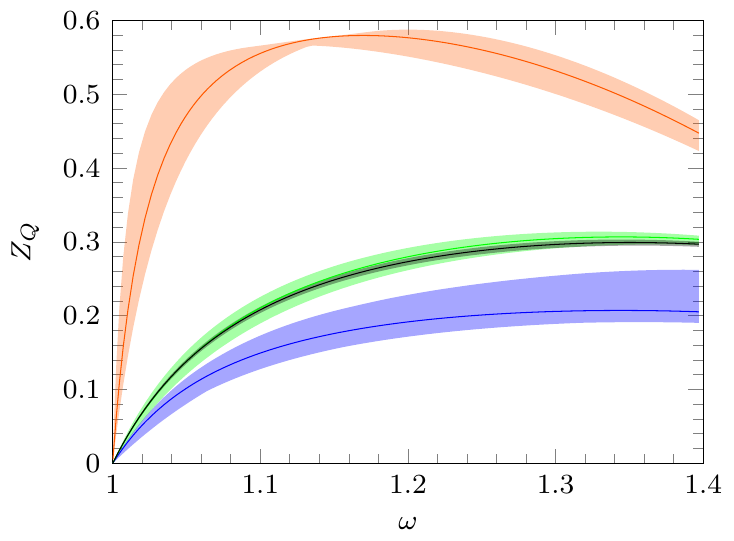}\  \ \includegraphics[scale=0.75]{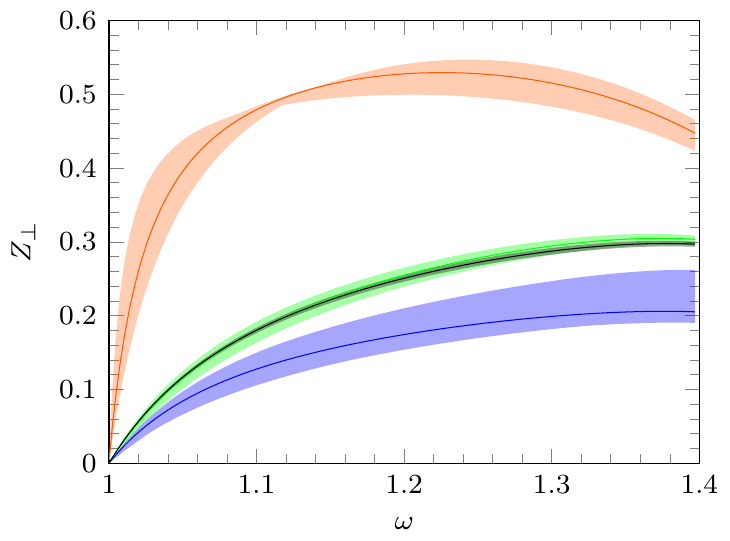}
\caption{ Same as Fig.~\ref{fig:dsstarasi}  but for the $\bar B_s\to D_s\tau\bar\nu_\tau$ transition.}
\label{fig:dsasi}
\end{center}
\end{figure}

We start by showing, in Table~\ref{tab:ratiosDs}, the values for the
 semileptonic decay widths $\Gamma_\tau=\Gamma(\bar B_s\to D^{(*)}_s\tau\bar\nu_\tau)$ 
and   $\Gamma_{\ell}=\Gamma(\bar B_s\to D^{(*)}_s\ell\bar\nu_{\ell})$, with
$\ell=e,\mu$, and the corresponding  ${\cal R}_{D^{(*)}_s}$ ratios,
evaluated within the SM and the three NP extensions,  L Fit 7 of Ref~\cite{Murgui:2019czp}, R S7a scenario of Ref.~\cite{Mandal:2020htr} and the L $R_2$ leptoquark model of  Ref.~\cite{Shi:2019gxi}, considered in this study.  Our results for the SM ratios are compatible with those obtained using a dispersive matrix approach in~\cite{Martinelli:2022xir}. Both use HPQCD lattice data, but our values make use of the updated values from~\cite{Harrison:2023dzh}; we find that the agreement is closer if we base our analysis on the same HPQCD inputs~\cite{McLean:2019qcx,Harrison:2021tol} as used in~\cite{Martinelli:2022xir}\footnote{The updated HPQCD results in~\cite{Harrison:2023dzh} have been used in the dispersive matrix method for $\bar B\to D^*$ semileptonic decays in~\cite{Martinelli:2023fwm}, but not yet used for $\bar B_s\to D_s^*$.}.
For the first two NP models, we clearly see the ratios deviate from the SM prediction\footnote{The LQCD results in Refs.~\cite{McLean:2019qcx} and \cite{Harrison:2023dzh} are ${\cal R}^{\rm SM}_{D_s}=0.2993 (46)$ and ${\cal R}^{\rm SM}_{D_s^*}=0.265(9)$, which are in excellent agreement with the prediction quoted in Table~\ref{tab:ratiosDs} obtained with the HQET parameterization of the $\bar B_s \to D^{(*)}_s$ form-factors.  
}. Their central values are higher than SM ones, with the highest one corresponding always to L Fit 7, which leads to ratios around 5$\sigma$ above the SM predictions.  The results are similar to those obtained in Ref.~\cite{Penalva:2022vxy} for the analogous $\bar B\to D^{(*)}$ decays (see Table 3 of that reference). In the L $R_2$ case, ${\cal R}_{D_s^*}$ is larger than the SM value while ${\cal R}_{D_s}$ is lower and compatible within errors.

\begin{table}[h!]
\begin{center}Distribution
\begin{tabular}{c|c|ccccc}\hline\hline
                        &&  SM  & L Fit 7 \cite{Murgui:2019czp}& R S7a 
			\cite{Mandal:2020htr}& L $R_2$ \cite{Shi:2019gxi}
                       \\\hline\tstrut
 &$\Gamma_{e(\mu)}$ & $0.93^{+0.07}_{-0.08}$\\ \tstrut 
$\bar B_s\to D_s$ &$\Gamma_\tau$ &$0.27\pm0.01$ &$0.36\pm0.02$ &$0.306^{+0.062}_{-0.019}$&$0.259^{+0.022}_{-0.015}$\\ \tstrut
& ${\cal R}_{D_s}$ &$0.295^{+0.015}_{-0.011}$&$0.387^{+0.024}_{-0.019}$&$0.329^{+0.067}_{-0.018}$
&$0.279^{+0.026}_{-0.015}$ 
 \\   \hline\tstrut
 &$\Gamma_{e(\mu)}$ & $1.91^{+0.13}_{-0.10}$\\ \tstrut 
$\bar B_s\to D^*_s$ &$\Gamma_\tau$& $0.507^{+0.018}_{-0.015}$&$0.605^{+0.025}_{-0.023}$&$0.581^{+0.031}_{-0.033}$&$0.554^{+0.020}_{-0.018}$\\ \tstrut
& ${\cal R}_{D^*_s}$ &
$0.265^{+0.009}_{-0.010}$&$0.316^{+0.014}_{-0.015}$&$0.304^{+0.016}_{-0.019}$&
$0.290\pm0.019$\\\hline\hline
\end{tabular}
\end{center}
\caption{Semileptonic decay widths $\Gamma_\tau=\Gamma(\bar B_s\to D^{(*)}_s\tau\bar\nu_\tau)$ 
and  $\Gamma_{e(\mu)}=\Gamma[\bar B_s\to D^{(*)}_s\, e(\mu)\bar\nu_{e(\mu)}]$ (in units of  $10\times |V_{cb}|^2 {\rm ps}^{-1}$)
and  ratios ${\cal R}_{D^{(*)}_s} =\Gamma(\bar B_s\to D^{(*)}_s\tau\bar\nu_\tau)
/\Gamma[\bar B_s\to D^{(*)}_s\, e(\mu)\bar\nu_{e(\mu)}]$ 
obtained using the SM-HQSS form factors,  the NP model L Fit 7 (R S7a) of Ref~\cite{Murgui:2019czp} (\cite{Mandal:2020htr}), which only includes left- (right-)handed 
neutrino NP operators and the L $R_2$ leptoquark model of Ref.~\cite{Shi:2019gxi}. Errors induced by the uncertainties in
the form-factors and  Wilson coefficients are added in quadrature.}
\label{tab:ratiosDs}
\end{table}

In Figs.~\ref{fig:dsstarasi} and \ref{fig:dsasi} we show now the values for
the $n_0(\omega)$ function introduced in Eq.~(\ref{eq:n0w}), which contains all the dynamical information of the $d\Gamma_{\rm SL}/d\omega$ differential decay width, and the set of tau spin, angular and spin-angular asymmetries introduced in Eq.~(\ref{eq:coeff}). Most of the observables allow for a clear distinction between SM and L Fit 7 results, the exception being the CM longitudinal spin asymmetry $\langle P_L^{\rm CM}\rangle$ for the $\bar B_s\to D_s^*$ decay. In fact, these observables also differentiate between L Fit 7 and the other two NP scenarios. With few exceptions, notably the $Z_Q$ and $Z_\perp$ asymmetries for the 
$\bar B_s\to D_s$ decays, the R S7a and L $R_2$ NP scenarios tend to agree within errors and they are closer to the SM, especially in the case of the L $R_2$ model.

\begin{figure}
\begin{center}
\includegraphics[scale=0.75]{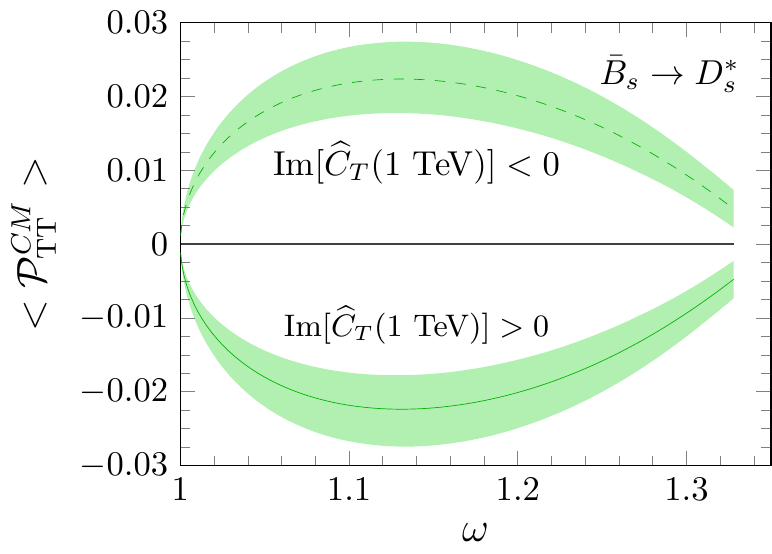}\
\includegraphics[scale=0.75]{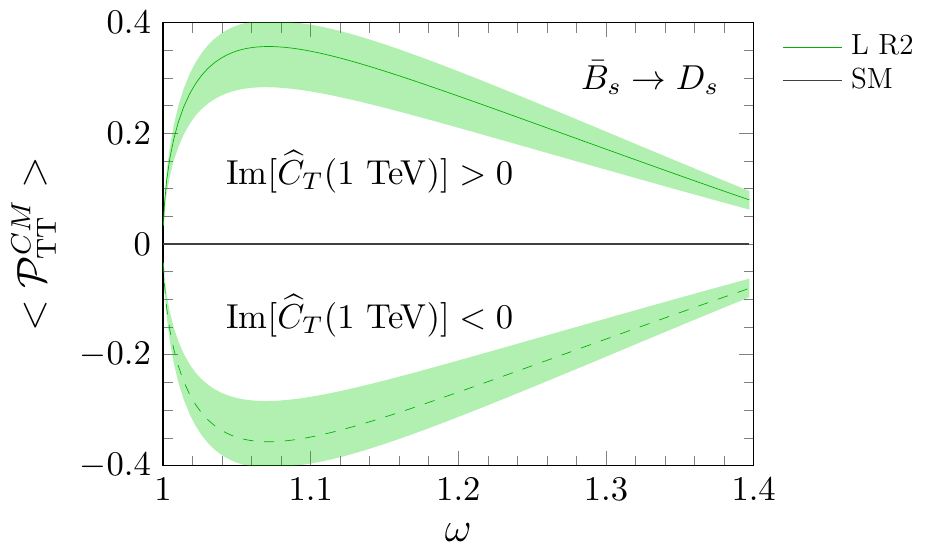}\ 
\includegraphics[scale=0.75]{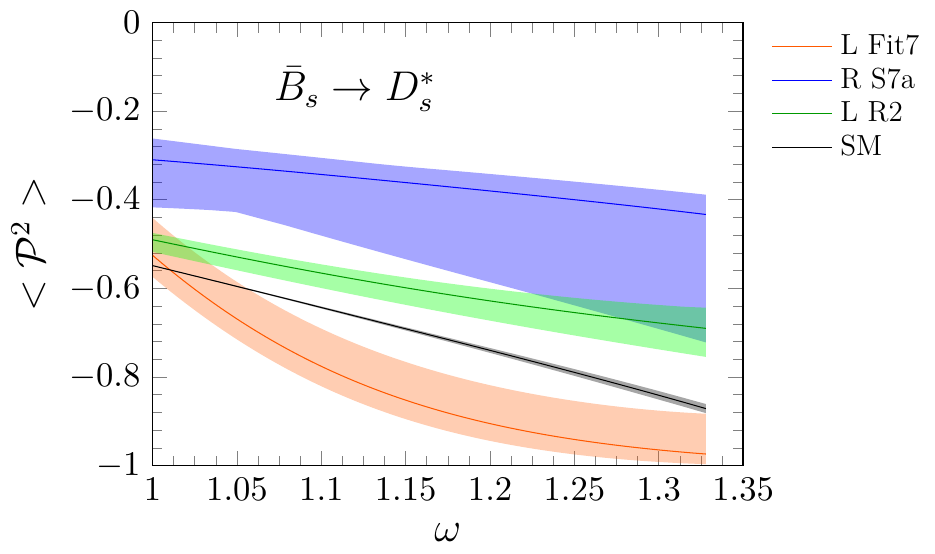}\ 
\caption{ Upper panels:  $\langle P^{\rm CM}_{TT}\rangle(\omega)$ for the  $\bar B_s\to D^*_s$ (left) and $\bar B_s\to D_s$ (right) decays evaluated with the L $R_2$ leptoquark model of Ref.~\cite{Shi:2019gxi}. Bottom panel: comparison of the $\langle P^2\rangle(\omega)$ distribution obtained in the SM and the NP extensions L Fit 7~\cite{Murgui:2019czp}, R S7a~\cite{Mandal:2020htr} and
L $R_2$~\cite{Shi:2019gxi}.}
\label{fig:ptt}
\end{center}
\end{figure}

As already mentioned, none of the observables shown so far is sensitive to CP breaking terms. 
To measure those one needs to analyze the CP violating triple product asymmetries that  involve the decay of the $H_c$ hadron~\cite{Duraisamy:2013pia,  Duraisamy:2014sna, Ligeti:2016npd, Bhattacharya:2020lfm,Boer:2019zmp,Hu:2020axt}, or otherwise to be able to fully establish the tau three-momentum. In the latter case, one has access to the    
$\langle P^{\rm CM}_{TT}\rangle(\omega)$ observable,  which gives the component of the CM tau-polarization vector 
along an axis perpendicular to the hadron-tau plane (see Eqs. (3.14), (3.24) and (3.25) of Ref.~\cite{Penalva:2021gef}). 
Among the different NP extensions considered in this work, only the L $R_2$ leptoquark model of Ref.~\cite{Shi:2019gxi}, with complex Wilson coefficients,  can generate a nonzero value for the $\langle P^{\rm CM}_{TT}\rangle(\omega)$ distribution.  In this NP model, the two nonzero WCs $C^S_{LL}$ and $C^T_{LL}$ are given, at the bottom-mass scale appropriate for the present calculation, in terms of just the value of $ C^T_{LL}$ at the  1 TeV scale, where
$C^S_{LL}(1\,{\rm TeV})= 4C^T_{LL}(1\,{\rm TeV})$, and the corresponding evolution matrix (see Ref.~\cite{Shi:2019gxi}). 
The best fit of the WCs to the $\bar B$-meson LFUV signatures does not fix the sign of the imaginary part of $C^T_{LL}(1\,{\rm TeV})$. 
Contrary to the other observables considered so far, $\langle P^{\rm CM}_{TT}\rangle(\omega)$ is linear in this imaginary part
and thus its measurement would break this degeneracy. The results for $\langle P^{\rm CM}_{TT}\rangle(\omega)$, 
using both possible signs for ${\rm Im} [C^T_{LL} (1\, {\rm TeV})]$, are shown in the upper  panels of Fig.~\ref{fig:ptt} for the $\bar B_s\to D^*_s$ (left) and $\bar B_s\to D_s$ (right) decays respectively. We see that the absolute value of this distribution is around one order of magnitude larger for the pseudoscalar than for the vector decay modes. An observation of a nonzero $\langle P^{\rm CM}_{TT}\rangle(\omega)$ value will be a clear indication of the existence of NP beyond the SM and CP violation.

In the bottom panel of Fig.~\ref{fig:ptt} we show the degree of polarization of the tau 
\bea
\langle P^2\rangle(\omega)=-\langle P^2_L+P_T^2+P^2_{TT}\rangle(\omega)
\eea
which is a Lorentz invariant quantity. As shown in Ref.~\cite{Penalva:2021gef}, this is exactly $-1$ for $0^-\to 0^-$ transitions, reflecting the fact that for such decays the outgoing taus are fully polarized. Thus we  only present the results for the $\bar B_s\to D^*_s$ decay. As seen from the figure this observable, which is sensitive to CP-odd terms in the effective Hamiltonian, discriminates very efficiently between different NP models and the SM.
\subsubsection{Distributions of charged tau decay products}

\begin{figure}
\begin{center}
\includegraphics[scale=0.6]{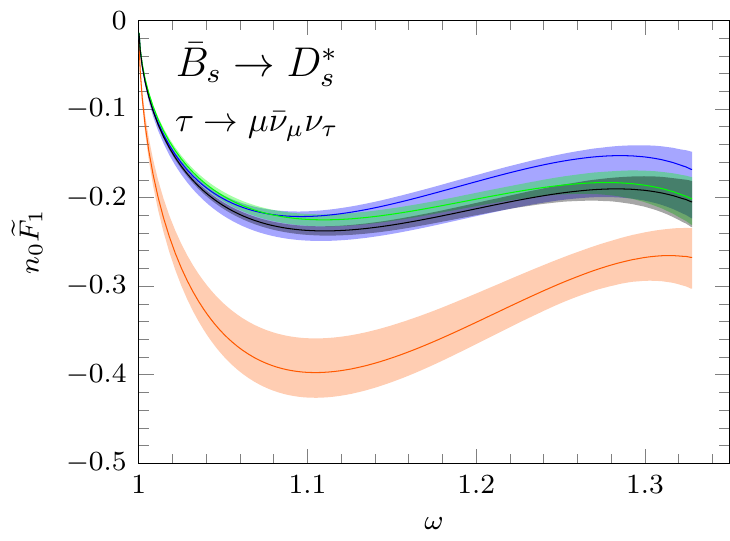}\
\includegraphics[scale=0.6]{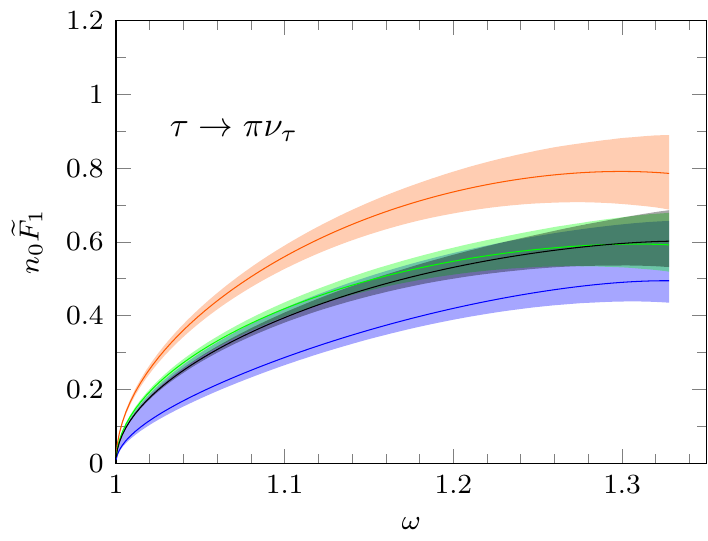}\ 
\includegraphics[scale=0.6]{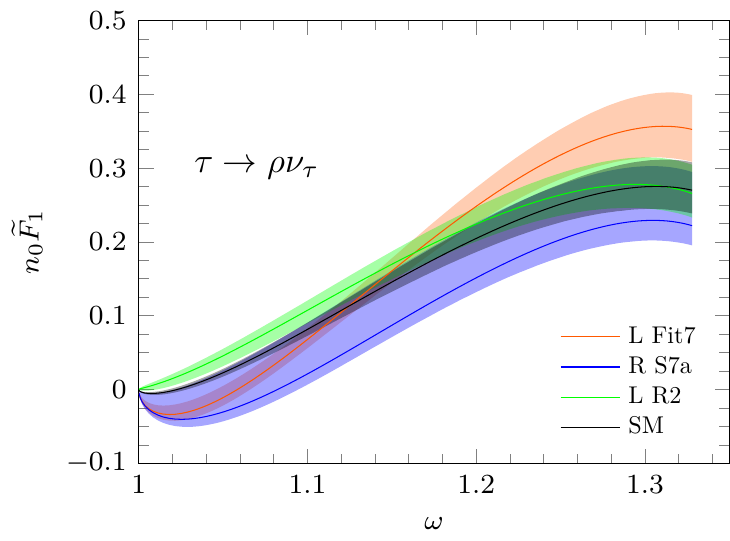}\\
\includegraphics[scale=0.6]{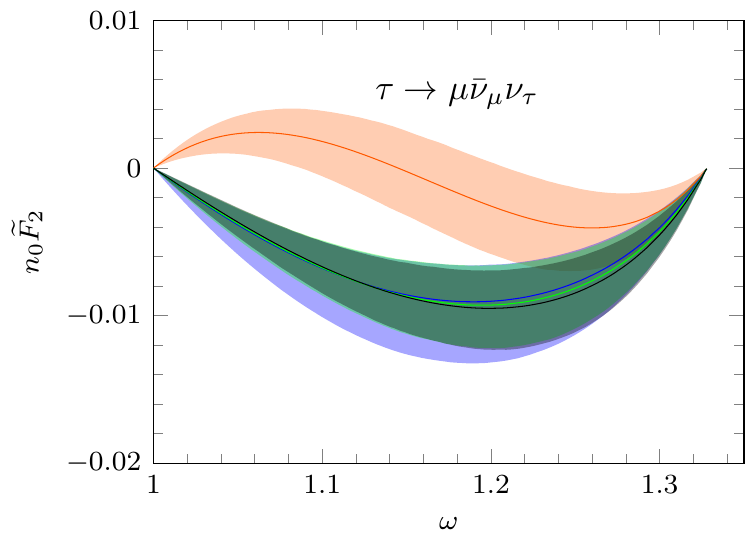}\
\includegraphics[scale=0.6]{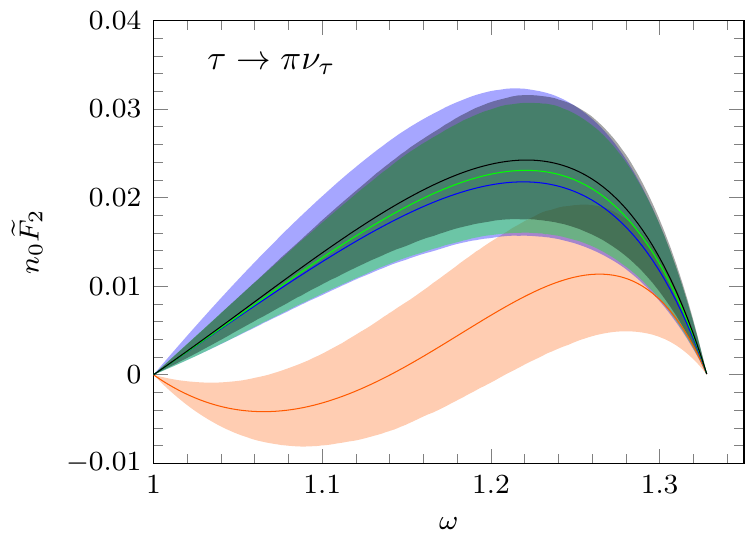}\ 
\includegraphics[scale=0.6]{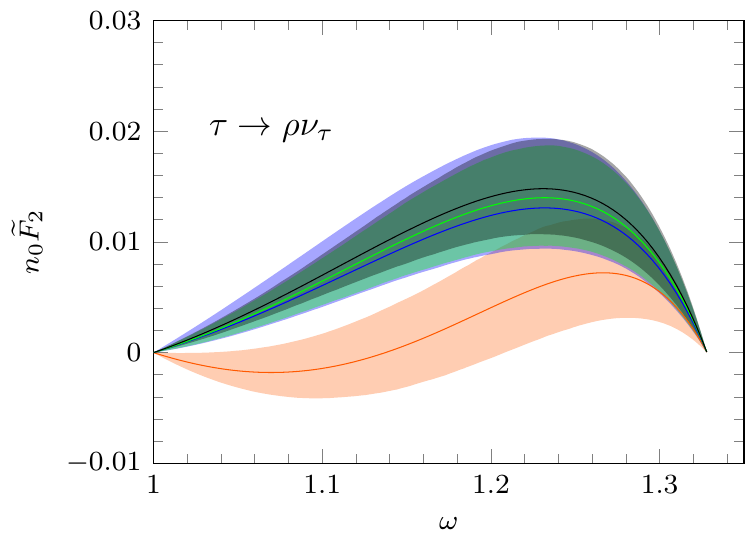}
\caption{ Distributions $[n_0\widetilde F_1^d](\omega)$ and $[n_0\widetilde F_2^d](\omega)$  obtained from $d\Gamma_d/(d\omega\, d\cos\theta_d)$ (Eq.~\eqref{eq:visible-distr_theta_d}) for the tau hadronic and leptonic $\bar B_s\to D^*_s\tau^-(\pi^-\nu_\tau,\rho^-\nu_\tau,\mu^-\bar\nu_\mu\nu_\tau)\bar\nu_\tau$ sequential decays. }
\label{fig:dsstardsn0fs} 
\end{center}
\end{figure}

\begin{figure}
\begin{center}
\includegraphics[scale=0.6]{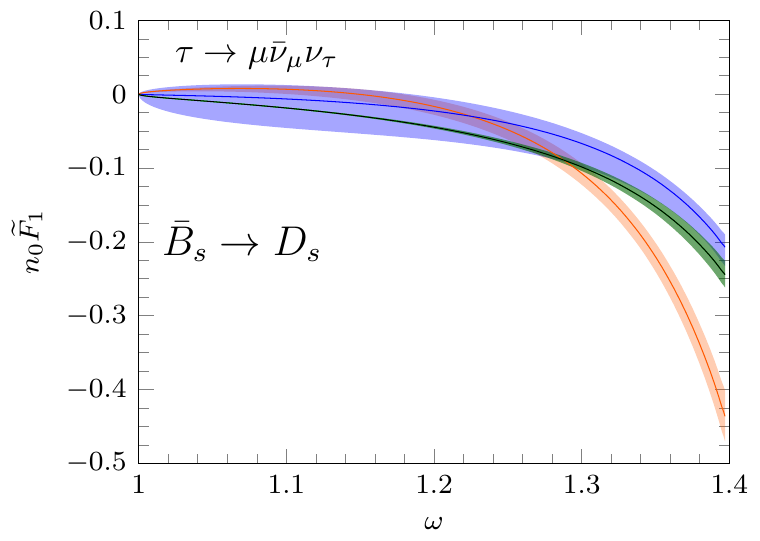}\
\includegraphics[scale=0.6]{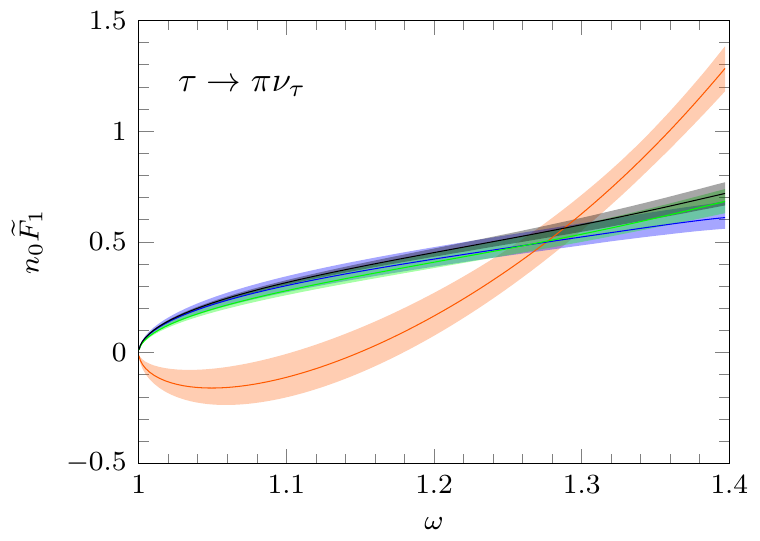}\ 
\includegraphics[scale=0.6]{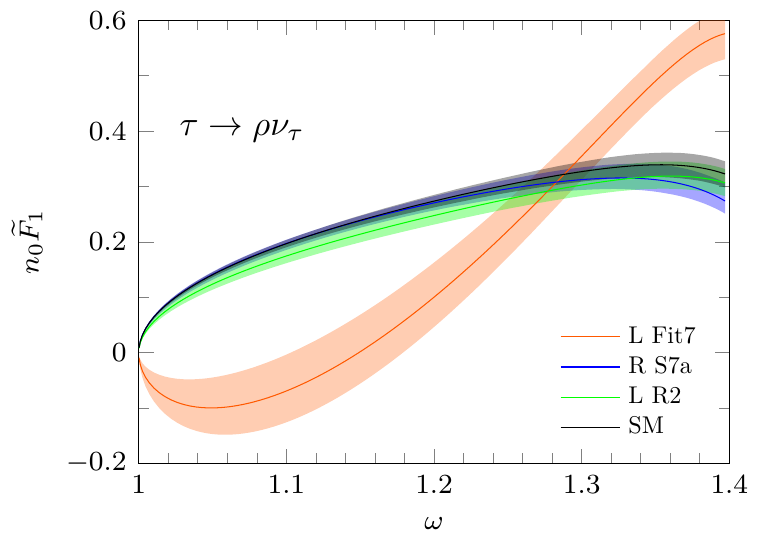}\\
\includegraphics[scale=0.6]{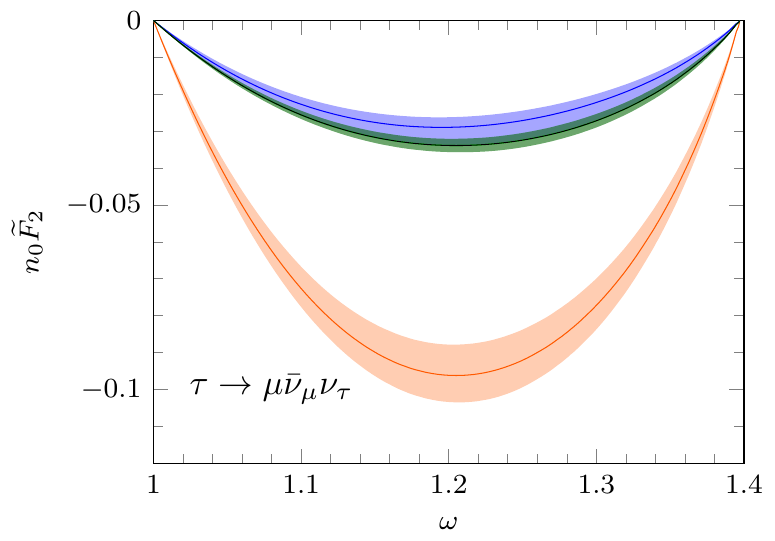}\
\includegraphics[scale=0.6]{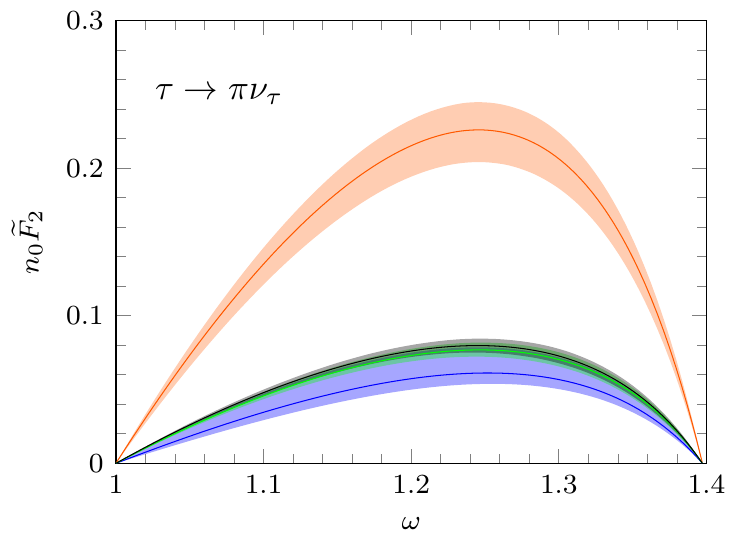}\ 
\includegraphics[scale=0.6]{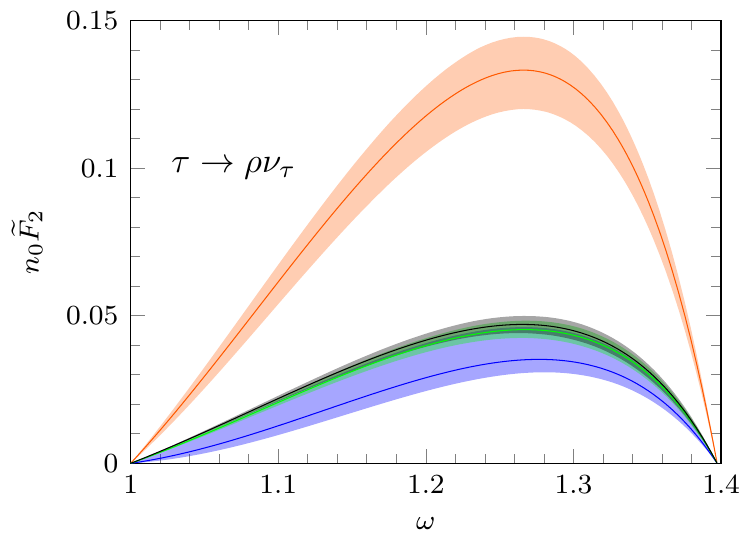}
\caption{ Same as Fig.~\ref{fig:dsstardsn0fs} for the $\bar B_s\to D_s\tau^-(\pi^-\nu_\tau,\rho^-\nu_\tau,\mu^-\bar\nu_\mu\nu_\tau)\bar\nu_\tau$ sequential decays. }
\label{fig:dsn0fs}
\end{center}
\end{figure}

In Figs.~\ref{fig:dsstardsn0fs} and \ref{fig:dsn0fs}, we give the products
$n_0(\omega)\widetilde F_{1,2}^d(\omega)$ (Eqs.~\eqref{eq:F1} and \eqref{eq:F2}) that can be obtained from the measurement of the double differential decay width $d\Gamma_d/(d\omega\, d\cos\theta_d)$ corresponding to the $\bar B_s\to D_s^{(*)}\tau^-(\pi^-\nu_\tau, \rho^-\nu_\tau, \mu^-\bar\nu_\mu\nu_\tau)\bar\nu_\tau$ sequential decays\footnote{The spin analyzing power makes the pion tau-decay mode a better candidate than the leptonic or rho modes for the extraction of information on the spin and spin-angular asymmetries (see discussion of Eq.~(2.11) of Ref.~\cite{Penalva:2022vxy}).}. In most cases, with the main exception being the $\tau\to \rho\nu_\tau$ decay mode for the $\bar B_s\to D_s^*$ decay, the predictions from the L Fit 7 model are clearly distinguishable from the ones obtained in the SM and the other two NP scenarios. The SM and the latter two NP models give results that agree within errors.

\begin{figure}
\begin{center}
\includegraphics[scale=0.6]{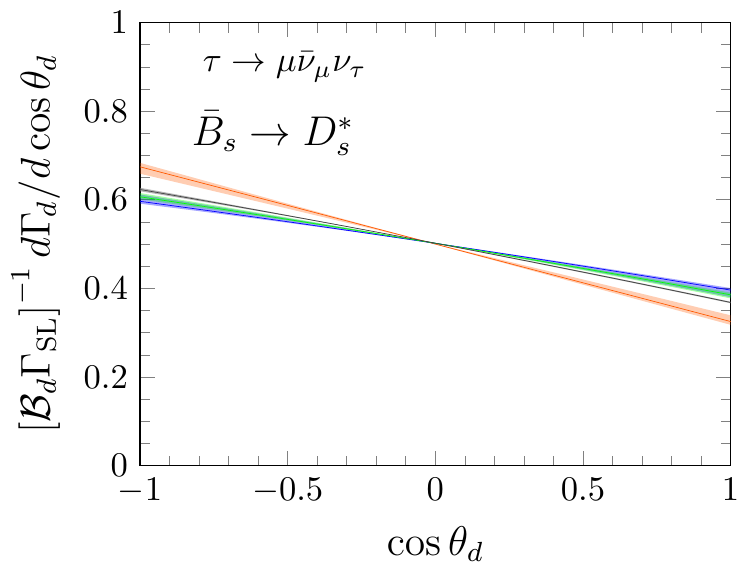}\
\includegraphics[scale=0.6]{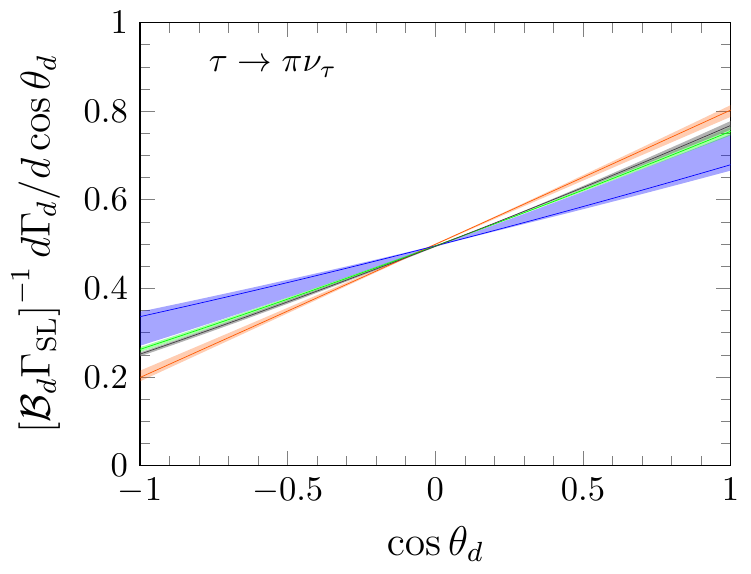}\
\includegraphics[scale=0.6]{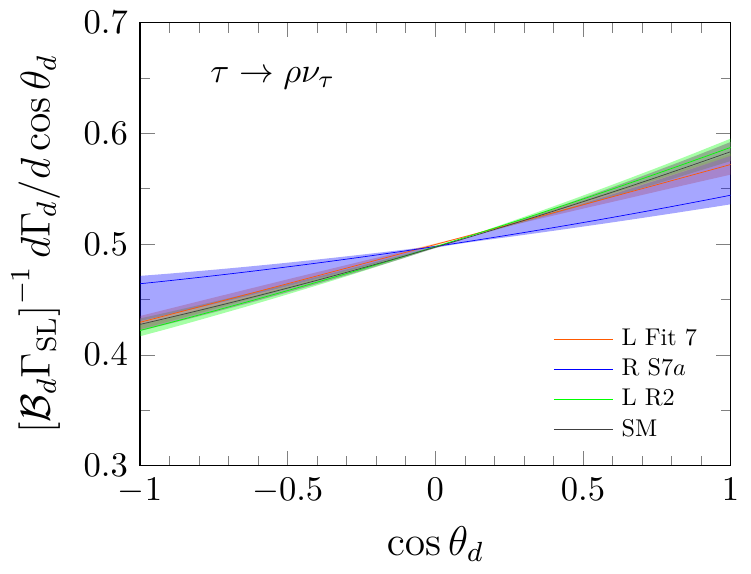}\\
\includegraphics[scale=0.6]{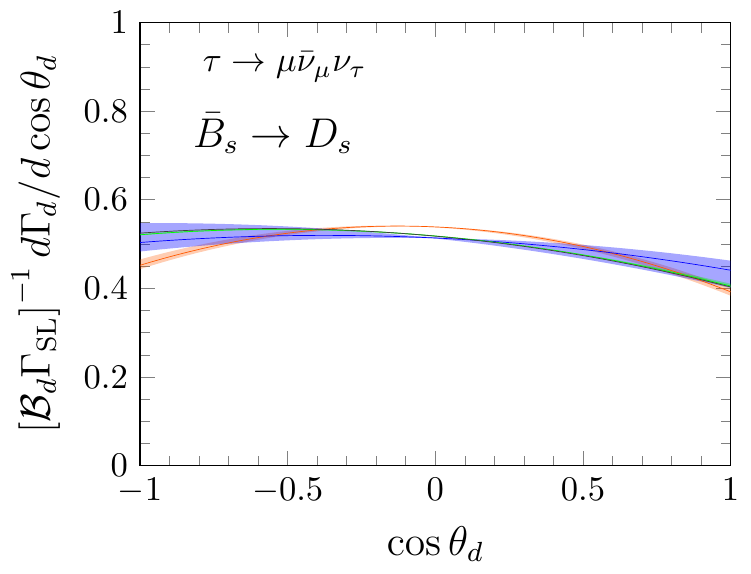}\
\includegraphics[scale=0.6]{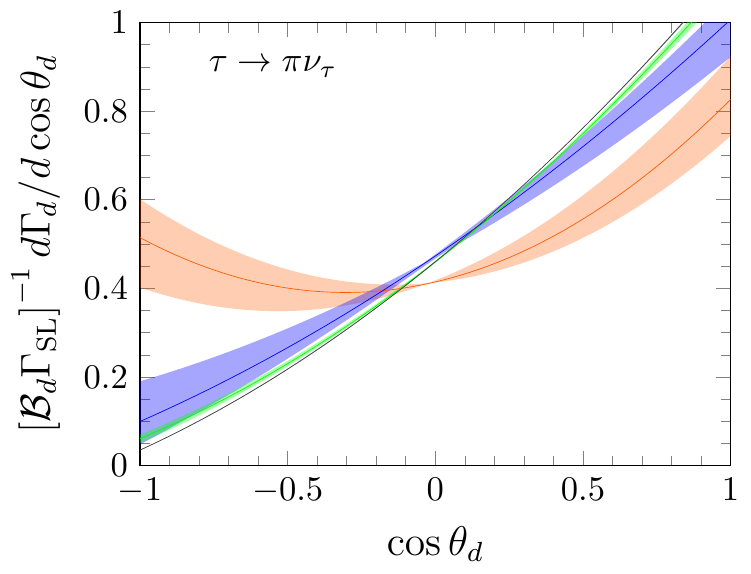}\ 
\includegraphics[scale=0.6]{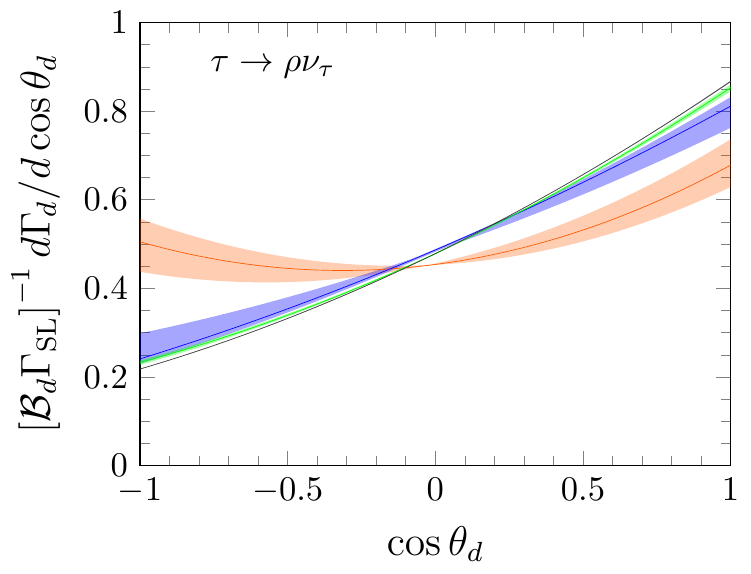}
\caption{ The $\omega$-integrated $d\Gamma_d/d\cos\theta_d$ distributions for the $\bar B_s\to D_s^{*}\tau^-(\pi^-\nu_\tau,\rho^-\nu_\tau,\mu^-\bar\nu_\mu\nu_\tau)\bar\nu_\tau$ (top) and $\bar B_s\to D_s\tau^-(\pi^-\nu_\tau,\rho^-\nu_\tau,\mu^-\bar\nu_\mu\nu_\tau)\bar\nu_\tau$ (bottom) sequential decays. Units of $[{\cal B}_d\Gamma_{\rm SL}]$.}
\label{fig:dstardsang}
\end{center}
\end{figure}

A similar situation is seen in Fig.~\ref{fig:dstardsang}, where we display the normalized $[{\cal B}_d\Gamma_{\rm SL}]^{-1}d\Gamma_d/d\cos\theta_d$ angular distribution for the 
$\bar B_s\to D_s^{(*)}\tau^-(\pi^-\nu_\tau,\rho^-\nu_\tau,\mu^-\bar\nu_\mu\nu_\tau)\bar\nu_\tau$ sequential decays. Again, with the exception of the $\rho$ channel for the $
\bar B_s\to D_s^*$ decay, we see that the  L Fit 7 NP scenario of Ref.~\cite{Murgui:2019czp} can be distinguished from the SM and the other two NP scenarios. This is most clearly seen for forward and backward angles of the pion and rho mesons from the hadronic $\tau$-decay modes in the 
parent $\bar B_s\to D_s$ semileptonic decay. As for the R S7a scenario of Ref.~\cite{Mandal:2020htr} and L $R_2$ Fit of
Ref.~\cite{Shi:2019gxi}, their corresponding distributions are compatible with the SM and among themselves within errors. In fact, for the L $R_2$ model, the central values are very close to the SM ones. 
These behaviors derive from the ones seen for $\widetilde F^d_{12}(\omega)$ in Figs.~\ref{fig:dsstardsn0fs} and \ref{fig:dsn0fs} and they are also seen in  
the corresponding $\widehat F^d_{1,2}$ coefficients  that we give in Tables~\ref{tab:hatfsDsDsstar1} and \ref{tab:hatfsDsDsstar2} for the leptonic and two hadronic $\tau$-decay channels, respectively. These latter coefficients are obtained after integrating over $\omega$ the $\widetilde F^d_{1,2}(\omega)$ functions, as indicated in Eq.~\eqref{eq:Gcos}, and depend on the tau-decay mode.
For L Fit 7, we generally find that one coefficient, or both, is always very different from SM and other NP model values. For the R S7a scenario, they are compatible with SM, within errors, and  they are very close to the SM ones in the L $R_2$ case.

\begin{table}[t]
\begin{center}
\begin{tabular}{c|c|cc}\hline\hline\tstrut
&& $\widehat F_{1}^{\mu}$& $\widehat F_{2}^{\mu}$\\ \hline \tstrut
&SM &$-0.0608^{+0.0006}_{-0.0005}$&$-0.0360\pm0.0006$\\ \tstrut
$\bar B_s\to D_s$&L fit 7&$-0.030^{+0.008}_{-0.011}$&$-0.0777^{+0.0021}_{-0.0005}$\\ \tstrut
&R fit S7a&$-0.03^{+0.02}_{-0.04}$&$-0.028\pm0.003$\\  \tstrut
&L $R_2$&$-0.0579^{+0.0024}_{-0.0018}$&$-0.0368^{+0.0022}_{-0.0013}$
\\ \hline \tstrut
&SM &$-0.128^{+0.003}_{-0.002}$&$-0.0042\pm0.0010$\\ \tstrut
$\bar B_s\to D^*_s$&L Fit 7&$-0.175^{+0.014}_{-0.008}$&$-0.0001\pm0.0011$\\ \tstrut
&R S7a&$-0.100^{+0.005}_{-0.016}$&$-0.0036^{+0.0010}_{-0.0017}$\\\tstrut
&L $R_2$&$-0.111^{+0.004}_{-0.007}$&$-0.0038\pm0.0010$\\  \hline\hline
\end{tabular}
\caption{ Predictions for the angular moments $\widehat F^{\mu}_{1,\,2}$ 
    for the
    $\bar B_s\to D^{(*)}_s\tau(\mu\bar\nu_\mu\nu_\tau)
    \bar\nu_\tau$ sequential decay evaluated in the SM and the same NP scenarios 
    considered in Table~\ref{tab:ratiosDs}.}
   \label{tab:hatfsDsDsstar1}
   \end{center}
\end{table}
\begin{table}[t]
\begin{center}
\begin{tabular}{c|c|cc|cc}\hline\hline\tstrut
&& $\widehat F_{1}^\pi$&$\widehat F_{2}^\pi$
&$\widehat F_{1}^\rho$&$\widehat F_{2}^\rho$\\ \hline \tstrut
&SM &$0.5443^{+0.0013}_{-0.0015}$&$0.0794\pm0.0014$
&$0.3247^{+0.0007}_{-0.0008}$&$0.0428\pm0.0008$\\ \tstrut
$\bar B_s\to D_s$&L fit 7&$0.16^{+0.11}_{-0.08}$&$0.171^{+0.002}_{-0.007}$
&$0.09^{+0.06}_{-0.05}$&$0.0918^{+0.0016}_{-0.0046}$\\ \tstrut
&R fit S7a&$0.45^{+0.05}_{-0.09}$&$0.053^{+0.012}_{-0.007}$
&$0.285^{+0.015}_{-0.053}$&$0.026^{+0.008}_{-0.004}$\\\tstrut  
&L $R_2$&$0.519^{+0.006}_{-0.011}$&$0.080^{+0.003}_{-0.005}$
&$0.310^{+0.003}_{-0.006}$&$0.0431^{+0.0013}_{-0.0024}$\\
\hline \tstrut
&SM &$0.258\pm0.006$&$0.010^{+0.003}_{-0.002}$
&$0.078\pm0.007$&$0.0055\pm0.0014$\\ \tstrut
$\bar B_s\to D^*_s$&L fit 7&$0.302^{+0.009}_{-0.016}$&$0.001\pm0.003$
&$0.071^{+0.007}_{-0.006}$&$0.0006\pm0.0016$\\ \tstrut
&R fit S7a&$0.171^{+0.067}_{-0.012}$&$0.008^{+0.004}_{-0.002}$
&$0.040^{+0.035}_{-0.007}$&$0.0042^{+0.0024}_{-0.0012}$\\ \tstrut
&L $R_2$&$0.246\pm0.007$&$0.009^{+0.003}_{-0.002}$
&$0.083^{+0.006}_{-0.007}$&$0.0047\pm0.0014$\\ \hline\hline
\end{tabular}
    \caption{ Predictions for the angular moments $\widehat F^{\pi,\rho}_{1,\,2}$ 
    for the
    $\bar B_s\to D^{(*)}_s\tau(\pi\nu_\tau,\,\rho\nu_\tau)
    \bar\nu_\tau$ sequential decays evaluated in the SM and the same NP scenarios 
    considered in Table~\ref{tab:ratiosDs}.}
   \label{tab:hatfsDsDsstar2}
   \end{center}
\end{table}

\begin{figure}
\begin{center}
\includegraphics[scale=0.6]{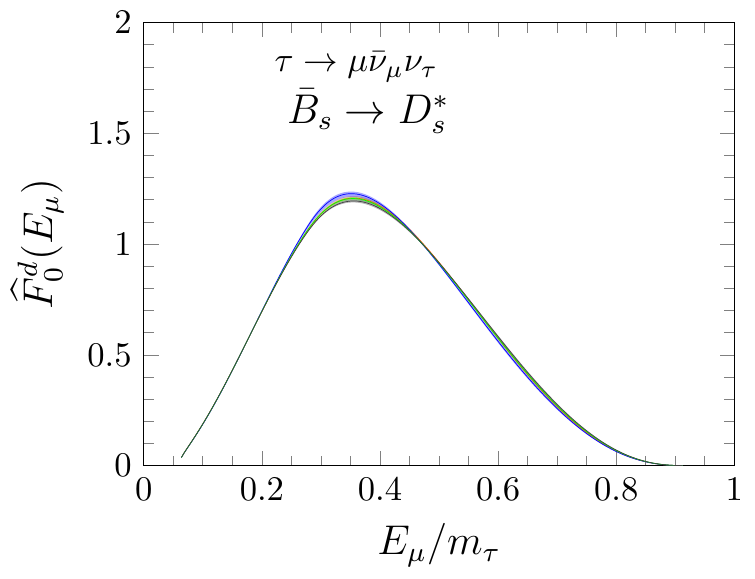}\
\includegraphics[scale=0.6]{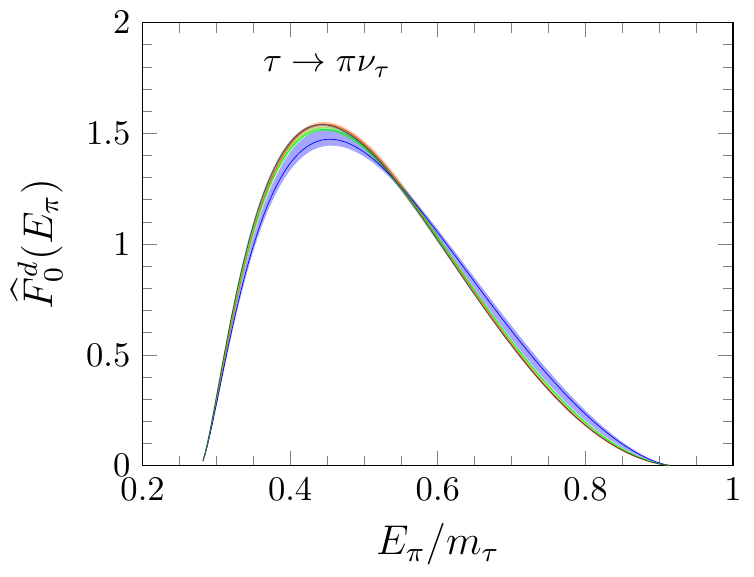}\
\includegraphics[scale=0.6]{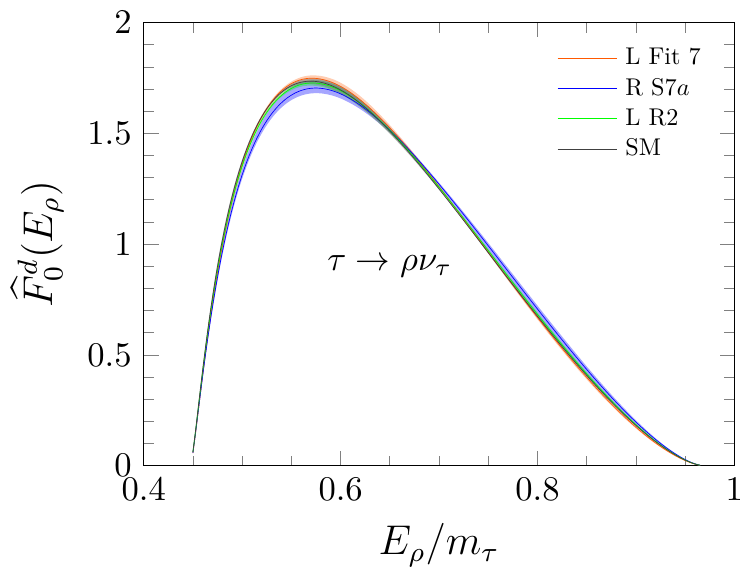}\\
\includegraphics[scale=0.6]{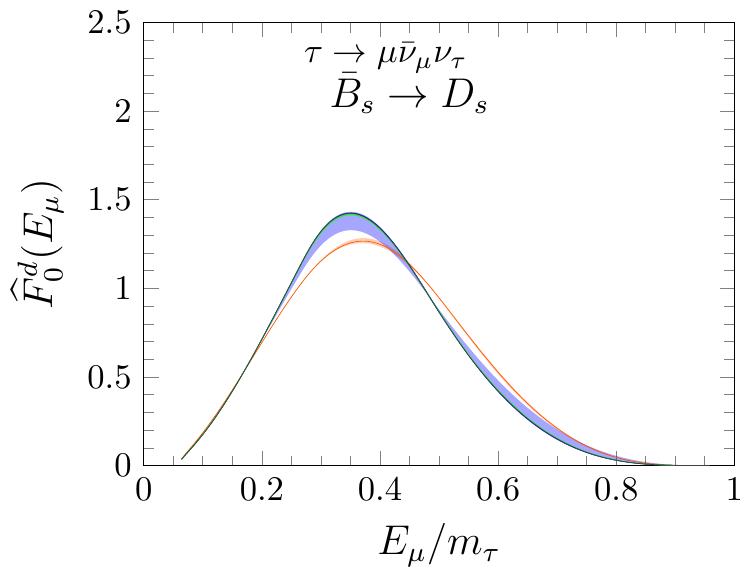}\
\includegraphics[scale=0.6]{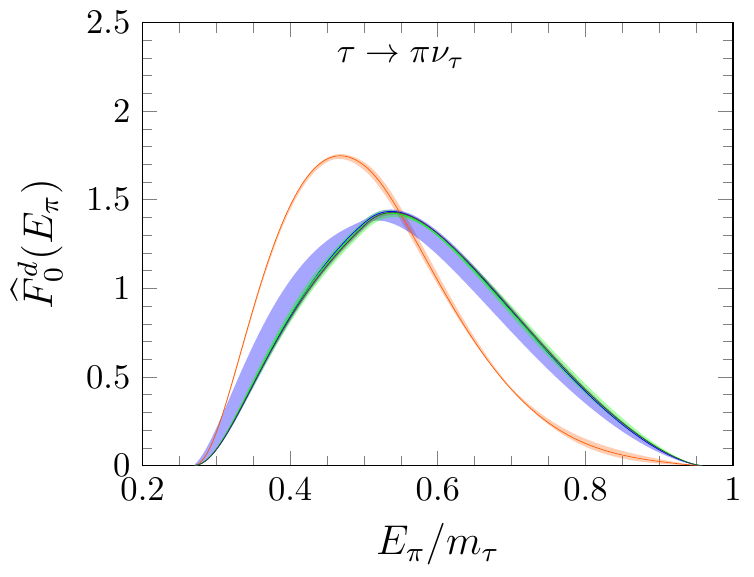}\ 
\includegraphics[scale=0.6]{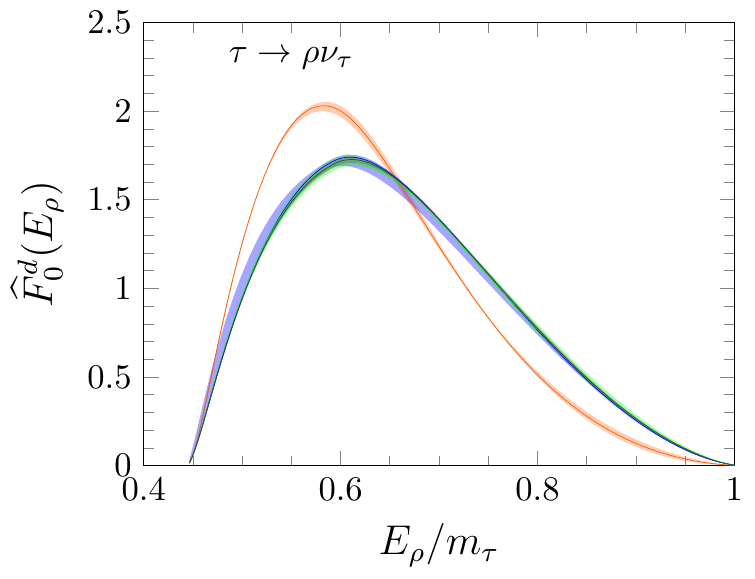}
\caption{ $\widehat F^d_0$ distribution (Eq.~(\ref{eq:ener})) for the $\bar B_s\to D_s^{(*)}\tau^-(\pi^-\nu_\tau,\rho^-\nu_\tau,\mu^-\bar\nu_\mu\nu_\tau)\bar\nu_\tau$ sequential decays.}
\label{fig:dstardsener}
\end{center}
\end{figure}

Finally, in Fig~\ref{fig:dstardsener}, we present the results for the dimensionless distribution
\be
\widehat F^d_0(E_d)=\frac{m_\tau}{2{\cal B}_d\Gamma_{\rm SL}}\frac{d\Gamma_d}{dE_d},
\label{eq:ener}
\ee
which contains all the relevant information on the $d\Gamma_d/dE_d$ energy differential decay width. For all three tau-decay channels considered. It is normalized  as
\be
\frac{1}{m_\tau}\int_{E_d^{\rm min}}^{E_d^{\rm min}} dE_d \widehat F^d_0(E_d)
= \frac12,
\ee
but  its energy dependence is still affected by the CM $\tau$ longitudinal polarization 
$\langle P^{\rm CM}_L\rangle(\omega)$. However, as seen in Fig~\ref{fig:dstardsener}, for the $\bar B_s\to D^*_s$ parent decay,  all NP scenarios considered are compatible with SM predictions, and among themselves, within uncertainties, while for the $\bar B_s\to D_s$, the distribution obtained from the L Fit 7 NP model of Ref.~\cite{Murgui:2019czp} can be distinguished from all other predictions.

\section{Summary}
\label{sec:summary}
We have used the results of the lattice evaluation of the SM form factors for the
$\bar B_s\to D_s$~\cite{McLean:2019qcx}  and the SM and tensor form factors for the $\bar B_s\to D_s^{*}$~\cite{Harrison:2023dzh} semileptonic decays, together with their NLO HQET expansions in Ref.~\cite{Bernlochner:2017jka}, to 
obtain in addition the scalar and pseudoscalar form factors of both transitions and the $\bar B_s\to D_s$ tensor matrix element, all of them  also needed for an analysis of NP effects on both semileptonic decays. We have compared results evaluated 
within the SM and three different NP extensions that have been previously used in the study of other CC $b\to c$ transitions. We find effects similar to those
 obtained for the SU(3)-analogue $\bar B\to D^{(*)}$  decays. We have evaluated the
 corresponding ${\cal R}_{D_s}$ and ${\cal R}_{D_s^{*}}$ ratios which, as in the  $\bar B\to D^{(*)}$ case, should be the easiest LFUV observable to measure. We have also analyzed the role that different tau asymmetries in the $\bar B_s\to D_s^{(*)}\tau^-\bar\nu_\tau$ decay could play, 
 not only in establishing the existence of NP, but also in distinguishing between different NP extensions of the SM. We have studied partially integrated  angular and energy distributions of the charged particle produced in the subsequent $\tau^-
 \to\pi^-\nu_\tau,\,\rho^-\nu_\tau,e^-(\mu^-)\bar\nu_{e(\mu)}\nu_\tau$  decays. The latter differential decay widths have a better statistics than the asymmetries themselves and they could also help in establishing the presence of NP beyond the SM.
 
 If NP is responsible for LFUV  it should  show up in $\bar B_s\to D_s^{(*)}$ semileptonic decays at the same level as for the $\bar B\to D^{(*)}$ ones. The analysis of this transition, as well of other CC $b\to c$ mediated decays, could then help in establishing or ruling out LFUV.

\section*{Acknowledgements}
 N.P. thanks Physics and Astronomy  at the University of Southampton for hospitality during the making of this  
  work  and a Generalitat Valenciana grant CIBEFP/2021/32. This research has 
 been supported  by   the Spanish Ministerio de Ciencia e Innovaci\'on (MICINN)
and the European Regional Development Fund (ERDF) under contracts PID2020-112777GB-I00 and 
PID2019-105439GB-C22, the EU STRONG-2020 project under the program H2020-INFRAIA-2018-1, 
grant agreement no. 824093 and by  Generalitat Valenciana under contract PROMETEO/2020/023. 

\appendix
\section{Mean values and covariance matrices of the  $a_i^{F}$ coefficients  in Eqs.~(\ref{eq:f0p_newpar}),(\ref{eq:def2z}) and (\ref{eq:newparDsstar}).} 
 As discussed in the main text,  we have changed the parametrizations  in Refs.~\cite{McLean:2019qcx,Harrison:2023dzh} and adopted  new ones in order to facilitate  the fitting of the form factors to their HQSS expressions.  Statistical details of the new coefficients   are collected here in Tables \ref{tab:f0fp}--\ref{tab:ht123ha123v}. 
  For each entry in the tables below, we provide three significant digits but neglect order $10^{-5}$ or smaller.

\begin{table}[h]
\begin{ruledtabular}
\begin{center}
\begin{tabular}{cccccc}
 &$a^0_0$& $a^0_1$&$a^0_2$&$a^+_0$&$a^+_1$
\\ \hline \tstrut
&$0.674\pm 0.009$&$-0.238\pm0.218$&$-0.13\pm1.61$&$0.764\pm0.018$&$-3.04\pm0.43$\\\hline \tstrut
$a_0^0$&$1.00$&$0.117$&\hspace{-.25cm}$-0.074$&$0.567$&\hspace{-.25cm}$-0.011$\\
$a_1^0$&&$1.00$&\hspace{-.25cm}$-0.062$&$0.421$&\hspace{-.25cm}$-0.030$\\
$a_2^0$&&&$1.00$&\hspace{-.25cm}$-0.145$&$0.229$\\
$a_0^+$&&&&$1.00$&\hspace{-.25cm}$-0.726$\\
$a_1^+$&&&&&$1.00$\\
\end{tabular}
    \caption{Central values and errors (first row) of the  $a_i^{0,+}$ coefficients 
    of the new $f_{+,0}$
    parametrization  introduced in Eq.~(\ref{eq:f0p_newpar})
     and their corresponding correlation matrix. Note that $a^+_2$ is fixed  
through the condition $f_0(0)=f_+(0)$. }
   \label{tab:f0fp}
   \end{center}
   \end{ruledtabular}
\end{table}
\begin{table}[h]
\begin{ruledtabular}
\begin{center}
\begin{tabular}{ccccc}
 & $h_{A_1}$&$h_{A_2}$&$h_{A_3}$&$h_V$
\\ \hline \tstrut
$a_0$&$0.907\pm0.010$&\hspace{-.25cm}$-0.333\pm 0.129$&\hspace{.25cm}$1.14\pm0.13$&\hspace{-.05cm}$1.25\pm0.04$\\
$a_1$&\hspace{-.25cm}$-1.01\pm0.09$&\hspace{-.25cm}$-0.066\pm0.639$&$-0.649\pm0.539$&\hspace{-.25cm}$-1.51\pm0.26$\\
$a_2$&$0.379\pm0.435$&$0.065\pm0.948$&$-0.200\pm0.853$&$0.507\pm0.752$\\
$a_3$&$0.275\pm0.817$&$0.007\pm0.989$&$-0.101\pm0.961$& $0.373\pm0.938$\\
\end{tabular}
    \caption{Central values and errors of the $a_i^F$ coefficients of the new  parametrization ($(\omega-1)$ expansion)  introduced in Eq.~(\ref{eq:newparDsstar}) for the $h_{A_1}, h_{A_2}, h_{A_3}$ and $h_V$ form factors. }
   \label{tab:ha123vdstar_newpar}
   \end{center}
   \end{ruledtabular}
\end{table}
\begin{table}[h]
\begin{ruledtabular}
\begin{center}
\begin{tabular}{cccc}
 & $h_{T_1}$&$h_{T_2}$&$h_{T_3}$
\\ \hline \tstrut
$a_0$&$0.933\pm0.015$&\hspace{-.25cm}$-0.152\pm0.045$&\hspace{-.25cm}$-0.027\pm0.149$\\
$a_1$&\hspace{-.25cm}$-1.09\pm0.10$&$0.231\pm0.286$&$0.106\pm0.661$\\
$a_2$&$0.512\pm0.473$&$0.449\pm0.892$&$0.019\pm0.939$\\
$a_3$&$0.138\pm0.858$&$0.244\pm0.979$&$0.005\pm0.988$\\
\end{tabular}
    \caption{Central values and errors of the $a_i^F$ coefficients of the new  parametrization ($(\omega-1)$-expansion)  introduced in Eq.~(\ref{eq:newparDsstar}) for the $h_{T_1}, h_{T_2}$ and $h_{T_3}$  form factors. }
   \label{tab:ht123dstar_newpar}
   \end{center}
   \end{ruledtabular}
\end{table}
\begin{table}[h]
\begin{ruledtabular}
\begin{center}
\begin{tabular}{cccccccccccccc}
 & $a_0^{{A_1}}$&$a_1^{{A_1}}$&$a_2^{{A_1}}$&$a_3^{{A_1}}$
& $a_0^{{A_2}}$&$a_1^{{A_2}}$&$a_2^{{A_2}}$&$a_3^{{A_2}}$& $a_0^{{3}}$&$a_1^{{A_3}}$&$a_2^{{A_3}}$&$a_3^{{A_3}}$\\ \hline \tstrut
$a_0^{{A_1}}$&   1.00  & $-0.160$&  \hspace{.25cm}0.0523 & $-0.0128$  & \hspace{.25cm}0.0144 & $-0.0177$ & $-0.0037$ & $-0.0008$  & \hspace{.25cm}0.0469  &$-0.0055$ &  \hspace{.25cm}0.0039  & \hspace{.25cm}0.0041 \\
$a_1^{{A_1}}$&   &   1.00&  $-0.6252$&   \hspace{.25cm}0.2511 & $-0.2650$&   \hspace{.25cm}0.1381&   \hspace{.25cm}0.0124&   \hspace{.25cm}0.0011 &  \hspace{.25cm}0.3039 & $-0.1410$&  $-0.0879$ & $-0.0342$\\
$a_2^{{A_1}}$&   &&   1.00&  $-0.7071$ &  \hspace{.25cm}0.0971&  $-0.1797$&  $-0.0297$&  $-0.0108$ & $-0.1326$ &  \hspace{.25cm}0.2761&   \hspace{.25cm}0.0374 & $-0.0025$ \\
$a_3^{{A_1}}$&   &&&   1.00 &  \hspace{.25cm}0.0646&  $-0.0750$&   \hspace{.25cm}0.0113&   \hspace{.25cm}0.0141 & $-0.0550$ &  \hspace{.25cm}0.0006&   \hspace{.25cm}0.1622 &  \hspace{.25cm}0.0921\\
$a_0^{{A_2}}$&  &&&&  1.00&  $-0.3542$&  $-0.0309$&  $-0.0108$ & $-0.8365$ &  \hspace{.25cm}0.4847&  $-0.0390$ & $-0.0365$ \\
$a_1^{{A_2}}$&   &&&&&   1.00&  $-0.2141$&  $-0.0479$ &  \hspace{.25cm}0.1783 & $-0.5943$&   \hspace{.25cm}0.0635 &  \hspace{.25cm}0.0510\\
$a_2^{{A_2}}$&   &&&&&&1.00&  $-0.0478$ &  \hspace{.25cm}0.0824 & $-0.0981$&  $-0.1070$ & $-0.0465$\\
$a_3^{{A_2}}$& &&&&&&&   1.00 &  \hspace{.25cm}0.0310 & $-0.0212$&  $-0.0592$ & $-0.0289$ \\
$a_0^{{A_3}}$& &&&&&&&&  1.00 & $-0.5695$&   \hspace{.25cm}0.1206 &  \hspace{.25cm}0.0622 \\
$a_1^{{A_3}}$&  &&&&&&&&&  1.00&  $-0.4675$ & $-0.1211$\\
$a_2^{{A_3}}$&  &&&&&&&&&&   1.00 & $-0.1682$\\
$a_3^{{A_3}}$&  &&&&&&&&&&&  1.00\\

\end{tabular}
    \caption{Correlation matrix for the $(w-1)$-expansion coefficients of   the $h_{A_1},\,h_{A_2}$ and $h_{A_3}$ form factors. }
   \label{tab:ha123}
   \end{center}
   \end{ruledtabular}
\end{table}
\begin{table}[h]
\begin{ruledtabular}
\begin{center}
\begin{tabular}{ccccc}
 & $a_0^{{V}}$&$a_1^{{V}}$&$a_2^{{V}}$&$a_3^{{V}}$\\ \hline \tstrut
 $a_0^{{V}}$ &1.00&  $-0.526$&   \hspace{.25cm}0.251  & \hspace{.25cm}0.0672\\
 $a_1^{{V}}$ &&   1.00&  $-0.720$&  $-0.0557$\\
 $a_2^{{V}}$ &&&  1.00 & $-0.307$\\
 $a_3^{{V}}$ &&&&   1.00\\
\end{tabular}
    \caption{Correlation matrix for the $(w-1)$-expansion coefficients of the $h_{V}$ form factor. }
   \label{tab:hvhv}
   \end{center}
   \end{ruledtabular}
\end{table}
\begin{table}[h]
\begin{ruledtabular}
\begin{center}
\footnotesize{\begin{tabular}{cccccccccccccc}
 & $a_0^{{A_1}}$&$a_1^{{A_1}}$&$a_2^{{A_1}}$&$a_3^{{A_1}}$
& $a_0^{{A_2}}$&$a_1^{{A_2}}$&$a_2^{{A_2}}$&$a_3^{{A_2}}$& $a_0^{{A_3}}$&$a_1^{{A_3}}$&$a_2^{{A_3}}$&$a_3^{{A_3}}$\\ \hline \tstrut
$a_0^{{V}}$&   \hspace{.25cm}0.0695&  $-0.0034$&  $-0.0023$&  $-0.0029$&   \hspace{.25cm}0.0187&  $-0.0044$&  $-0.0054$&  $-0.0024$&  $-0.0022$&   \hspace{.25cm}0.0072&  $-0.0024$&  $-0.0011$\\
 $a_1^{{V}}$ & \hspace{.25cm}0.0151&   \hspace{.25cm}0.0181&   \hspace{.25cm}0.0034&  $-0.0045$&   \hspace{.25cm}0.0149&   \hspace{.25cm}0.0137&   \hspace{.25cm}0.0015&   \hspace{.25cm}0.0003&  $-0.0120$&   \hspace{.25cm}0.0054&   \hspace{.25cm}0.0023&   \hspace{.25cm}0.0013\\
$a_2^{{V}}$&  $-0.0094$&  $-0.0045$&   \hspace{.25cm}0.0107&   \hspace{.25cm}0.0057&  $-0.0018$&   \hspace{.25cm}0.0137&   \hspace{.25cm}0.0093&   \hspace{.25cm}0.0041&  $-0.0043$&  $-0.0021$&   \hspace{.25cm}0.0060&   \hspace{.25cm}0.0031\\
 $a_3^{{V}}$ &$-0.0032$&  $-0.0022$&   \hspace{.25cm}0.0040&   \hspace{.25cm}0.0028& $ -0.0017$&   \hspace{.25cm}0.0051&   \hspace{.25cm}0.0038&   \hspace{.25cm}0.0017&  $-0.0009$&  $-0.0009$&   \hspace{.25cm}0.0023&   \hspace{.25cm}0.0011\\
\end{tabular}}
    \caption{Correlation matrix for the $(w-1)$-expansion coefficients of $h_V$ and   $h_{A_1},\,h_{A_2}$ and $h_{A_3}$  form factors. }
   \label{tab:ha123hv}
   \end{center}
   \end{ruledtabular}
\end{table}

\begin{table}[h]
\begin{ruledtabular}
\begin{center}
\begin{tabular}{ccccccccccccc}
 & $a_0^{{T_1}}$&$a_1^{{T_1}}$&$a_2^{{T_1}}$&$a_3^{{T_1}}$
& $a_0^{{T_2}}$&$a_1^{{T_2}}$&$a_2^{{T_2}}$&$a_3^{{T_2}}$& $a_0^{{T_3}}$&$a_1^{{T_3}}$&$a_2^{{T_3}}$&$a_3^{{T_3}}$\\ \hline \tstrut
$a_0^{{T_1}}$&1.00&  $-0.230$&   \hspace{.25cm}0.0740  &$-0.0056$ &  \hspace{.25cm}0.0593 & $-0.0300$ & \hspace{.25cm}0.0044 &  \hspace{.25cm}0.0025  & \hspace{.25cm}0.0252 &$-0.0101$ &  \hspace{.25cm}0.0030 &  \hspace{.25cm}0.0027\\
$a_1^{{T_1}}$&&  1.00 &$-0.633$&  \hspace{.25cm}0.208 &  \hspace{.25cm}0.0328  & \hspace{.25cm}0.0912 & $-0.0629$ & $-0.0285$ &  \hspace{.25cm}0.304&  $-0.154$ & $-0.0689$ & $-0.0272$\\
$a_1^{{T_1}}$&&&1.00&  $-0.645$&  $-0.0483$&  \hspace{.25cm}0.0569 &  \hspace{.25cm}0.112 &  \hspace{.25cm}0.0456&  $-0.136$&   \hspace{.25cm}0.293&   \hspace{.25cm}0.0586 &  \hspace{.25cm}0.0134\\
$a_1^{{T_1}}$&&&&1.00&   0.0025 & $-0.0379$&   \hspace{.25cm}0.0511&   \hspace{.25cm}0.0306 & $-0.0511$&   \hspace{.25cm}0.0546 &  \hspace{.25cm}0.0738&   \hspace{.25cm}0.0389\\
$a_0^{{T_2}}$&&&&&1.00&  $-0.471$&   \hspace{.25cm}0.130&   \hspace{.25cm}0.0347 &  \hspace{.25cm}0.0446&  $-0.0400$&  $-0.0140$&  $-0.0051$\\
$a_1^{{T_2}}$&&&&&&1.00&  $-0.589$&  $-0.0819$&   \hspace{.25cm}0.0828 &  \hspace{.25cm}0.0552  & \hspace{.25cm}0.0017 & $-0.0022$\\
$a_1^{{T_2}}$&&&&&&&1.00&  $-0.0957$&  $-0.0427$ &  \hspace{.25cm}0.135 &  \hspace{.25cm}0.0491 &  \hspace{.25cm}0.0192\\
$a_1^{{T_2}}$&&&&&&&&1.00&  $-0.0212$&   \hspace{.25cm}0.0615 &  \hspace{.25cm}0.0239  & \hspace{.25cm}0.0097\\
$a_0^{{T_3}}$&&&&&&&&&1.00&  $-0.573$&   \hspace{.25cm}0.0128  & \hspace{.25cm}0.0227\\
$a_1^{{T_3}}$&&&&&&&&&&1.00 & $-0.336$ & $-0.111$\\
$a_1^{{T_3}}$&&&&&&&&&&&1.00 & $-0.0548$\\
$a_1^{{T_3}}$&&&&&&&&&&&&1.00\\
\end{tabular}
    \caption{Correlation matrix for the $(w-1)$-expansion coefficients of the $h_{T_1},\,h_{T_2}$ and $h_{T_3}$ form factors. }
   \label{tab:ht123ht123}
   \end{center}
   \end{ruledtabular}
\end{table}

\begin{table}[h]
\begin{ruledtabular}
\begin{center}
\footnotesize{\begin{tabular}{ccccccccccccc}
 & $a_0^{{T_1}}$&$a_1^{{T_1}}$&$a_2^{{T_1}}$&$a_3^{{T_1}}$
& $a_0^{{T_2}}$&$a_1^{{T_2}}$&$a_2^{{T_2}}$&$a_3^{{T_2}}$& $a_0^{{T_3}}$&$a_1^{{T_3}}$&$a_2^{{T_3}}$&$a_3^{{T_3}}$\\ \hline \tstrut
$a_0^{{A_1}}$&  \hspace{.05cm}0.289 &$-0.0368$& \hspace{.25cm}0.0078 & $-0.0004$ &$-0.0098$ & $-0.0112$ & $-0.0003$ & $-0.0004$ &  \hspace{.25cm}0.0005 & $-0.0001$ & 
  \hspace{.25cm}0.0018 &  \hspace{.25cm}0.0010\\
$a_1^{{A_1}}$& $-0.0123$  & \hspace{.05cm}0.210&  $-0.0882$&  $-0.0290$ &  \hspace{.25cm}0.0170 &  \hspace{.25cm}0.0175  & \hspace{.25cm}0.0028 &  \hspace{.25cm}0.0022  & \hspace{.25cm}0.0534 & $-0.0048$ &
 $-0.0028$ & $-0.0015$\\
$a_2^{{A_1}}$& $-0.0018$ & $-0.0940$ &  \hspace{.05cm}0.185&  $-0.0099$ & $-0.0121$ &  \hspace{.25cm}0.0043&   \hspace{.25cm}0.0040 &  \hspace{.25cm}0.0016 & $-0.0047$ &  \hspace{.25cm}0.0373 & 
 \hspace{.25cm}0.0256 &  \hspace{.25cm}0.0126\\
$a_3^{{A_1}}$& $-0.0016$ & $-0.0243$ & $-0.0214$ &  \hspace{.25cm}0.0842 &  \hspace{.25cm}0.0008 &  \hspace{.25cm}0.0083 & $-0.0120$ & $-0.0075$ & $-0.0006$ &  \hspace{.25cm}0.0219&  $-0.0178$&  $-0.0116$\\
$a_0^{{A_2}}$&  \hspace{.25cm}0.0095 & $-0.0794$&   \hspace{.25cm}0.0079 &  \hspace{.25cm}0.0066 &  \hspace{.25cm}0.0053 & $-0.0095$ & $-0.0054$ & $-0.0034$ & $-0.189$ &  \hspace{.25cm}0.0284 &  \hspace{.25cm}0.0124&   \hspace{.25cm}0.0052\\
$a_1^{{A_2}}$&  \hspace{.25cm}0.0000 &  \hspace{.25cm}0.0200 & $-0.0534$&  $-0.0228$ & $-0.0012$ & $-0.0048$ &  \hspace{.25cm}0.0113  & \hspace{.25cm}0.0059 & $-0.0112$ & $-0.110$& $ -0.0261$&  $-0.0079$\\
$a_2^{{A_2}}$& $-0.0019$ & $-0.0079$  & \hspace{.25cm}0.0034  & \hspace{.25cm}0.0051 & $-0.0027$&  $-0.0012$ &  \hspace{.25cm}0.0042 &  \hspace{.25cm}0.0020 &  \hspace{.25cm}0.0102&  $-0.0312$&  $-0.0100$&  $-0.0037$\\
$a_3^{{A_2}}$& $-0.0009$ & $-0.0050$&   \hspace{.25cm}0.0040&   \hspace{.25cm}0.0040 & $-0.0013$&  $-0.0003$&   \hspace{.25cm}0.0016 &  \hspace{.25cm}0.0007 &  \hspace{.25cm}0.0052&  $-0.0111$&  $-0.0041$&  $-0.0016$\\
$a_0^{{A_3}}$&   \hspace{.25cm}0.0179 &  \hspace{.25cm}0.0881&  $-0.0170$&  $-0.0105$ &  \hspace{.25cm}0.0023&   \hspace{.25cm}0.0100&   \hspace{.25cm}0.0052 &  \hspace{.25cm}0.0034 &  \hspace{.05cm}0.205&  $-0.0254$&  $-0.0119$&  $-0.0050$\\
$a_1^{{A_3}}$&  \hspace{.25cm}0.0004 & $-0.0160$&   \hspace{.25cm}0.0690&   \hspace{.25cm}0.0316 &  \hspace{.25cm}0.0011&   \hspace{.25cm}0.0125&  $-0.0105$ & $-0.0061$ & $-0.0541$&   \hspace{.05cm}0.119&   \hspace{.25cm}0.0260&	\hspace{.25cm}0.0074\\
$a_2^{{A_3}}$& $-0.0051$&  $-0.0287$&   \hspace{.25cm}0.0593&   \hspace{.25cm}0.0050 & $-0.0047$&   \hspace{.25cm}0.0051&  $-0.0023$ & $-0.0018$ & $-0.0031$&   \hspace{.25cm}0.0178&   \hspace{.25cm}0.0131&	\hspace{.25cm}0.0063\\
$a_3^{{A_3}}$& $-0.0020$&  $-0.0138$&   \hspace{.25cm}0.0270&   \hspace{.25cm}0.0007 & $-0.0026$&   \hspace{.25cm}0.0020&  $-0.0005$ & $-0.0005$ & $-0.0002$&   \hspace{.25cm}0.0029&   \hspace{.25cm}0.0054&	\hspace{.25cm}0.0029\\
$a_0^{{V}}$&  \hspace{.25cm}0.0694 & $-0.0022$&  $-0.0003$&  $-0.0009$ &  \hspace{.25cm}0.0304&   \hspace{.25cm}0.0021&  $-0.0006$ & $-0.0002$ &  \hspace{.25cm}0.0043&   \hspace{.25cm}0.0132&   \hspace{.25cm}0.0036&	\hspace{.25cm}0.0013\\
$a_1^{{V}}$&  \hspace{.25cm}0.0041&   \hspace{.25cm}0.0141&   \hspace{.25cm}0.0027& $ -0.0028$ & $-0.0012$&   \hspace{.25cm}0.0127&  $-0.0025$ & $-0.0014$ & $-0.0106$&   \hspace{.25cm}0.0030&   \hspace{.25cm}0.0033&	\hspace{.25cm}0.0016\\
$a_2^{{V}}$& $-0.0075$&  $-0.0023$ &  \hspace{.25cm}0.0091&   \hspace{.25cm}0.0049 & $-0.0039$&   \hspace{.25cm}0.0018&   \hspace{.25cm}0.0062 &  \hspace{.25cm}0.0026 &  \hspace{.25cm}0.0032&  $-0.0043$&  $-0.0029$&  $-0.0014$\\
$a_3^{{V}}$& $-0.0026$ & $-0.0012$ &  \hspace{.25cm}0.0034&   \hspace{.25cm}0.0025 & $-0.0015$&   \hspace{.25cm}0.0005&   \hspace{.25cm}0.0028 &  \hspace{.25cm}0.0012 &  \hspace{.25cm}0.0018&  $-0.0019$&  $-0.0015$&  $-0.0008$\\

\end{tabular}}
    \caption{Correlation matrix for the $(w-1)$-expansion between  the $h_{T_1},\,h_{T_2}$ and $h_{T_3}$ and the $h_{A_1},\,h_{A_2},\,h_{A_3}$ and $h_V$ coefficients. }
   \label{tab:ht123ha123v}
   \end{center}
   \end{ruledtabular}
\end{table}

\bibliography{B2Dbib}

\end{document}